%% file: bodyorder.tex
\documentclass[reqno]{tran-l}

\input{external/preamble.tex}
\numberwithin{equation}{section}
\numberwithin{theorem}{section}
\usepackage{geometry}
\usepackage{xcolor}
\usepackage{esint}
\newcommand{\cco}[1]{\textcolor{orange}{\tt \footnotesize [CO: #1]}}

\newcommand{\jt}[1]{\textcolor{magenta}{\tt \footnotesize [JT: #1]}}

\newcommand{\asTB}{\textup{(\textbf{\hyperlink{TB}{TB}})}}
\newcommand{\asEP}{\textup{(\textbf{\hyperlink{EP}{EP}})}}

\newcommand{\s}{{\bm u}}

\newcommand{\z}{Z}

\makeatletter
\let\@fnsymbol\@arabic
\makeatother

\makeatletter
\newcommand{\leqnomode}{\tagsleft@true\let\veqno\@@leqno}
\newcommand{\reqnomode}{\tagsleft@false\let\veqno\@@eqno}
\makeatother

\begin{document}



\title
    [Body-Order Approximations of an Electronic Structure Potential Energy Landscape]
    {Rigorous Body-Order Approximations of an Electronic Structure Potential Energy Landscape}

\author{Jack Thomas}
\author{Huajie Chen}
\author{Christoph Ortner}

\maketitle\thispagestyle{empty}

\vspace{-1em}

\begin{abstract}
    We show that the local density of states (LDOS) of a wide class of 
    tight-binding models has a weak body-order expansion. Specifically, 
    we prove that the resulting body-order expansion for analytic observables
    such as the electron density or the energy has an exponential 
    rate of convergence both at finite Fermi-temperature as well as 
    for insulators at zero Fermi-temperature. We discuss potential consequences 
    of this observation for modelling the potential energy landscape, as 
    well as for solving the electronic structure problem.
\end{abstract}
\let\thefootnote\relax\footnote{
    \textit{Date:}~\today.\\
    \textit{2020 Mathematics Subject Classification:} \texttt{65E05}; \texttt{74E15}; \texttt{81V45}; \texttt{81V70}.
    \textit{Addresses:}~\texttt{chen.huajie@bnu.edu.cn}. School of Mathematical Sciences, Beijing Normal University, China.\\
    \texttt{ortner@math.ubc.ca}. Department of Mathematics, University of British Columbia, Vancouver, Canada.\\
    \texttt{j.thomas.1@warwick.ac.uk}. Mathematics Institute, Zeeman Building, University of Warwick, UK.
}

\section{Introduction}
\label{sec:intro}
%
%
%
An atomistic \textit{potential energy landscape} (PEL) is a mapping assigning energies $E(\bm{r})$, or local energy contributions, to atomic structures 
$\bm{r} = \{\bm{r}_\ell\}_{\ell \in\Lambda} \in (\mathbb R^d)^\Lambda$,
where $\Lambda$ is a general (possibly infinite) index set. High-fidelity models are provided by the Born--Oppenheimer PEL associated with {\it ab initio} electronic structure models such as tight-binding, Kohn--Sham density functional theory, Hartree--Fock, or even lower level quantum chemistry models~\cite{ 
   ParrYang1994,
   bk:finnis,
       HohenbergKohn1964, KohnSham1965,
       Thomas1927,
       Hartree1928
}. But even now, the high computational cost of electronic structure models severely limits their applicability in material modelling to thousands of atoms for static and hundreds of atoms for long-time dynamic simulations. 

There is a long and successful history of using surrogate models for the simulation of materials, devised to remain computationally tractable but capture as much detail of the reference \textit{ab initio} PEL as possible. 
Empirical interatomic potentials are purely phenomenological and are able to capture a minimal subset of desired properties of the PEL, severely limiting their transferability~\cite{Stillinger1985-os, Daw1984-vr}. The rapid growth in computational resources, increased both the desire and the possibility to match as much of an {\it ab inito} PEL as possible. 
A continuous increase in the complexity of parameterisations since the 1990s~\cite{Ercolessi1994-ry, Baskes1992-oh, Bazant1997-qb} has over time naturally led to a new generation of ``machine-learned interatomic potentials'' employing universal approximators instead of empirical mechanistic models. 
Early examples include symmetric polynomials~\cite{Braams2009, Shapeev2016-pd}, artificial neural networks~\cite{Behler2007-ng} and kernel methods~\cite{Bartok2010-mv}. A striking case is the Gaussian approximation potential for Silicon~\cite{Bartok2018-fk}, capturing the vast majority of the PEL of Silicon of interest for material applications.

The purpose of the present work is, first, to rigorously evaluate some of the implicit or explicit assumptions underlying this latest class of interatomic potential models, as well as more general models for atomic properties. Specifically, we will identify natural modelling parameters as {\em approximation parameters} and rigorously establish convergence. 
Secondly, our results indicate that nonlinearities are an important feature, highlighting some superior theoretical properties. Finally, unlike existing nonlinear models, we will identify explicit low-dimensional nonlinear parameterisations yet prove that they are systematic. 
In addition to justifying and supporting the development of new models for general atomic properties, our results establish generic properties of \textit{ab initio} models that have broader consequences, e.g. for the study of the mechanical properties of atomistic materials \cite{EhrlacherOrtnerShapeev2016, ChenNazarOrtner2019, ChenOrtnerThomas2019:locality, Thomas2020:scTB}.
The application of our results to the construction and analysis of practical parameterisations (approximation schemes) that exploit our results will be pursued elsewhere.

Our overarching principle is to search for representations of properties of {\it ab initio} models in terms of {\em simple components}, where ``simple'' is of course highly context-specific. To illustrate this point, let us focus on modelling the potential energy landscape (PEL), which motivated this work in the first place. Pragmatically, we require that these {\em simple components} are easier to analyse and manipulate analytically or to fit than the PEL. For many materials (at least as long as Coulomb interaction does not play a role) the first step is to decompose the PEL into site energy contributions, 
\begin{equation}
    \label{eq:site-energy-intro}
    E({\bm r}) = \sum_{\ell \in \Lambda} E_\ell({\bm r}),
\end{equation}
where one assumes that each $E_\ell$ is {\em local}, i.e., it depends only weakly on atoms far away. In previous works we have made this rigorous for the case of tight-binding models of varying complexity~\cite{ChenOrtner16, ChenLuOrtner18, ChenOrtnerThomas2019:locality, Thomas2020:scTB}.
In practise, one may therefore truncate the interaction by admitting only those atoms ${\bm r}_k$ with $r_{\ell k} < r_{\rm cut}$ as arguments. Typical cutoff radii range from 5\AA~to 8\AA, which means that on the order 30 to 100 atoms still make important contributions.  Thus the site energy $E_\ell$ is still an extremely high-dimensional object and short of identifying low-dimensional features it would be practically impossible to numerically approximate it, due to the curse of dimensionality. 

A classical example that illustrates our search for such low-dimensional features is the embedded atom model (EAM)~\cite{Daw1984-vr}, which assigns to each atom $\ell \in \Lambda$ a site energy 
\[
    E^{\rm eam}_\ell({\bm r}) = \sum_{k \neq \ell} \phi(r_{\ell k})
            + F\big( {\textstyle \sum_{k \neq \ell} \rho(r_{\ell k})} \big).
\]
While the site energy $E^{\rm eam}_\ell$ remains {\em high-dimensional}, the representation is in terms of three {\em one-dimensional} functions $\phi, \rho, F$ which are easily represented for example in terms of splines with relatively few parameters. Such a low-dimensional representation significantly simplifies parameter estimation, and vastly improves generalisation of the model outside a training set. 
Unfortunately, the EAM model and its immediately generalisations~\cite{Baskes1992-oh} have limited ability to capture a complex \textit{ab initio} PEL. 
Still, this example inspires our search for representations of the PEL involving parameters that are 
\begin{itemize}
    \item low-dimensional,
    \item short-ranged.
\end{itemize}

Following our work on locality of interaction~\cite{ChenOrtner16, ChenLuOrtner18, ChenOrtnerThomas2019:locality, Thomas2020:scTB} we will focus on a class of tight-binding models as the {\it ab initio} reference model. These can be seen either as discrete approximations to density functional theory~\cite{bk:finnis} or alternatively as electronic structure toy models sharing many similarities with the more complex Kohn--Sham DFT and Hartree--Fock models. 

To control the dimensionality of representations, a natural idea is to to consider a body-order expansion, 
\begin{align} 
    E_\ell(\bm{r}) &\approx V_0 +
    \sum_{k\not=\ell} V_1(\bm{r}_{\ell k}) 
    + 
    \sum_{
    \above
        {k_1,k_2\not=\ell}
        {k_1 < k_2}
    }
    V_2(\bm{r}_{\ell k_1}, \bm{r}_{\ell k_2})
    + \cdots +
    \sum_{
    \above
        {k_1,\dots, k_N\not=\ell}
        {k_1 < \cdots < k_N}
    }
    V_N\big( \bm{r}_{\ell k_1}, \dots, \bm{r}_{\ell k_N} ),
    \label{eq:body-order-expansion-intro}
\end{align}
where 
$\bm{r}_{\ell k} \coloneqq \bm{r}_k - \bm{r}_\ell$
and we say that 
$V_n(\bm{r}_{\ell k_1}, \dots, \bm{r}_{\ell k_n})$
is an \textit{$(n+1)$-body potential} modelling the interaction of a centre atom $\ell$ and $n$ neighbouring atoms $\{k_1,\dots,k_n\}$. This expansion was traditionally truncated at body-order {\it three} ($N = 2$) due to the exponential increase in computational cost with $N$. 
However, it was recently demonstrated by Shapeev's moment tensor potentials (MTPs)~\cite{Shapeev2016-pd} and Drautz' atomic cluster expansion (ACE)~\cite{Drautz2019:cluster_expansion} that a careful reformulation leads to models with at most linear $N$-dependence. Indeed, algorithms proposed in~\cite{Shapeev2016-pd, BachmayrEtAl2019:cluster_expansion} suggest that the computational cost may even be $N$-independent, but this has not been proven. 
Even more striking is the fact that the MTP and ACE models which are both {\em linear models} based on a body-ordered approximation, currently appear to outperform the most advanced nonlinear models in regression and generalisation tests~\cite{Zuo2020-ba, 2020-pace1}.

These recent successes are in stark contrast with the ``folklore'' that body-order expansions generally converge slowly, if at all~\cite{Drautz2019:cluster_expansion,Stillinger1985-os,Drautz2004,Biswas1987,Haliciogli1988}. The fallacy in those observations is typically that they implicitly assume a vacuum cluster expansion (cf.~\S\ref{sec:cluster}).
Indeed, our first set of main results in \S~\ref{sec:linear} will be to demonstrate that a rapidly convergent body-order {\em approximation} can be constructed if one accounts for the chemical environment of the material. We will precisely characterise the convergence of such an approximation as $N \to \infty$, in terms of the Fermi-temperature and the band-gap of the material. 

In the simplest scheme we consider, we achieve this by considering atomic properties $[O(\Ham)]_{\ell \ell}$, where $\Ham$ is a tight-binding Hamiltonian and $O$ an analytic function. Approximating $O$ by a polynomial on the spectrum $\sigma(\Ham)$ results in an approximation of the atomic property $[p(\Ham)]_{\ell \ell}$, which is naturally ``body-ordered''.  
To obtain quasi-optimal approximation results, naive polynomial approximation schemes (e.g. Chebyshev) are suitable only in the simplest scenarios. For the insulating case we leverage potential theory techniques which in particular yield quasi-optimal approximation rates on unions of disconnected domains. Our main results are obtained by converting these into approximation results on atomic properties, analysing their qualitative features, and taking care to obtain sharp estimates in the zero-Fermi-temperature limit.

These initial results provide strong evidence for the accuracy of a linear body-order approximation in relatively simple scenarios, and would for example be useful in a study of the mechanical response of single crystals with a limited selection of possible defects. 
However, they come with limitations that we discuss in the main text. In response, we consider a much more general framework, generalizing the theory of bond order potentials~\cite{Horsfield1996}, that incorporates our linear body-ordered model as well as a range of nonlinear models. We will highlight a specific nonlinear construction with significantly improved theoretical properties over the linear scheme.



For both the linear and nonlinear body-ordered approximation schemes we prove that they inherit regularity, symmetries and locality of the original quantity of interest. 

Finally, we consider the case of self-consistent tight-binding models such as DFTB~\cite{
    Elstner2014,
    Koskinen2009,
    Seifert2012
}. In this case the highly nonlinear charge-equilibration leads {\it in principle} to arbitrarily complex intermixing of the nuclei information, and thus arbitrarily high body-order. However, our results on the body-ordered approximations for linear tight-binding models mean that each iteration of the self-consistent field (SCF) iteration can be expressed in terms of a low body-ordered and local interaction scheme. This leads us to propose a self-similar compositional representation of atomic properties that is highly reminiscent of recurrant neural network architectures. Each ``layer'' of this representation remains ``simple'' in the sense that we specified above.

\section{Results}

\subsection{Preliminaries}
\subsubsection{Tight Binding Model}
\newcommand{\asSC}{\textbf{(SC)}}


We suppose $\Lambda$ is a finite or countable index set. For $\ell \in \Lambda$, we denote the \textit{state} of atom $\ell$ by $\s_\ell = (\bm r_\ell, v_\ell, \z_\ell)$ where $\bm r_\ell \in \mathbb R^d$ denotes the position, $v_\ell$ the effective potential, and $\z_\ell$ the atomic species of $\ell$. Moreover, we define $\bm r_{\ell k} \coloneqq \bm r_k - \bm r_\ell$, $r_{\ell k}\coloneqq |\bm r_{\ell k}|$, and $\bm u_{\ell k} \coloneqq ( \bm r_{\ell k}, v_\ell, v_k, \z_\ell, \z_k)$. 
For functions $f$ of the relative atomic positions $\s_{\ell k}$, the gradient denotes the gradient with respect to the spatial variable: 
$\nabla f(\s_{\ell k}) \coloneqq \nabla\big( \xi \mapsto f(( \xi, v_\ell, v_k, \z_\ell, \z_k ))\big)\big|_{\xi = \bm r_{\ell k}}$.
The whole configuration is denoted by $\s = (\bm r, v, \z) = (\{\bm r_\ell\}_{\ell\in\Lambda}, \{v_\ell\}_{\ell\in\Lambda}, \{\z_\ell\}_{\ell\in\Lambda})$.

For a given configuration $\bm u$, the \hypertarget{TB}{tight binding Hamiltonian} takes the following form: 
%
%
\begin{assumption}{\textbf{(TB)}}
For $\ell, k \in \Lambda$ {and $\ctNumorbitals$ atomic orbitals per atom}, we suppose that 
\begin{align}\label{eq:TB_Ham}
	\Ham(\s)_{\ell k} 
	= h\big( \bm{u}_{\ell k} \big) 
	+ \sum_{m \not\in \{\ell, k\}} t(\bm{u}_{\ell m}, \bm{u}_{km}) 
	+  \delta_{\ell k} v_\ell \mathrm{Id}_{\ctNumorbitals}
\end{align}
where $h$ and $t$ {have values in $\mathbb R^{\ctNumorbitals\times\ctNumorbitals}$, are independent of the effective potential $v$, and} are continuously differentiable with
\begin{align}\label{eq:TB_local}
	\left| h(\bm u_{\ell k}) \right| 
	+ \left| \nabla h(\bm u_{\ell k}) \right| 
	&\leq h_0 e^{-\gamma_0 \, r_{\ell k}}, 
	\qquad \text{and}\\ 
	\left| t(\bm u_{\ell m},\bm u_{km}) \right|
	+\left| \nabla t(\bm u_{\ell m},\bm u_{km}) \right|
	%
	%
	&\leq h_0 e^{-\gamma_0 (r_{\ell m} + r_{km})}, \label{eq:TB_local2}
\end{align}
for some $h_0,\gamma_0>0$.

Moreover, we suppose the Hamiltonian satisfies the following symmetries:
\begin{itemize}
    \item $h(\s_{\ell k}) = h(\s_{k\ell})^\mathrm{T}$ and $t(\s_{\ell m}, \s_{km}) = t(\s_{k m}, \s_{\ell m})^\mathrm{T}$
    for all $\ell, k, m\in \Lambda$,
    \item For orthogonal transformations $Q \in \mathbb R^{d \times d}$, there exist orthogonal $D^\ell(Q) \in \mathbb R^{\ctNumorbitals\times\ctNumorbitals}$ such that 
    $\Ham( Q \s ) = D(Q) \Ham(\s) D(Q)^\mathrm{T}$ 
    where $D(Q) = \mathrm{diag}( \{ D^\ell(Q) \}_{\ell\in\Lambda} )$ and $Q\s \coloneqq ( \{Q \bm r_\ell\}_{\ell \in \Lambda}, v, \z)$.
    %
\end{itemize}

\end{assumption}

\begin{remark} 
	\textit{(i)} The constants in \cref{eq:TB_local}-\cref{eq:TB_local2} are independent of the atomic sites $\ell, k, m \in\Lambda$. 
	\textit{(ii)} The Hamiltonian is symmetric and thus the spectrum is real. 
    \textit{(iii)} The operators $\Ham(\s)$ and $\Ham( Q\s)$ are similar, and thus have the same spectra.
	\textit{(iv)} The symmetry assumptions \cite{SlaterKoster1954} of \asTB~are justified in~\cite[Appendix~A]{ChenOrtner16}.
	%
	%
	%
	\textit{(v)} The entries of $\Ham(\s)_{\ell k} \in \mathbb R^{N_\mathrm{b} \times N_\mathrm{b}}$ will be denoted $\Ham(\s)_{\ell k}^{ab}$ for $1 \leq a,b \leq \ctNumorbitals$. When clear from the context, we drop the argument $(\s)$ in the notation.
\end{remark}

The assumptions \asTB~define a general \textit{three-centre} tight binding model, whereas, if $t \equiv 0$, a simplification made in the majority of tight binding codes, we say \asTB~is a \textit{two-centre} model \cite{bk:finnis}. 

%
The choice of potential in \asTB~defines a hierarchy of tight binding models. If $v = {\rm const}$, \asTB~defines a linear tight binding model, a simple yet common model
\cite{
    ChenLuOrtner18,
    ChenOrtner16,
    OrtnerThomas2020:pointdefs,
    ChenOrtnerThomas2019:locality
}. 
In this case, we implicitly assume that the Coulomb interactions have been screened, a typical assumption made in practice for a wide variety of materials \cite{Cohen1994,Papaconstantopoulos1997,Papaconstantopoulos2015,Mehl1996}. 
Supposing that $v$ is a function of a self-consistent electronic density, we arrive at a non-linear model such as DFTB~\cite{Koskinen2009,Elstner2014,Seifert2012}. Abstract variants of these nonlinear models have been analysed, for example, in \cite{Thomas2020:scTB,ELu10}. 
Through much of this article we will treat $\bm{r}, v$ as independent inputs into the Hamiltonian, but will discuss their connection and self-consistency in \S\ref{sec:nonlin}.

For a finite system $\s$, we consider \textit{analytic observables} of the density of states 
\cite{
    ChenLuOrtner18,
    Thomas2020:scTB
}: 
for functions 
$O \colon \mathbb R \to \mathbb R$
that can be analytically continued into an open neighbourhood of 
$\sigma\big( \Ham( \s ) \big)$, 
we consider
\[
    \mathrm{Tr}\, O
    \big(
        \Ham(\s) 
    \big) 
    = \sum_s O(\lambda_s) 
\]
where 
$(\lambda_s, \psi_s)$
are normalised eigenpairs of $\Ham(\s)$. Many properties of the system, including the particle number functional and Helmholtz free energy, may be written in this form 
\cite{
    ChenOrtner16,
    ChenLuOrtner18,
    OrtnerThomas2020:pointdefs,
    Thomas2020:scTB
}. 
By distributing these quantities amongst atomic positions, we obtain a well-known spatial decomposition
\cite{
    bk:finnis,
    Ercolessi2005,
    ChenOrtner16,
    ChenLuOrtner18
}, 
\begin{align}\label{eq:localQoI}
    \mathrm{Tr}\, O
    \big(
        \Ham(\s) 
    \big)  
    = \sum_{\ell \in\Lambda} O_\ell(\s)
    \quad \text{where} \quad
    {O_\ell(\s) \coloneqq 
    \mathrm{tr} \big[O\big(\Ham(\s)\big)_{\ell\ell}\big]
    = \sum_s O(\lambda_s) \big|[\psi_s]_\ell\big|^2.}
\end{align}
%
%
For infinite systems, we may define $O_\ell(\s)$ through the thermodynamic limit 
\cite{
    ChenOrtner16,
    ChenLuOrtner18
}
or via the holomorphic functional calculus; see \S\ref{pf:resolvent-calculus} for further details. 


When discussing derivatives of the local observables, we will simplify notation and write
\begin{align}
    \frac
        {\partial O_\ell(\s)}
        {\partial \s_m}
    \coloneqq \bigg( 
        \frac
            {\partial O_\ell(\s)}
            {\partial \bm{r}_m},
        \frac
            {\partial O_\ell(\s)}
            {\partial v_m}
    \bigg).
\end{align}

\subsubsection{Local Observables}
Although the results in this paper apply to general analytic observables, our primary interest is in applying them to two special cases:

A local observable of particular importance is the electron density: for inverse Fermi-temperature 
$\beta 
\in (0,+\infty]$ 
and fixed chemical potential $\mu$, we have
\begin{align}\label{eq:fd}
    \rho = F^\beta(\s) 
    \quad
    \text{where} \quad 
    F^\beta(z) \coloneqq \begin{cases}
        \big( 1 + e^{\beta(z - \mu)} \big)^{-1} &\text{if } \beta < \infty \\
        \chi_{(-\infty,\mu)}(z) + \frac{1}{2}\chi_{\{\mu\}}(z)
            &\text{if }\beta = \infty.
    \end{cases}
\end{align}
Throughout this paper $F^\beta(\s) \coloneqq \big(F_\ell^\beta(\s)\big)_{\ell \in \Lambda}$ will denote a vector and so \cref{eq:fd} reads $\rho_\ell = F^\beta_\ell(\s)$ for all $\ell\in\Lambda$.

In \S\ref{sec:nonlin}, we consider the case where the effective potential $v \colon \mathbb R^\Lambda \to \mathbb R^\Lambda$ is a function of the electron density \cref{eq:fd} which leads to the following self-consistent local observables: 
\begin{align}\label{eq:SC_observables}
\begin{cases}
    O^{\mathrm{sc}}_\ell(\s) \coloneqq O_\ell\big( \s(\rho^\star) \big) \\
    \rho^\star = F^\beta\big( \s(\rho^\star) \big) 
\end{cases}
\end{align}
where $\s(\rho) \coloneqq \big( \bm r, v(\rho), \z \big)$.

\begin{remark}
    All the results of this paper also hold for the off-diagonal entries of the density matrix
    ($\rho_{\ell k} \coloneqq \mathrm{tr} \, F^\beta\big(\Ham(\s)\big)_{\ell k}$)
    without any additional work. This fact will be clear from the proofs. It is likely though that additional properties related to the off-diagonal decay (near-sightedness) and spatial regularity further improve the ``sparsity'' of the density matrix. A complete analysis would go beyond the scope of this work.
\end{remark}

The second observable we are particularly interested in is the site energy, which allows us to decompose the total potential energy landscape into localised ``atomic'' contributions. In the grand potential model for the electrons, which is appropriate for large or infinite condensed phase systems~\cite{ChenLuOrtner18}, it is defined as
\begin{align}\label{eq:grand-pot}
    G^\beta_\ell(\s) \coloneqq \mathrm{tr}
    \big[
        G^\beta\big( 
            \Ham(\s) 
        \big)_{\ell\ell}
    \big] \quad
    \text{where} \quad
        G^\beta(z) \coloneqq 
        \begin{cases}
            \frac{2}{\beta} \log\big( 1 - F^\beta(z) \big)
                &\text{if } \beta < \infty \\
            2(z - \mu) \chi_{(-\infty, \mu)}(z) 
                &\text{if } \beta = \infty.
        \end{cases}
\end{align}
%
The total grand potential is defined as 
$\sum_\ell G^\beta_\ell(\s)$
\cite{
    ChenLuOrtner18,
    OrtnerThomas2020:pointdefs
}.

For $\beta < \infty$, the functions $F^\beta(\,\cdot\,)$ and $G^\beta(\,\cdot\,)$ are analytic in a strip of width $\pi \beta^{-1}$ about the real axis \cite[Lemma~5.1]{ChenOrtnerThomas2019:locality}. To define the zero Fermi-temperature observables, we assume that $\mu$ lies in a spectral gap ($\mu \not\in \sigma\big( \Ham(\s) \big)$; see \S\ref{sec:metal-insulator-defect}). In this case, $F^\beta(\,\cdot\,)$ and $G^\beta(\,\cdot\,)$ extend to analytic functions in a neighbourhood of $\sigma\big( \Ham(\s) \big)$ for all $\beta \in (0,\infty]$.

In order to describe the relationship between the various constants in our estimates and the inverse Fermi-temperature or spectral gap (in the case of insulators), we will state all of our results for $O^\beta = F^\beta$ or $G^\beta$. Other analytic quantities of interest can be treated similarly with constants depending, e.g., on the region of analyticity of the corresponding function $z\mapsto O(z)$.

\subsubsection{Metals, insulators, and defects}
\label{sec:metal-insulator-defect}
As we can see from \cref{eq:localQoI}, the structure of the spectrum $\sigma\big( \Ham(\s) \big)$ will have a key role in the analysis. Firstly, by \asTB, $\Ham(\s)$ is a bounded self-adjoint operator on $\ell^2(\Lambda \times \{1,\dots,N_\mathrm{b}\})$ and thus the spectrum is real and contained in some bounded interval. In order to keep the mathematical results general, we will not impose any further restrictions on the spectrum. However, to illustrate the main ideas, we briefly describe typical spectra seen in metals and insulating systems. 

\begin{figure}[tb]
	\centering
	\resizebox{\columnwidth} {!} {
		\input{diagrams/metal_insulator} }%
	\caption{Schematic plots of the spectrum $\sigma\big( \Ham(\s) \big)$ of a metal (top) and insulator (bottom).
	\label{fig:metal_insulator}}
\end{figure}
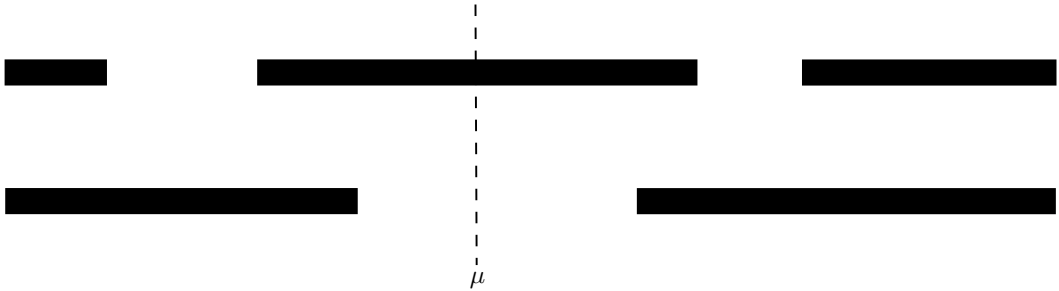

In the case where $\bm u$ describes a multi-lattice in $\mathbb R^d$ formed by taking the union of finitely many shifted Bravais lattices, the spectrum $\sigma\big( \Ham(\s) \big)$ is the union of finitely many continuous energy bands \cite{bk:kittel}. That is, there exist continuous functions, $\ep^\alpha \colon \overline{\mathrm{BZ}} \to \mathbb R$, on the \textit{Brillouin zone} $\mathrm{BZ}$, a compact connected subset of $\mathbb R^d$, such that
\[
    \sigma\big( \Ham(\s) \big) = \bigcup_{\alpha} \ep^\alpha( \mathrm{BZ} ).
\]
In particular, in this case, $\sigma\big( \Ham(\s) \big) = \sigma_{\mathrm{ess}}\big( \Ham(\s) \big)$ is the union of finitely many intervals on the real line. The \textit{band structure} $\{\ep^\alpha\}$ relative to the position of the chemical potential, $\mu$, determines the electronic properties of the system \cite{bk:sutton}. In metals $\mu$ lies within a band, whereas for insulators, $\mu$ lies between two bands in a \textit{spectral gap}. Schematic plots of these two situations are given in Figure~\ref{fig:metal_insulator}.  

We now consider perturbations of a reference configuration $\s^\mathrm{ref} = (\bm r^\mathrm{ref}, v^\mathrm{ref}, \z^\mathrm{ref})$ defined on an index set $\Lambda^\mathrm{ref}$, 
\begin{prop}[Perturbation of the Spectrum]
\label{prop:pert_spec}
    For $\delta, R_\mathrm{def} > 0$, there exists $\delta_0 > 0$ such that if $\s = (\bm r, v, \z)$ is a configuration defined on some index set $\Lambda$ satisfying 
    $\Lambda \setminus B_{R_\mathrm{def}} = \Lambda^\mathrm{ref} \setminus B_{R_\mathrm{def}}$, 
    $\Lambda \cap B_{R_\mathrm{def}}$ is finite, 
    $\z_k = \z_k^\mathrm{ref}$ for all $k \in \Lambda\setminus B_{R_\mathrm{ref}}$, 
    and $\sup_{k \in \Lambda \setminus B_{R_\mathrm{def}}} \big[ |\bm r_k - \bm r_k^\mathrm{ref}| + | v_k - v_k^\mathrm{ref} | \big] \leq \delta_0$, then 
    \[
        \Big| \sigma\big( \Ham(\s) \big) \setminus B_\delta\big( \sigma\big(\Ham(\s^\mathrm{ref})\big) \big) \Big| < \infty.
    \]
\end{prop}

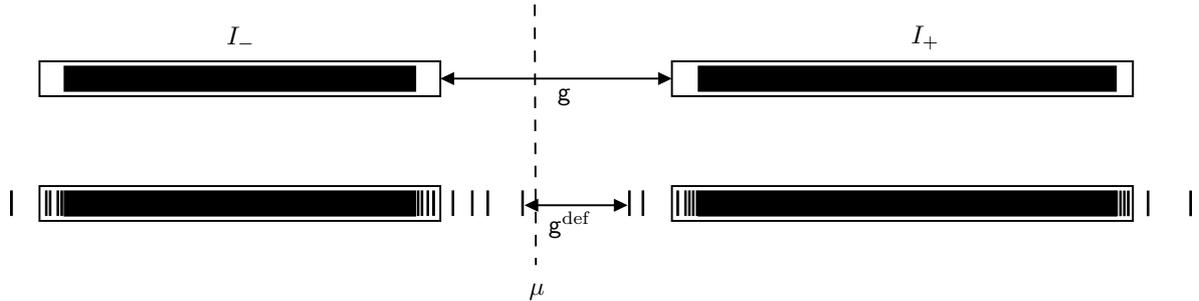
\begin{figure}[tb]
	\centering
	\resizebox{\columnwidth} {!} {
		\input{diagrams/def} }%
	\caption{Top: Schematic plot of the spectrum $\sigma\big( \Ham(\s^\mathrm{ref}) \big)$ for an insulating system, together with two compact intervals $I_-$ and $I_+$ as in \cref{eq:I} and the constant $\mathsf{g}$ from \cref{eq:g}. Bottom: The spectrum $\sigma\big( \Ham(\s) \big)$ after considering perturbations satisfying Proposition~\ref{prop:pert_spec}. While the edges of the spectrum may be accumulation points for a sequence of eigenvalues within the band gap, the number of such eigenvalues bounded away from the edges is finite. 
	\label{fig:def}}
\end{figure}

In particular, if $\s^\mathrm{ref}$ describes a multilattice, then, since local perturbations in the \textit{defect core} are of finite rank, the essential spectrum is unchanged and we obtain finitely many eigenvalues bounded away from the spectral bands. Moreover, a small global perturbation can only result in a small change in the spectrum. Again, a schematic plot of this situation is given in Figure~\ref{fig:def}.   

For the remainder of this paper, we consider the following notation:
\begin{definition}
\label{def:config}
    Suppose that $\s^\mathrm{ref}$ is a general reference configuration defined on $\Lambda^\mathrm{ref}$ and $\s$ is a configuration arising due to Proposition~\ref{prop:pert_spec}. Then, we define $I_-$ and $I_+$ to be compact intervals and $\{\lambda_j\}$ to be a finite set such that 
    \begin{align}\label{eq:I}
        \sigma\big( \Ham(\s^\mathrm{ref}) \big) \subset I_- \cup I_+,
        \qquad
        \sigma\big( \Ham(\s) \big) \subset I_- \cup \{\lambda_j\} \cup I_+
    \end{align}
    and $\max I_- \leq \mu \leq \min I_+$. Moreover, we define
    \begin{align}\label{eq:g}
        \mathsf{g} &\coloneqq \min I_+ - \max I_- \geq 0, 
        \qquad \text{and} \\ \label{eq:gdef} 
        \mathsf{g}^\mathrm{def} &\coloneqq \min I_+ \cup \{\lambda_j \colon \lambda_j \geq \mu\}  - \max I_- \cup \{\lambda_j \colon \lambda_j \leq \mu \}. 
    \end{align}
\end{definition}
The constants in Definition~\ref{def:config} are also displayed in Figure~\ref{fig:def}. The constant $\mathsf{g}$ in Definition~\ref{def:config} is slightly arbitrary in the sense that as long as $B_\delta\big( \sigma\big(\Ham(\s^\mathrm{ref})\big) \big) \subset I_- \cup I_+$ (where $\delta$ is the constant from Proposition~\ref{prop:pert_spec}), then there exists a finite set $\{\lambda_j\}$ as in \cref{eq:I}. Choosing smaller $\mathsf{g}$ reduces the size of the set $\{\lambda_j\}$. 

\subsection{Vacuum cluster expansion} 
\label{sec:cluster}
%
%
For a system of $M$ identical particles $X_1,\dots,X_M$, a maximal body-order $N$, and a permutation invariant energy $E = E(\{X_1,\dots,X_M\})$, we may consider the vacuum cluster expansion,
\begin{align}
\label{eq:classical-body-order}
    E(\{X_1,\dots,X_M\}) \approx \sum_{n=0}^{N} \sum_{1 \leq m_1 < \dots < m_n \leq M} V^{(n)}(X_{m_1},\dots,X_{m_n})
\end{align}
where the $n$-body interaction potentials $V^{(n)}$ are defined by considering all isolated clusters of $j \leq n$ atoms: 
\[
    V^{(n)}(X_{1},\dots,X_{n}) 
    = \sum_{j = 0}^n
    (-1)^{n-j} 
    \sum_{1 \leq m_1 < \dots < m_j \leq n} 
    E(\{X_{m_1},\dots,X_{m_j}\}).
\]
The expansion \cref{eq:classical-body-order} is exact for $N = M$. The vacuum cluster expansion is the traditional and, arguably, the most natural many-body expansion of a potential energy landscape. However, in many systems, it converges extremely slowly with respect to the body-order $N$ and is thus computationally impractical.
%
An intuitive explanation for this slow convergence is that, when defining the body-order expansion in this way, we are building an interaction law for a condensed or possibly even crystalline phase material from clusters in vacuum where the bonding chemistry is significantly different. Although this observation appears to be ``common knowledge'' we were unable to find references that provide clear evidence for it. However, some limited discussions and further references can be found in \cite{Drautz2019:cluster_expansion,Stillinger1985-os,Drautz2004,Biswas1987,Haliciogli1988}.
%

Our own approach employs an entirely different mechanism, which in particular incorporates environment information and leads to an exponential convergence of an $N$-body approximation. Technically, our approximation is not an expansion, that is, the $n$-body terms $V^{(n)}$ of the classical cluster expansion are replaced by terms that depend also on the highest body-order $N$. We will provide a more technical discussion contrasting our results with the vacuum cluster expansion in \S\ref{sec:concl:vacuum}.

\subsection{A general framework}
\label{sec:framework}
%
%
Before we consider two specific body-ordered approximations, we present a general framework which illustrates the key features needed for a convergent scheme:
    To that end, we introduce the \textit{local density of states} (LDOS) $D_\ell$ \cite{bk:finnis} which is the (positive) measure $D_\ell$ supported on $\sigma(\Ham)$ such that
    \begin{align}\label{eq:LDOS}
        \int x^n \mathrm{d}D_\ell(x) = \mathrm{tr}[\Ham^n]_{\ell\ell}, 
        \qquad \text{for } n \in \mathbb N_0.
    \end{align}
    Existence and uniqueness follows from the spectral theorem for normal operators (e.g.~see \cite[Theorem~6.3.3]{bk:spectral} or \cite{bk:Teschl}). 
    In particular, \cref{eq:localQoI} may be written as the integral
    $
        O_\ell(\s) = \int O \,\mathrm{d}D_\ell
    $.
    
    Then, on constructing a (possibly signed) unit measure $D_\ell^N$ with exact first $N$ moments (that is, $\int x^n \mathrm{d}D_\ell^N(x) = \mathrm{tr}[\Ham^n]_{\ell\ell}$ for $n = 1,\dots, N$), we may define the approximate local observable $O_\ell^N(\s) \coloneqq \int O \, \mathrm{d}D_\ell^N$, and obtain the following general error estimates: 
    \begin{align}
        \big|
            O_\ell(\s) - O_\ell^N(\s)
        \big|
        &= \inf_{P_N \in \mathcal P_N}
        \Big|
            \int \big( O - P_N \big) \mathrm{d}\big( D_\ell - D_\ell^N \big)
        \Big|\\
        &\leq  \big\|
            D_\ell - D_\ell^N
        \big\|_{\mathrm{op}}
        \inf_{P_N \in \mathcal P_N}
        \big\|
            O - P_N
        \big\|_{\infty}
       \label{eq:general_estimate}
    \end{align}
    %
    %
    where $\mathcal P_N$ denotes the set of polynomials of degree at most $N$, and $\|\,\cdot\,\|_{\mathrm{op}}$ is the operator norm on a function space $(\mathcal S, \|\,\cdot\,\|_\infty)$. For example, we may take $\mathcal S$ to be the set of functions analytic on an open set containing $\mathscr C$, a contour encircling $\mathrm{supp}\big( D_\ell - D_\ell^N \big)$, and consider 
    \[
        \| O \|_{\infty} \coloneqq \frac{\mathrm{len}(\mathscr C)}{2\pi} \|O\|_{L^\infty( \mathscr C )}.   
    \]
    Alternatively, we may consider $\mathcal S = L^\infty\big( \mathrm{supp}(D_\ell - D_\ell^N)  \big)$ leading to the total variation operator norm.
        %
    %
    Equation~\cref{eq:general_estimate} highlights the key generic features that are crucial ingredients in obtaining convergence results:
    \begin{itemize}
        \item \textit{Analyticity.} The potential theory results of \S\ref{pf:pot-theory} connect the asymptotic convergence rates for polynomial approximation to the size and shape of the region of analyticity of $O$.
        
        \item \textit{Spectral Pollution.} While ${\rm supp} D_\ell \subset \sigma(H)$, this need not be true for $D_\ell^N$. Indeed, if ${\rm supp} D_\ell^N$ introduces additional points within the band gap, thia may significantly slow the convergence of the polynomial approximation; cf. \S\ref{sec:concl:vacuum}.
        
        
        \item \textit{Regularity of $D_\ell^N$.} Roughly speaking, the first term of \cref{eq:general_estimate} measures how ``well-behaved'' $D_\ell^N$ is. In particular, if $D_\ell^N$ is positive, then this term is bounded independently of $N$, whereas, if $D_\ell^N$ is a general signed measure, then this factor contributes to the asymptotic convergence behaviour. 
    \end{itemize}

    In the following sections, we introduce linear (\S\ref{sec:linear}) and nonlinear (\S\ref{sec:bop}) approximation schemes that fit into this general framework. Moreover, in \S\ref{sec:concl:vacuum}, we also write the vacuum cluster expansion as an integral against an approximate LDOS.
    We investigate which of the requirements listed above fail to complements the intuitive explanation for the slow convergence of the vacuum cluster expansion.
    
    In the appendices, we review other approximation schemes that fit into this general framework such as the quadrature method (Appendix~\ref{app:quadrature}), numerical bond order potentials (Appendix~\ref{app:generalBOP}), and the kernel polynomial method (Appendix~\ref{app:KPM}).
    
    

\subsection{Linear body-ordered approximation}
\label{sec:linear}
We will construct two distinct but related many-body approximation models. To construct our first model we exploit the observation that polynomial approximations of an analytic function correspond to body-order expansions of an observable. 

An intuitive approach is to write the local observable in terms of its Chebyshev expansion and truncate to some maximal polynomial degree. The corresponding projection operator is a simple example of the kernel polynomial method (KPM) \cite{SilverRoder1994} and the basis for analytic bond order potentials (BOP) \cite{Pettifor1989}. We discuss in Appendix~\ref{app:KPM} that these schemes put more emphasis on the approximation of the local density of states (LDOS) and, in particular, exploit particular features of the Chebyshev polynomials to obtain a positive approximate LDOS. Since our focus is instead on the approximation of observables, we employ a different approach that is tailored to specific properties of the band structure and leads to superior convergence rates for these quantities.




For a finite set of interpolation points $X = \{x_j\}$, and a complex-valued function $O$ defined on $X$, we denote by $I_XO(z)$ the degree $|X|-1$ polynomial interpolant of $z\mapsto O(z)$ on $X$. This gives rise to the body-order approximation
%
\begin{align}\label{eq:interpolation}
    I_X O_\ell(\s) \coloneqq \mathrm{tr} \big[ 
        I_X O\big(\Ham(\s)\big)_{\ell\ell}
     \big].
\end{align}
We may connect \cref{eq:interpolation} to the general framework in \S\ref{sec:framework} by defining
    \[
        I_{X_N} O^\beta_\ell(\s^\mathrm{ref}) = \int O^\beta \mathrm{d}D_\ell^{N,\mathrm{lin}} 
        \qquad \text{where} \qquad 
        D_\ell^{N,\mathrm{lin}} \coloneqq \mathrm{tr}\sum_{j} \ell_j(\Ham)_{\ell\ell} \, \delta(\,\cdot\,- \ep_j)
    \]
    and $\ell_j$ are the node polynomials corresponding to an interpolation set $X_N = \{\ep_j\}$ (that is, $\ell_j(\ep_i) = \delta_{ij}$).
\begin{prop}
\label{prop}
    $I_X O_\ell(\s)$ has finite body-order. 
    More specifically, there exists $N \in \mathbb N$ and $(n+1)$-body potentials $V_{nN}$ for $n = 0,\dots,N-1$ such that
    \begin{align}\label{eq:prop-boby-order}
        I_X O_\ell(\s) = \sum_{n = 0}^{N-1} \sum_{\above 
                {k_1,\dots,k_n \not= \ell}
                {k_1 < \dots < k_n}
        } V_{nN}(\s_\ell; \s_{\ell k_1}, \dots, \s_{\ell k_n}).
    \end{align}
\end{prop}
\begin{proof}[Sketch of the Proof.]
    Since \cref{eq:interpolation} is a linear combination of the monomials $[\Ham^n]_{\ell\ell}$, it is enough to show that, for each $n \in \mathbb N$,
    \begin{align}\label{eq:monomial}
        [\Ham^{n}]_{\ell\ell} 
        = \sum_{\ell_1,\dots,\ell_{n-1}}
        \Ham_{\ell\ell_1}  \Ham_{\ell_1\ell_2} \cdots 
        \Ham_{\ell_{n-1}\ell}
    \end{align}
    has finite body order. 
    Each term in \cref{eq:monomial} depends on the central atom $\ell$, the $n-1$ neighbouring sites $\ell_1, \dots, \ell_{n-1}$, and the at most $n$ additional sites arising from the three-centre summation in the tight binding Hamiltonian \asTB. In particular, \cref{eq:interpolation} has body order at most $2(|X|-1)$. See \S\ref{sec:body-order} for a complete proof including an explicit definition of the $V_{nN}$.
\end{proof}

If one uses Chebyshev points as the basis for the body-ordered approximation \cref{eq:interpolation}, the rates of convergence depend on the size of the largest \textit{Bernstein ellipse} (that is, ellipses with foci points $\pm1$) contained in the region of analyticity of $z \mapsto O(z)$ \cite{bk:Trefethen2019}. This leads to a exponentially convergent body-order expansion in the metallic finite-temperature case (see \S~\ref{sec:projection} for the details).

However, the resulting estimates deteriorate in the zero-temperature limit. Instead, we apply results of potential theory to construct interpolation sets $X_N$ that are adapted to the spectral properties of the system (see \S\ref{pf:pot-theory} for examples) and \textit{(i)} do not suffer from spectral pollution, and \textit{(ii)} (asymptotically) minimise the total variation of $D_{\ell}^{N,\mathrm{lin}}$ which, in this context, is the Lebesgue constant \cite{bk:Trefethen2019} for the interpolation operator $I_{X_N}$. This leads to rapid convergence of the body-order approximation based on \eqref{eq:interpolation}. 
The interpolation sets $X_N$ depend only on the intervals $I_-, I_+$ from Definition~\ref{def:config} (see also Figure~\ref{fig:def}) and can be chosen independently of $\s^\mathrm{ref}$ as long as $B_\delta\big( \sigma\big( \Ham(\s^\mathrm{ref}) \big) \big) \subset I_- \cup I_+$.
%
%
%
\begin{theorem}
\label{thm:bodyOrder_interpolation}
Suppose $\bm u^\mathrm{ref}$ is given by Definition~\ref{def:config}.
Fix $0 < \beta \leq \infty$ and suppose that, either $\beta < \infty$ or $\mathsf{g}>0$. Then, there exist constants $\gamma_N  > 0$ and interpolation sets 
$X_N \subset I_- \cup I_+$ 
satisfying \cref{eq:prop-boby-order} such that
\begin{align*}
%
    \big|O^\beta_\ell(\bm u^\mathrm{ref}) - I_{X_N} O^\beta_\ell(\bm u^\mathrm{ref})\big| &\leq C_1 e^{- \gamma_N N},  \qquad \text{and}\\ 
    \Bigg|
        \frac
            {\partial O^\beta_\ell}
            {\partial \bm u_m} (\bm u^\mathrm{ref})
        - 
        \frac
            {\partial I_{X_N} O^\beta_\ell}
            {\partial \bm u_m} (\bm u^\mathrm{ref})
    \Bigg|
    &\leq C_2 e^{-\frac{1}{2} \gamma_N N} e^{-\eta\, r_{\ell m}},
\end{align*}
where $O^\beta = F^\beta$ or $G^\beta$ and $C_1, C_2>0$ are independent of $N$. The asymptotic convergence rate 
$\gamma \coloneqq \lim_{N\to\infty} \gamma_N$ 
is positive and exhibits the asymptotic behaviour
\begin{equation}
    C_1 \sim (\mathsf{g} + \beta^{-1})^{-1}, 
    \quad 
    C_2 \sim (\mathsf{g} + \beta^{-1})^{-3}, 
    \quad \text{and} \quad
    \gamma, \eta \sim \mathsf{g} + \beta^{-1} 
    \quad \text{ as } \mathsf{g} + \beta^{-1} \to 0.
\end{equation}
In this asymptotic relation, we assume that the limit $\mathsf{g}\to 0$ is approached symmetrically about the chemical potential $\mu$.
\end{theorem}
\begin{remark}
    {Higher derivatives may be treated similarly under the assumption that higher derivatives of the tight binding Hamiltonian \asTB~exist and are short ranged.}
\end{remark}

\subsubsection{The Role of the Point Spectrum} \label{sec:pointspec}
We now turn towards the important scenario when a localised defect is embedded within a homogeneous crystalline solid. Recall from \S~\ref{sec:metal-insulator-defect} (see in particular Fig.~\ref{fig:def}) that this gives rise to a discrete spectrum, which ``pollutes'' the band gap~\cite{OrtnerThomas2020:pointdefs}. 
Thus, the spectral gap is reduced and a naive application of Theorem~\ref{thm:bodyOrder_interpolation} leads to a reduction in the convergence rate of the body-ordered approximation. 
We now improve these estimates by showing that, away from the defect, we obtain improved pre-asymptotics, reminiscent of similar results for locality of interaction~\cite{ChenOrtnerThomas2019:locality}.

In the following, we fix $\s$ satisfying Definition~\ref{def:config}. While improved estimates may be obtained by choosing $\{\lambda_j\}$ as interpolation points, leading to asymptotic exponents that are independent of the defect, in practice, this requires full knowledge of the point spectrum. Since the point spectrum within the spectral gap depends on the whole atomic configuration, the approximate quantities of interest corresponding to these interpolation operators would no longer satisfy Proposition~\ref{prop}. 

\begin{remark}
    This phenomenon has been observed in the context of Krylov subspace methods for solving linear equations $Ax  = b$ where outlying eigenvalues delay the convergence by $O(1)$ steps without affecting the asymptotic rate \cite{Driscoll1998:krylov}. Indeed, since the residual after $n$ steps may be written as $r_n = p_n(A) r_0$ where $p_n$ is a polynomial of degree $n$, there is a close link between polynomial approximation and convergence of Krylov methods.  
\end{remark}

On the other hand, we may use the exponential localisation of the eigenvectors corresponding to isolated eigenvalues to obtain pre-factors that decay exponentially as $|\bm{r}_\ell| \to \infty$:

\begin{theorem}
\label{thm:bodyOrder_interpolation_discrete}
Suppose $\s$ satisfies Definition~\ref{def:config} with $\mathsf{g}>0$. Fix $0<\beta \leq \infty$ and suppose that, if $\beta = \infty$, then $\mathsf{g}^\mathrm{def}>0$, and let $C_1,C_2, \gamma_N, \gamma, \eta$, and $X_N \subset I_-\cup I_+$ be given by Theorem~\ref{thm:bodyOrder_interpolation}. Then,
\begin{align}
    \big|O^\beta_\ell(\s) - I_{X_N} {O}^\beta_\ell(\s)\big| 
    &\leq 
    C_1 e^{-\gamma_N N}
    +
    C_3 e^{-\gamma_\mathrm{CT}  
    |\bm{r}_\ell| }
    e^{-\frac{1}{2}\gamma_N^{\mathrm{def}} N} \\
    \Bigg|
        \frac
            {\partial O^\beta_\ell}
            {\partial \bm u_m} (\bm u)
        - 
        \frac
            {\partial I_{X_N} O^\beta_\ell}
            {\partial \bm u_m} (\bm u)
    \Bigg| 
    &\leq 
    \Big( C_2 e^{-\frac{1}{2}\gamma_N N}
    +
    C_4 e^{-\gamma_\mathrm{CT} 
    |\bm{r}_\ell|}
    e^{-\frac{1}{2}\gamma_N^{\mathrm{def}} N} \Big) e^{-\eta \,
    r_{\ell m}}\label{eq:lin-derivatives}
\end{align}
where $O^\beta = F^\beta$ or $G^\beta$ and $C_3, C_4>0$ are independent of $N$. The asymptotic convergence rate 
$\gamma^\mathrm{def} \coloneqq \lim_{N\to\infty} \gamma^\mathrm{def}_N$ 
is positive and we have
\begin{equation}
    \gamma^\mathrm{def} \sim \mathsf{g}^\mathrm{def} + \beta^{-1}
    \quad \text{ as } \mathsf{g}^\mathrm{def} + \beta^{-1} \to 0,
    \qquad \text{and} \qquad  
    \gamma_\mathrm{CT}, \eta \sim \mathsf{g} + \beta^{-1} 
    \quad \text{ as } \mathsf{g} + \beta^{-1} \to 0.
\end{equation}
In these asymptotic relations, we assume that the limits $\mathsf{g}^\mathrm{def},\mathsf{g}\to 0$ are approached symmetrically about the chemical potential $\mu$.
\end{theorem}

In practice, Theorem~\ref{thm:bodyOrder_interpolation_discrete} means that, for atomic sites $\ell$ away from the defect-core, the observed pre-asymptotic error estimates may be significantly better than the asymptotic convergence rates obtained in Theorem~\ref{thm:bodyOrder_interpolation}.


\begin{remark}[Locality]
    (i) By Theorem~\ref{thm:bodyOrder_interpolation_discrete}, and the locality estimates for the exact observables $O_\ell^\beta$ \cite{ChenOrtnerThomas2019:locality}, we immediately obtain corresponding locality estimates for the approximate quantities:
    \begin{equation}
     \left|
        \frac
            {\partial I_{X_N} O^\beta_\ell(\s) }
            {\partial \s_m} 
    \right|
    \lesssim e^{- \eta \,r_{\ell m}}.
\end{equation}

    (ii) We investigate another type of locality in Appendix~\ref{app:locality} where we show that various truncation operators result in approximation schemes that only depend on a small atomic neighbourhood of the central site. An exponential rate of convergence as the truncation radius tends to infinity is obtained.
\end{remark}

\subsection{A non-linear representation}
\label{sec:bop}
%
%
The method presented in \S\ref{sec:linear} approximates local quantities of interest by approximating the integrand $O \colon \mathbb C \to \mathbb C$ with polynomials. 
As we have seen, this leads to approximation schemes that are linear functions of the spatial correlations $\{[\Ham^n]_{\ell\ell}\}_{n \in \mathbb N}$. In this section, we construct a non-linear approximation related to bond-order potentials (BOP) \cite{Finnis2007, Horsfield1996, Drautz:2020} and show that the added non-linearity leads to improved asymptotic error estimates that are independent of the discrete spectra lying within the band gap. In this way, the nonlinearity captures ``spectral information'' from $\Ham$ rather than only approximating $O \colon \mathbb C \to \mathbb C$ without reference to the Hamiltonian. 

    Applying the recursion method \cite{Haydock1972:recursion, Haydock1972:recursionII}, a reformulation of the Lanczos process \cite{Lanczos1950}, we obtain a tri-diagonal (Jacobi) operator $T$ on $\ell^2(\mathbb N_0)$ whose spectral measure is the LDOS $D_\ell$ \cite{Teschl:jacobioperators} (see \S\ref{sec:recursion} for the details). 
We then truncate $T$ by taking the principal $\frac{1}{2}(N+1) \times \frac{1}{2}(N+1)$ submatrix $T_{\frac{1}{2}(N-1)}$ and define
\begin{align}\label{eq:Theta}
    \ctNonlin\big( \Ham_{\ell\ell}, [\Ham^2]_{\ell\ell}, \dots, [\Ham^{N}]_{\ell\ell} \big)
    \coloneqq O^\beta(T_{\frac{1}{2}(N-1)})_{00}
    = \int O^\beta \mathrm{d}D_\ell^{N,\mathrm{nonlin}},
\end{align}
where $D_\ell^{N,\mathrm{nonlin}} = \sum_s [\psi_s]_{0}^2 \delta(\,\cdot - \lambda_s)$ is a spectral measure for $T_{\frac{1}{2}(N-1)}$ (that is, $(\lambda_s,\psi_s)$ are normalised eigenpairs of $T_{\frac{1}{2}(N-1)}$). By showing that the first $N$ moments of $D_\ell^{N,\mathrm{nonlin}}$ are exact, we are able to apply \cref{eq:general_estimate} to obtain the following error estimates. The asymptotic behaviour of the exponent in these estimates follows by proving that the spectral pollution of $D_\ell^{N,\mathrm{nonlin}}$ in the band gap is sufficiently mild.
%
\begin{theorem}\label{thm:BOP}
    Suppose $\s$ satisfies Definition~\ref{def:config}. Fix $0 < \beta \leq \infty$ and suppose that, if $\beta = \infty$, then $\mathsf{g}, \mathsf{g}^\mathrm{def} >0$. Then, for $N$ odd, there exists an open set $U \subset \mathbb C^{N}$ such that \cref{eq:Theta} extends to an analytic function $\ctNonlin \colon U \to \mathbb C$, such that 
    \begin{align}
        \Big|
            O^\beta_\ell(\s) - \ctNonlin\big( \Ham_{\ell\ell}, [\Ham^2]_{\ell\ell}, \dots, [\Ham^{N}]_{\ell\ell} \big)
        \Big|
        &\lesssim e^{-\gamma_N N} 
        %
            %
        %
    \end{align}
where $O^\beta = F^\beta$ or $G^\beta$. The asymptotic convergence rate $\gamma \coloneqq \lim_{N\to\infty} \gamma_N$ is positive and $\gamma \sim \mathsf{g} + \beta^{-1}$ as $\mathsf{g} + \beta^{-1} \to 0$. 
\end{theorem}

\begin{remark}
    It is important to note that $\ctNonlin \colon U \to \mathbb C$ can be constructed without knowledge of $\Ham$ because, as we have seen, if the discrete eigenvalues are known {\em a priori}, then Theorem~\ref{thm:BOP} is immediate from Theorem~\ref{thm:bodyOrder_interpolation_discrete} by adding finitely many additional interpolation points on the discrete spectrum. 
    
    In particular, the fact that $\Theta$ is a {\em material-agnostic} nonlinearity has potentially far-reaching consequences for material modelling.
\end{remark}

\begin{remark}[Quadrature Method] 
Alternatively, we may use the sequence of orthogonal polynomials \cite{Freud:OP} corresponding to $D_\ell$ as the basis for a Gauss quadrature rule to evaluate local observables. This procedure, called the {\em Quadrature Method} \cite{Nex1978,Haydock1984:comparison}, 
is a precursor of the bond order potentials. Outlined in Appendix~\ref{app:quadrature}, we show that it produces an alternative scheme also satisfying Theorem~\ref{thm:BOP}. 
\end{remark}

\begin{remark}[Convergence of Derivatives] 
\label{rem:derivatives}
    In this more complicated nonlinear setting, obtaining results such as \cref{eq:lin-derivatives} is more subtle. 
   We require an additional assumption on $D_\ell$, which we believe maybe be typically satisfied, but we currently cannot justify it and have therefore postponed this discussion to Appendix~\ref{app:derivatives}.
    We briefly mention, however, that if $D_\ell$ is absolutely continuous (e.g., in periodic systems), we obtain
    \[
        \bigg|
            \frac{\partial}{\partial \s_m}
            \Big(
                O^\beta_\ell(\s) 
                - \ctNonlin\big( \Ham_{\ell\ell}, [\Ham^2]_{\ell\ell}, \dots, [\Ham^{N}]_{\ell\ell} \big)
            \Big)
        \bigg|
        \lesssim e^{-\frac{1}{2}\gamma_N N} e^{-\eta \,r_{\ell m}}.
    \]
\end{remark}

%

\subsection{The vacuum cluster expansion revisited}
\label{sec:concl:vacuum}
%
%
For $\ell\in \Lambda$, we denote by $\Ham\big|_{\ell;K}$ the Hamiltonian matrix corresponding to the finite subsystem $\{\ell\}\cup K \subset \Lambda$: for $k_1,k_2 \in \{\ell\} \cup K$,
\begin{align}
\label{eq:restriction}
    \big[\Ham\big|_{\ell;K}\big]_{k_1k_2} \coloneqq 
    h(\s_{k_1 k_2}) + \sum_{m \in \{ \ell \} \cup K} t(\s_{k_1 m}, \s_{k_2 m}) + \delta_{k_1k_2} v_{k_1} \mathrm{Id}_{\ctNumorbitals}.
\end{align} 

For an observable $O$, the vacuum cluster expansion as detailed in \S\ref{sec:cluster} is constructed as follows:
\begin{align}
    O_\ell^{N,\mathrm{vac}}(\s) &\coloneqq \sum_{n=0}^{N-1} 
    \sum_{\above
        {k_1,\dots,k_n \not= \ell}
        {k_1 < \dots < k_n}
    }
    V^{(n)}(\s_\ell; \s_{\ell k_1},\dots, \s_{\ell k_n})
    \qquad \text{where} \label{revisited:vacuum1}\\ 
    V^{(n)}(\s_\ell; \s_{\ell k_1}, \dots, \s_{\ell k_n}) 
    &= 
    \sum_{K \subseteq \{k_1,\dots,k_n\}}
    (-1)^{n - |K|}
    O\big(\Ham\big|_{\ell;K}\big)_{\ell\ell}. \label{revisited:vacuum2}
\end{align}
%
%
%
Therefore, on defining the \textit{spectral measure}  
$
    D_{\ell;K} \coloneqq \sum_s \delta\big( \,\cdot\, - \lambda_s(K) \big) |[\psi_s(K)]_{\ell}|^2
$
where $\big(\lambda_s(K), \psi_s(K) \big)$ the are normalised eigenpairs of $\Ham\big|_{\ell;K}$, we may write the vacuum cluster expansion as in \S\ref{sec:framework}:
\begin{align}\label{eq:vac_measure}
    O^{N,\mathrm{vac}}_\ell(\s) = \int O \, \mathrm{d}D_{\ell}^{N,\mathrm{vac}}
    \qquad \text{where} \qquad 
    D_{\ell}^{N,\mathrm{vac}} \coloneqq \sum_{n=0}^{N-1} 
    \sum_{\above
        {k_1,\dots,k_n \not= \ell}
        {k_1 < \dots < k_n}
    }
    \sum_{K \subseteq \{k_1,\dots,k_n\}}
    (-1)^{n-|K|}D_{\ell;K}. 
\end{align}

While $D_{\ell}^{N,\mathrm{vac}}$ is a generalised signed measure (with values in $\mathbb R \cup \{\pm \infty\}$), all moments are finite:
\begin{align}
    \int x^j \, \mathrm{d}D_{\ell}^{N,\mathrm{vac}}(x) 
    %
    %
    &= \sum_{
    \above
        {\ell_1,\dots,\ell_{j-1}}
        {|\{ \ell,\ell_1,\dots,\ell_{j-1}\}| \leq N}
    }
    \Ham_{\ell \ell_1} \Ham_{\ell_1 \ell_2} \dots \Ham_{\ell_{j-1}\ell}. \label{eq:vac_moments}
\end{align}
Equation~\cref{eq:vac_moments} follows from the proof of Proposition~\ref{prop}, see \cref{pf:body-order3}. In particular, the first $N$ moments of $D_\ell^{N,\mathrm{vac}}$ are exact. Therefore, we may apply the general error estimate \cref{eq:general_estimate} and describe the various features of $D_\ell^{N,\mathrm{vac}}$ which provide mathematical intuition for the slow convergence of the vacuum cluster expansion:
\begin{itemize}
    \item \textit{Spectral Pollution.} When splitting the system up into arbitrary subsystems as is the case in the vacuum cluster expansion, one expects significant spectral pollution in the band gaps, leading to a reduction in the convergence rate,
    
    \item \textit{Regularity of $D_\ell^{N,\mathrm{vac}}$.} The approximate LDOS is a linear combination of countably many Dirac deltas and does not have bounded variation. Moreover, $D_\ell^{N,\mathrm{vac}}$ has values in $\mathbb R \cup\{\pm\infty\}$.
\end{itemize}


\subsection{Self-consistency}
\label{sec:nonlin}
%
%
%

Throughout this section, we suppose that the effective potential is a function of a self-consistent electron density: that is, \cref{eq:fd} becomes the following nonlinear equation:
\begin{align}\label{eq:SC}
    \rho^\star = F^\beta\big( \s(\rho^\star) \big)
\end{align}
where $\s(\rho) \coloneqq \big(\bm r, v(\rho), \z\big)$. \hypertarget{EP}{We shall assume that the effective potential satisfies the following:}
\begin{assumption}{\textbf{(EP)}}
    We suppose that $v \colon \mathbb R^\Lambda \to \mathbb R^\Lambda$ is twice continuously differentiable with 
    \[
        \big| \nabla v(\rho)_{\ell k} \big|
        %
        %
        \leq C e^{-\gamma_v \,r_{\ell k}}
    \]
    for some $\gamma_v > 0$.
\end{assumption}

\begin{remark}
    \textit{(i)} For a smooth function $\widetilde{v} \colon \mathbb R \to \mathbb R$, the effective potential 
    $v(\rho)_\ell \coloneqq \widetilde{v}(\rho_\ell)$
    satisfies \asEP. This leads to the simplest abstract nonlinear tight binding models discussed in \cite{ELu10, Thomas2020:scTB}.
    \textit{(ii)} The (short-ranged) Yukawa potential defined by
    $v(\rho)_\ell \coloneqq \sum_{m \not= \ell} \frac{\rho_m - Z_m}{r_{\ell m}} e^{-\tau \, r_{\ell m}}$ 
    (for some $\tau > 0$) also fits into this general framework. This setting already covers many important modelling scenarios and also serves as a crucial stepping stone towards charge equilibration under full Coulomb interaction, which goes beyond the scope of the present work.
\end{remark}

The main result of this section is the following: if there exists a self-consistent solution $\rho^\star$ to \cref{eq:SC}, then we can approximate $\rho^\star$ with self-consistent solutions to the following approximate self-consistency equation:
\begin{align}\label{eq:SC-approx}
    \rho_{N} = I_N{F}^\beta\big( \s(\rho_N) \big),
\end{align}
for sufficiently large $N$. The operator $I_N F^\beta$ is a linear body-ordered approximation of the form we analyzed in detail in \S~\ref{sec:linear}.

To do this, we require a natural \hypertarget{STAB}{stability assumption} on the electronic structure problem, which was employed for example in~\cite{ELu10, ELu2012,  Thomas2020:scTB}:
\begin{assumption}{\textbf{(STAB)}}
    The \emph{stability operator} $\mathscr L(\rho)$ is the Jacobian of 
    $\rho \mapsto F^\beta\big( \s(\rho) \big)$.
    We say electron densities $\rho^\star$ solving \cref{eq:SC} are \emph{stable} if $I - \mathscr L(\rho^\star)$ is invertible as a bounded linear operator $\ell^2 \to \ell^2$.
\end{assumption}

\begin{remark}[Stability]
    \textit{(i)} The stability condition of Theorem~\ref{thm:nonlin} is a minimal starting assumption that naturally arises from the analysis 
    \cite{
        ELu10, 
        ELu2012,
        Thomas2020:scTB
    }. 
    For example, if $\rho$ is a stable self-consistent electron density, then there exists $\phi^{(m)} \in \ell^2(\Lambda)$ such that \cite{Thomas2020:scTB}:
    \[
        \frac{\partial \rho_\ell}{\partial \s_m} 
        = \Big[ \big( I - \mathscr L(\rho) \big)^{-1} \phi^{(m)} \Big]_\ell.
    \]
    \textit{(ii)} As noted in \cite{ELu10} (in a slightly simpler setting), the stability condition of Theorem~\ref{thm:nonlin} is automatically satisfied for multi-lattices with $\nabla v$ positive semi-definite. In fact, in this case the stability operator is negative semi-definite. 
\end{remark}

\begin{theorem}\label{thm:nonlin}
    For $\s$ satisfying Definition~\ref{def:config}, suppose that $\rho^\star$ is a corresponding stable self-consistent electron density.
    Then, for $N$ sufficiently large, there exist self-consistent solutions $\rho_{N}$ of \cref{eq:SC-approx} such that
    \begin{align}
        \big\|
            \rho_{N} - \rho^\star
        \big\|_{\ell^\infty}
        \leq C e^{-\gamma_N N}, 
    \end{align}
    where $\gamma_N$ are the constants from Theorem~\ref{thm:bodyOrder_interpolation} applied to $\s(\rho^\star)$.
\end{theorem}

\begin{corollary}\label{cor:sc-observables}
    Suppose that $\rho^\star$ and $\rho_N$ are as in Theorem~\ref{thm:nonlin} and denote by $O^\mathrm{sc}_\ell(\bm u) \coloneqq O_\ell\big( \s(\rho^\star) \big)$ a self-consistent local observable as in \cref{eq:SC_observables}. Then, 
    \[
        \big| O_\ell^\mathrm{sc}(\s) - I_N O_\ell\big( \s(\rho_N) \big) \big|
        \leq C e^{-\gamma_N N},
    \]
    where $\gamma_N$ are the constants from Theorem~\ref{thm:bodyOrder_interpolation} applied to $\s(\rho^\star)$.
\end{corollary}


In order for this result to be of any practical use, we need to solve the non-linear equation \cref{eq:SC-approx} for the electron density via a self-consistent field (SCF) procedure. Supposing we have the electron density $\rho^{i}$ and corresponding state $\s^{i} \coloneqq \s(\rho^{i})$ after $i$ iterations, we diagonalise the Hamiltonian $\Ham(\s^{i})$ and hence evaluate the output density $\rho^\mathrm{out} = I_N{F}^\beta(\s^{i})$. At this point, since the simple iteration $\rho^{i+1} = \rho^{\mathrm{out}}$ does not converge in general, a mixing strategy, possibly combined with Anderson acceleration \cite{Chupin2020}, is used in order to compute the next iterate. The analysis of such mixing schemes is a major topic in electronic structure and numerical analysis in general and so we only present a small step in this direction. 



\begin{prop}[Stability]\label{lem:stable}
    The approximate electron densities $\rho_N$ arising from Theorem~\ref{thm:nonlin} are \emph{stable} in the following sense: $I - \mathscr L_N(\rho_N) \colon \ell^2 \to \ell^2$ is an invertible bounded linear operator where $\mathscr L_N$ is the Jacobian of $\rho \mapsto I_NF^\beta\big( \s(\rho) \big)$.  Moreover, $\big( I - \mathscr L_N(\rho_N) \big)^{-1}$ is uniformly bounded in $N$ in operator norm.
\end{prop}

\begin{theorem}\label{cor:SCF}
    For $\s$ satisfying Definition~\ref{def:config}, suppose that $\rho_{N}$ is a corresponding approximate self-consistent electron density stable in the sense of Proposition~\ref{lem:stable}. For fixed $\rho^0$, we define $\{\rho^i\}_{i = 0}^\infty$ via the Newton iteration 
    \[
        \rho^{i+1} = \rho^i 
        - \big(
            I - {\mathscr L}_N(\rho^i)
        \big)^{-1} 
        \Big(
            \rho^i - I_N{F}^\beta\big(\s(\rho^i)\big)
        \Big).
    \]
    Then, for $\|\rho^0 - \rho_{N}\|_{\ell^\infty}$ sufficiently small, the Newton iteration converges quadratically to $\rho_{N}$.
\end{theorem}

A more thorough treatment of these SCF results is beyond the scope of this work. See
\cite{
    Cances2020:SCF,
    Herbst2020:SCF,
    Levitt:screening
} 
for recent results in the context of Hartree-Fock and Kohn-Sham density functional theory. For a recent review of SCF in the context density functional theory, see \cite{Woods2019}. 

\begin{remark}
    It is clear from the proofs of Theorems~\ref{thm:nonlin}~and~\ref{cor:SCF} that as long as the approximate scheme $F^{\beta,N}$ satisfies
    \[
        \left|F^\beta_\ell(\s) - F^{\beta,N}_\ell(\s)\right|
        \lesssim e^{-\gamma_N N}
        \qquad \text{and} \qquad
        \left| \frac{\partial F^\beta_\ell(\s)}{\partial v_m} - \frac{\partial F^{\beta,N}_\ell(\s)}{\partial v_m} \right|
        \lesssim e^{-\frac{1}{2}\gamma_N N}e^{-\eta r_{\ell m}},
    \]
    then we may approximate \cref{eq:SC} with approximate self-consistent solutions $\rho_N = F^{\beta,N}\big(\s(\rho_N)\big)$. In particular, as long as we have the estimate from Remark~\ref{rem:derivatives} (see Appendix~\ref{app:derivatives} for the technical details), then we may use the nonlinear approximation scheme $\ctNonlin$ from Theorem~\ref{thm:BOP} in Theorems~\ref{thm:nonlin}~and~\ref{cor:SCF}. In this case, we obtain error estimates that are (asymptotically) independent of the discrete spectrum.
\end{remark}

\section{Conclusions and Discussion}
The main result of this work is a sequence of rigorous results about body-ordered approximations of a wide class of properties extracted from tight-binding models for condensed phase systems, the primary example being the potential energy landscape. Our results demonstrate that exponentially fast convergence can be obtained, provided that the chemical environment is taken into account. In the spirit of our previous results on the locality of interaction \cite{ChenOrtner16,Thomas2020:scTB,ChenOrtnerThomas2019:locality}, these provide further theoretical justification --- albeit qualitative --- for widely assumed properties of atomic interactions. More broadly, our analysis illustrates how to construct general low-dimensional but systematic representations of high-dimensional complex properties of atomistic systems. Our results, as well as potential generalisations, serve as a starting point towards a rigorous end-to-end theory of multi-scale and coarse-grained models, including but not limited to machine-learned potential energy landscapes. 

In the following paragraphs will make further remarks on the potential applications of our results, and on some apparent limitations of our analysis.

\subsection{Representation of atomic properties}
Our initial motivation for studying the body-order expansion was to explain the (unreasonable?) success of machine-learned interatomic potentials~\cite{Behler2007-ng,Bartok2010-mv,Shapeev2016-pd}, and our remarks will focus on this topic, however in principle they apply more generally. 

Briefly, given an {\it ab initio} potential energy landscape (PEL) $E^{\rm QM}$ for some material one formulates a parameterised interatomic potential 
\[
    E(\{\bm r_\ell\}_\ell) = 
    \sum_{\ell} \varepsilon({\bm \theta}, \{ {\bm r}_{\ell k} \}_{k \neq \ell} )
\]
and then ``learn'' the parameters ${\bm \theta}$ by fitting them to observations of the reference PEL $E^{\rm QM}$. A great variety of such parameterisations exist, including but not limited to neural networks~\cite{Behler2007-ng}, kernel methods~\cite{Bartok2010-mv} and symmetric polynomials~\cite{Shapeev2016-pd, BachmayrEtAl2019:cluster_expansion, Drautz2019:cluster_expansion}. 
Symmetric polynomials are linear regression schemes where each basis function has a natural body-order attached to it. It is particularly striking that for very low body-orders of four to six these schemes are able to match and often outperform the more complex nonlinear regression schemes~\cite{Shapeev2016-pd, 2020-pace1, Zuo2020-ba}. Our analysis in the previous sections provides a partial explanation for these results, by justifying why one may expect that a reference {\it ab initio} PEL intrinsically has a low body-order. Moreover, classical approximation theory can now be applied to the body-ordered components as they are finite-dimensional to obtain new approximation results where the curse of dimensionality is alleviated.

Our results on nonlinear representations are less directly applicable to existing MLIPs, but rather suggest new directions to explore. Still, some connections can be made. The BOP-type construction of \S~\ref{sec:bop},
\begin{equation} \label{eq:bop:conclusion}
    \ctNonlin\big( \Ham_{\ell\ell}, [\Ham^2]_{\ell\ell}, \dots, [\Ham^{N}]_{\ell\ell} \big)
\end{equation}
points towards a blending of machine-learning and BOP techniques that have not been explored to the best of our knowledge. A second interesting connection is to the overlap-matrix based fingerprint descriptors (OMFPs) introduced in~\cite{Zhu2016-ih} where a global spectrum for a small subcluster is used as a descriptor, while \eqref{eq:bop:conclusion} can be understood as taking the projected spectrum as the descriptor. Thus, Theorem~\ref{thm:BOP} suggests (1) an interesting modification of OMFPs which comes with guaranteed completeness to describe atomic properties; and (2) a possible pathway towards proving completeness of the original OMFPs.

Finally, our self-consistent representation of \S~\ref{sec:nonlin} motivates how to construct compositional models, reminiscent of artificial neural networks, but with minimal nonlinearity that is moreover physically interpretable. Although we did not pursue it in the present work, this is a particularly promising starting point to incorporate meaningful electrostatic interaction into the MLIPs framework.

\subsection{Linear body-ordered approximation: the preasymptotic regime}
\label{sec:concl:preasymptotics}
Possibly the most significiant limitation of our analysis of the linear body-ordered approximation scheme is that the estimates deteriorate when defects cause a pollution of the point spectrum. Here, we briefly demonstrate that this appears to be an asymptotic effect, while in the pre-asymptotic regime this deterioration is not noticable. 

To explore this we choose a union of intervals $E \supseteq \sigma(\Ham)$ and a polynomial $P_N$ of degree $N$ and note
\begin{align}
    \label{eq:err-estimate}
    \Big|\big[O(\Ham) - P_N(\Ham)\big]_{\ell\ell}\Big| 
    \leq \big\| O(\Ham) - P_N(\Ham) \big\|_{\ell^2 \to \ell^2}
    = \big\| O - P_N \big\|_{L^\infty(\sigma(\Ham))}
    \leq \big\| O - P_N \big\|_{L^\infty(E)}.
\end{align}
We then construct interpolation sets (Fej\'er sets) such that the corresponding polynomial interpolant gives the optimal asymptotic approximation rates (for details of this construction, see \S\ref{pf:pot-theory}-\S\ref{sec:optimal}). We then contrast this with a best $L^\infty(E)$-approximation, and with the nonlinear approximation scheme from Theorem~\ref{thm:BOP}. We will observe that the non-linearity leads to improved asymptotic but comparable pre-asymptotic approximation errors.

\begin{figure}[htb]
    \begin{subfigure}[b]{.49\textwidth}
        \centering
        \includegraphics[width=\textwidth]{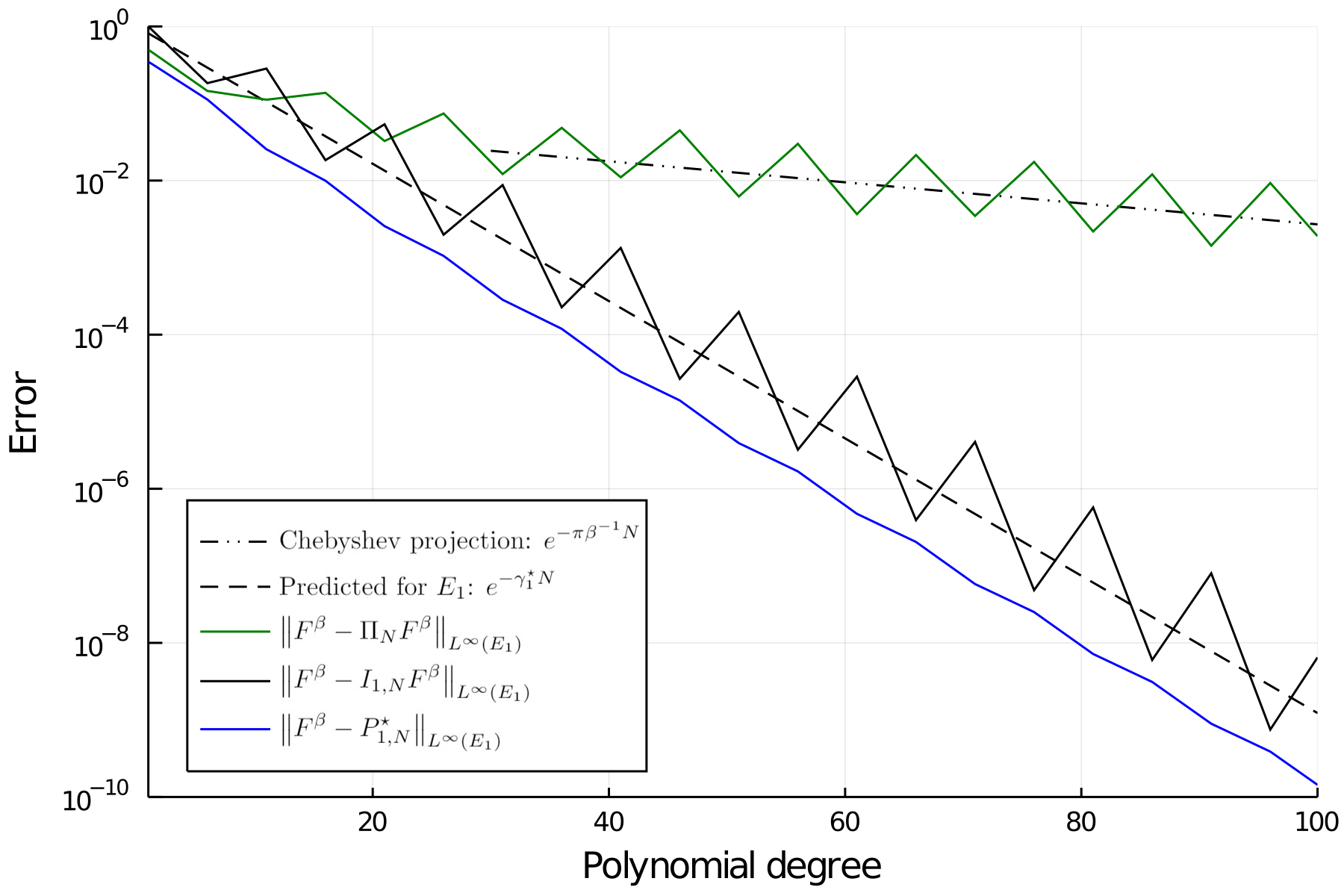}
       \caption{$E_1 = [-1,a]\cup[b,1]$.}
        \label{fig:pre_asymptotics_1}
    \end{subfigure}
    \begin{subfigure}[b]{.49\textwidth}
        \centering
        \includegraphics[width=\textwidth]{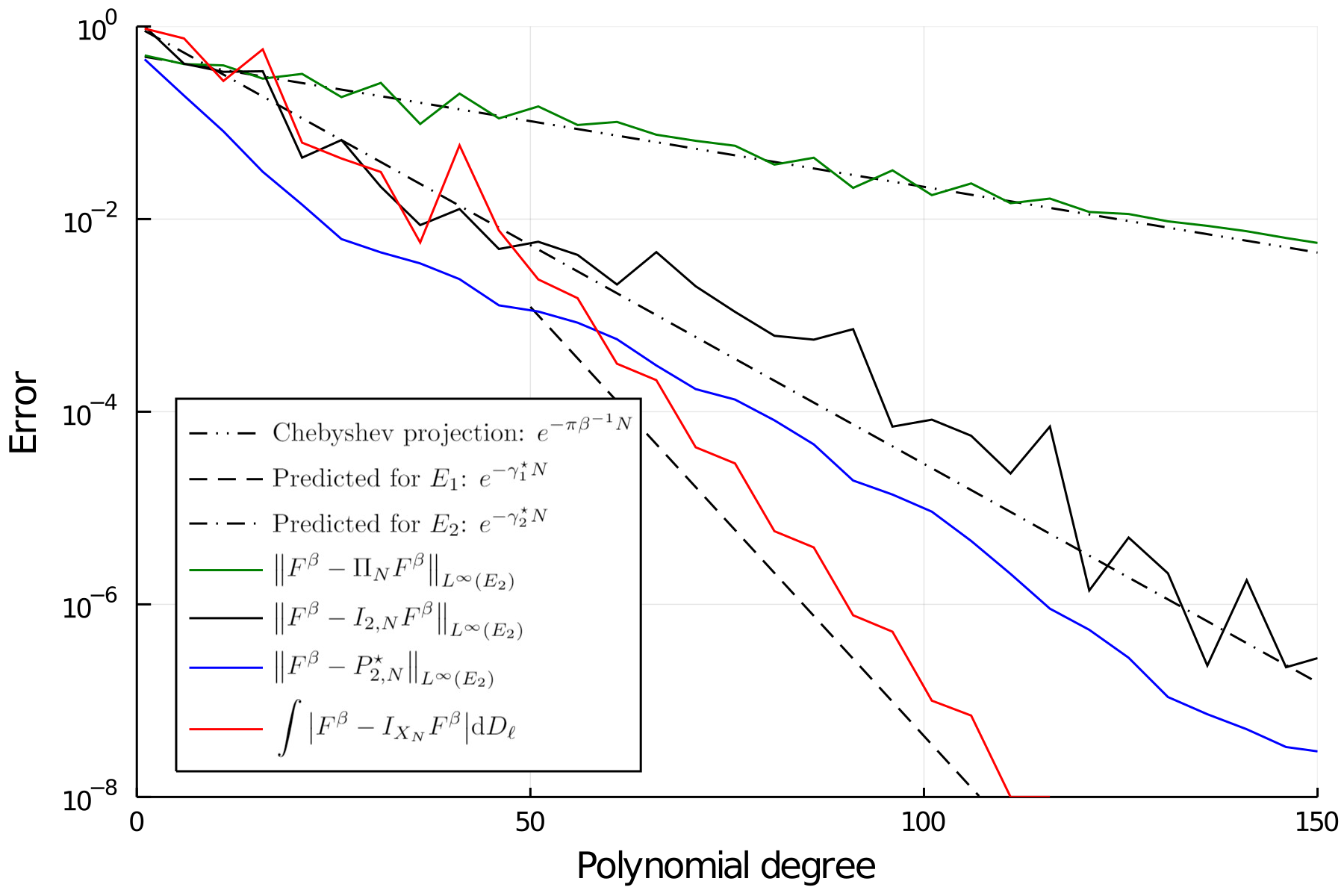}
        \caption{$E_2 = [-1,a]\cup[c,d]\cup[b,1]$ and $\{\lambda_j\} = \{c, \tfrac{c+d}{2}, d\}$.}
        \label{fig:pre_asymptotics_2}
    \end{subfigure}
     \caption{Approximation errors for Chebyshev projection (green), polynomial interpolation in Fej\'er sets on $E_j$ (black), best $L^\infty(E_j)$ polynomial approximation (blue), and, for $j=2$, errors in the nonlinear approximation scheme (red). We also plot the corresponding predicted asymptotic rates (from \cref{eq:Bernstein_error}, \cref{eq:optimal_poly_approx}, and Theorem~\ref{thm:bodyOrder_interpolation}). 
     Here, we only plot data points for $N \in \{1,6,11,16,\dots\}$.}
     \label{fig:pre_asymptotics}
\end{figure}

As a representative scenario we consider the Fermi-Dirac distribution $F^\beta(z) = (1 + e^{\beta z})^{-1}$ with $\beta = 100$ and both the ``defect-free'' case 
$E_1 \coloneqq [-1,a]\cup[b,1]$ and 
$E_2 \coloneqq [-1,a] \cup [c,d] \cup [b,1]$  
with the parameters $a = -0.2, b = 0.2, c = -0.06$, and $d = -0.03$. 
Then, for fixed polynomial degree $N$ and $j \in \{1,2\}$, we construct the $(N+1)$-point Fej\'er set for $E_j$ and the corresponding polynomial interpolant $I_{j,N}F^\beta$. 
Moreover, we consider a polynomial $P^\star_{j,N}$ of degree $N$ minimising the right hand side of \cref{eq:err-estimate} for $E = E_j$. Then, in Figure~\ref{fig:pre_asymptotics}, we plot the errors 
$\| F^\beta - I_{j,N} F^\beta \|_{L^\infty(E_j)}$
and 
$\| F^\beta - P_{j,N}^\star \|_{L^\infty(E_j)}$
for both $j = 1$ (Fig.~\ref{fig:pre_asymptotics_1}) and $j=2$ (Fig.~\ref{fig:pre_asymptotics_2}) against the polynomial degree $N$ together with the theoretical asymptotic convergence rates for best $L^\infty(E_j)$ polynomial approximation \cref{eq:optimal_poly_approx}.

What we observe is that, as expected, introducing the interval $[c,d]$ into the approximation domain drastically affects the asymptotic convergence rate and the errors in the approximation based on interpolation. While the best approximation errors follow the asymptotic rate for larger polynomial degree, it appears that, pre-asymptotically, the errors are significantly reduced. We also see that the approximation errors are significantly better than the general error estimate $\| F^\beta - \Pi_N F^\beta \|_{L^\infty} \lesssim e^{-\pi \beta^{-1} N}$ where $\Pi_N$ is the Chebyshev projection operator (see \S\ref{sec:projection}). 

Moreover, in Figure~\ref{fig:pre_asymptotics_2}, we plot the errors when using a nonlinear approximation scheme satisfying Theorem~\ref{thm:BOP}. In this simple experiment, we consider the Gauss quadrature rule 
$\ctNonlin \coloneqq \int I_{X_N} F^\beta \mathrm{d}D_\ell$
where $X_N$ are the zeros of the degree $N+1$ orthogonal polynomial 
(see Appendix~\ref{app:quadrature}) with respect to 
$
    \mathrm{d}D_\ell(x) \coloneqq 
    \big( 
        \chi_{E_0}(x)  + \sum_{j} \delta(x - \lambda_j) 
    \big) \mathrm{d}x, 
$
for a finite set $\{\lambda_j\} \subset [c,d]$. While $D_\ell$ does not correspond to a physically relevant Hamiltonian, the same procedure may be carried out for any measure supported on $E_1$ with $\mathrm{supp}\, D_\ell \cap [c,d]$ finite. Then plotting the upper bounds $\int | F^\beta - I_{X_{N}}F^\beta | \mathrm{d}D_\ell$, we observe improved asymptotic convergence rates that agree with that of the ``defect-free'' case from Figure~\ref{fig:pre_asymptotics_1}. However, the improvement is only observed in the asymptotic regime which corresponds to body-orders never reached in practice.

\section*{Acknowledgements}

We gratefull acknowledge stimulating discussions with Simon Etter.
 
HC is supported by the Natural Science Foundation of China under grant 11971066.
CO is supported by EPSRC Grant EP/R043612/1, Leverhulme Research Project Grant RPG-2017-191 and 
by the Natural Sciences and Engineering Research Council of Canada (NSERC) [funding reference number IDGR019381].
JT is supported by EPSRC as part of the MASDOC DTC, Grant No. EP/HO23364/1. 
 
\section{Proofs}
\label{sec:proofs}

\subsection{Preliminaries}

Here, we introduce the concepts needed in the proofs of the main results. 

\subsubsection{Hermite Integral Formula}
\label{pf:Hermite}

For a finite interpolation set $X\subset \mathbb C$, we let $\ell_X(z) \coloneqq \prod_{x\in X}(z - x)$ be the correpsonding \textit{node polynomial}. 

For fixed $z\in \mathbb C \setminus X$, we suppose that $O$ is analytic on an open neighbourhood of $X\cup\{z\}$. Then, for a simple closed positively oriented contour (or system of contours) $\mathscr C$ contained in the region of analyticity of $O$, encircling $X$, and avoiding $\{z\}$, we have 
\begin{align}\label{eq:Hermite:prelim}
    I_X O(z)
    = \frac{1}{2\pi i} 
        \oint_{\mathscr C}
        \frac
            {\ell_X(\xi) - \ell_X(z)}
            {\ell_X(\xi)}
        \frac
            {O(\xi)}
            {\xi - z}
        \mathrm{d}\xi.
\end{align}

If, in addition, $\mathscr C$ encircles $\{z\}$, then 
\begin{align}\label{eq:Hermite:prelim-2}
    O(z) - I_X O(z)
    = \frac{1}{2\pi i} 
        \oint_{\mathscr C}
        \frac
            {\ell_X(z)}
            {\ell_X(\xi)}
        \frac
            {O(\xi)}
            {\xi - z}
        \mathrm{d}\xi.
\end{align}
The proof of these facts is a simple application of Cauchy's integral formula, \cite{BakNewman10:complexanalysis,bk:Trefethen2019}.

\subsubsection{Resolvent Calculus} \label{pf:resolvent-calculus}

Given a configuration $\s$, we consider the Hamiltonian
$\Ham = \Ham(\s)$ 
and functions $O$ analytic in some neighbourhood of the spectrum $\sigma(\Ham)$. We define $O(\Ham)$ via the holomorphic functional calculus \cite{bk:spectral}:
\begin{align}\label{eq:holomorphic}
    O(\Ham) \coloneqq -\frac{1}{2\pi i}\oint_{\mathscr C} O(z) (\Ham - z)^{-1}\mathrm{d}z
\end{align}
where $\mathscr C$ is a simple closed positively oriented contour (or system of contours) contained in the region of analyticity of $O$ and encircling the spectrum $\sigma(\Ham)$.

The following Combes--Thomas resolvent estimate \cite{CombesThomas1973} will play a key role in the analysis:
\begin{lemma}[Combes-Thomas]\label{lem:CT}
    Suppose that $\s$ satisfies Definition~\ref{def:config} and $z \in \mathbb C$ is contained in a bounded set with
    $\mathrm{dist}\left(z, \sigma\big(\Ham(\s)\big)\right)>0$ 
    and 
    $\mathfrak{d}\coloneqq \mathrm{dist}\left(z, \sigma\big(\Ham(\s^\mathrm{ref})\big)\right) > \delta$. 
    Then, there exists a constant $C>0$ such that
    \begin{gather*}
        \left|
            \left[(\Ham(\s) - z)^{-1}\right]_{\ell k}
        \right|
        \leq C_{\ell k} e^{-\gamma_\mathrm{CT} r_{\ell k}}, 
        \qquad \text{where}\\
        C_{\ell k} \coloneqq 2 \mathfrak{d}^{-1} + C  e^{-\gamma_\mathrm{CT} (|\bm r_\ell| + |\bm r_k| - |\bm r_{\ell k}| )}
    \end{gather*}
    and $\gamma_\mathrm{CT} \coloneqq c \min\{1, \mathfrak{d}\}$ and $c>0$ depends on $h_0, \gamma_0, d$ and $\min_{\ell\not=k} r_{\ell k}$. 
\end{lemma}
\begin{proof}
    A proof with $\gamma_{\mathrm{CT}}$ depending instead on $\mathrm{dist}\left(z, \sigma\big(\Ham(\s)\big)\right)$ can be found in \cite{ChenOrtner16}. A low-rank update formula leads to the improved ``defect-independent'' result \cite{ChenOrtnerThomas2019:locality} where the exponent only depends on the distance between $z$ and the reference spectrum. See \cite{Thomas2020:scTB} for an explicit description of $\gamma_\mathrm{CT}$ in terms of the constants $\gamma_0, d$ and the non-interpenetration constant $\min_{\ell\not=k} r_{\ell k}$.
\end{proof}

A key observation for arguments involving forces (or more generally, derivatives of the analytic quantities of interest) is that the Combes-Thomas estimate allows us to bound derivatives of the resolvent operator:
\begin{lemma}\label{lem:der_resolvent}
    Suppose that $z \in \mathbb C$ with $\mathfrak{d}\coloneqq \mathrm{dist}\left(z, \sigma\big(\Ham(\s)\big)\right) > 0$. Then, 
    \[
        \left|
        \frac
            {\partial\left[(\Ham(\s) - z)^{-1}\right]_{\ell k}}
            {\partial \s_m}
        \right|
        \leq 4 h_0 \mathfrak{d}^{-2} e^{-\frac{1}{2}\min\{\gamma_0, \gamma_{\mathrm{CT}}\} (r_{\ell m} + r_{m k})}
    \]
    where $\gamma_\mathrm{CT}$ is the Combes-Thomas constant from Lemma~\ref{lem:CT} and $\gamma_0$ is the constant from~\asTB.
\end{lemma}
\begin{proof}
    This result can be found in the previous works 
    \cite{
        ChenOrtner16,
        ChenLuOrtner18,
        ChenOrtnerThomas2019:locality
    },
    but we give a brief sketch for completeness.
    
    Derivatives of the resolvent have the following form:
    \begin{align}\label{eq:der-resolvent}
        \frac
            {\partial (\Ham(\s) - z)^{-1}}
            {\partial \s_m}
        = - (\Ham(\s) - z)^{-1}
        \frac
            {\partial \Ham(\s)}
            {\partial \s_m}
        (\Ham(\bm{r},v) - z)^{-1}.
    \end{align}
    The result follows by applying the Combes-Thomas resolvent estimates together with the fact that the Hamiltonian is short-ranged~\asTB. 
    
    Assuming that the Hamiltonian has higher derivatives that are also short-ranged, higher order derivatives of the resolvent can be treated similarly \cite{ChenOrtner16}.
\end{proof}

\subsubsection{Local Observables}
Firstly, we note that $F^\beta(\,\cdot\,)$ is analytic away from the simple poles at $\pi \beta^{-1} (2\mathbb{Z} + 1)$. Moreover, 
$G^\beta(\,\cdot\,)$
can be analytically continued onto the open set
$\mathbb C \setminus \left\{ \mu + i r \colon r\in\mathbb R, |r| \geq \pi\beta^{-1} \right\}$ \cite{ChenOrtnerThomas2019:locality}.
Therefore, we may consider \cref{eq:holomorphic} with 
$O = F^\beta$
or 
$G^\beta$ 
and a contour $\mathscr C_\beta$ encircling $\sigma\left(\Ham\right)$ and avoiding 
$\mathbb C \setminus \left\{ \mu + i r \colon r\in\mathbb R, |r| \geq \pi\beta^{-1} \right\}$. 
Therefore, we may choose $\mathscr C_\beta$ so that the constant $\mathfrak{d}$, from Lemma~\ref{lem:CT}, is proportional to $\beta^{-1}$. Moreover, if there is a spectral gap, the constant $\mathfrak{d}$ is uniformly bounded below by a positive constant multiple of $\mathsf{g}$ as $\beta \to \infty$.

In the case of insulators at zero Fermi-temperature, we take $\mathscr C_\infty$ encircling 
$\sigma\left(\Ham(\s)\right)\cap(-\infty, \mu)$ 
and avoiding the rest of the spectrum. Therefore, we may choose $\mathscr C_\infty$ so that the constant $\mathfrak{d}$, from Lemma~\ref{lem:CT}, is proportional to $\mathsf{g}$.

Following \cite[Lemma~4]{ChenOrtner16}, we can conclude that 
$\sigma(\Ham) \subset [\underline{\sigma},\overline{\sigma}]$ 
for some 
$\underline{\sigma}, \overline{\sigma}$ 
depending on $h_0, \gamma_0, v, d$ and $\min_{\ell\not=k} r_{\ell k}$. This means that, the contours $\mathscr C_\beta$ can be chosen to have finite length and, when applying Lemma~\ref{lem:CT}, we have
$\gamma_\mathrm{CT} = c \min\{1, \max\{ \beta^{-1}, \mathsf{g} \}\}$.

Moreover, for all $0 < \mathsf{b} < \pi$ and bounded sets $A_\beta \subset A \subset  \mathbb C$ such that 
\[
    \mathrm{dist}(z, \{ \mu + i r \colon r\in \mathbb R, |r| \geq \pi \beta^{-1}\}) \geq \mathsf{b}\beta^{-1} \qquad \text{for all } z\in A_\beta,
\]
both $F^\beta(\,\cdot\,)$ and $G^\beta(\,\cdot\,)$ are uniformly bounded on $A_\beta$ independently of $\beta$ \cite[Lemma~5.2]{ChenOrtnerThomas2019:locality}.

\subsubsection{Chebyshev Projection and Interpolation in Chebyshev Points}
\label{sec:projection}
We denote by $\{T_n\}$ the \textit{Chebyshev polynomials} (of the first kind) satisfying $T_n(\cos \theta) = \cos n\theta$ on $[-1,1]$ and, equivalently, the recurrence $T_{0} = 1, T_1 = x$, and $T_{n+1}(x) = 2x T_n(x) - T_{n-1}(x)$.

For $O$ Lipshitz continuous on $[-1,1]$, there exists an absolutely convergent Chebyshev series expansion: there exists $c_n$ such that $O(z) = \sum_{n=0}^\infty c_n T_n(z)$. For maximal polynomial degree $N$, the corresponding projection operator is denoted $\Pi_N O(z) \coloneqq \sum_{n=0}^N c_n T_n(z)$. This approach is a special case of the Kernel Polynomial Method (KPM) which we briefly review in Appendix~\ref{app:KPM}.

On the other hand, supposing that the interpolation set is given by the \textit{Chebyshev points} $X = \{ \cos \frac{j\pi}{N} \}_{0 \leq j \leq N}$, we may expand the polynomial interpolant $I_N O \coloneqq I_{X} O$ in terms of the Chebyshev polynomials: there exists $c_n^\prime$ such that $I_N O(z) = \sum_{n=0}^N c_n^\prime T_n(z)$.

For functions $O$ that can be analytically continued the \textit{Bernstein ellipse} $E_\rho \coloneqq \{ \frac{1}{2}( z + z^{-1} ) \colon |z| = \rho\}$ for $\rho> 1$, the corresponding coefficients $\{c_n\}$, $\{c^\prime_n\}$ decay exponentially with rate $\rho$. This leads to the following error estimates 
\begin{align}\label{eq:Bernstein_error}
    \| O - \Pi_N O \|_{L^\infty([-1,1])} + \| O - I_N O \|_{L^\infty([-1,1])} \leq 6 \|O\|_{L^\infty(E_\rho)} \frac{\rho^{-N}}{\rho - 1}.
\end{align}

For $O^\beta = F^\beta$ or $G^\beta$, these estimates give an exponential rate of convergence with exponent depending on $\sim \beta^{-1}$. Indeed, after scaling $\Ham$ so that the spectrum is contained in $[-1,1]$, we obtain
\begin{align}\label{eq:error_estimate_chebyshev}
    \Big|
        O_\ell^\beta(\s) - \Pi_N O_\ell^\beta(\s)     
    \Big|
    \leq \Big\| 
        O^\beta(\Ham) - \Pi_N O^\beta(\Ham)
    \Big\|_{\ell^2 \to \ell^2}
    \leq \| O^\beta - \Pi_N O^\beta \|_{L^\infty([-1,1])},
\end{align}
and we conclude by directly applying \cref{eq:Bernstein_error}. The same estimate also holds for $I_N$ (or any polynomial). 

For full details of all the statements made in this subsection, see \cite{bk:Trefethen2019}.

\subsubsection{Classical Logarithmic Potential Theory}
\label{pf:pot-theory}
In this section, we give a very brief introduction to classical potential theory in order to lay out the key notation. For a more thorough treatment, see \cite{Ransford1995} or \cite{ch:Levin2006, Saff2010, bk:Trefethen2019, phd:Etter2019}.

It can be seen from the Hermite integral formula \cref{eq:Hermite:prelim-2} that the approximation error for polynomial interpolation may be determined by taking the ratio of the size of the node polynomial $\ell_X$ at the approximation points to the size of $\ell_X$ along an appropriately chosen contour. Logarithmic potential theory provides an elegant mechanism for choosing the interpolation points so that the asymptotic behaviour of $\ell_X$ can be described.

We suppose that $E \subset \mathbb C$ is a compact set. 
We will see that choosing the interpolation nodes as to maximise the geometric mean of pairwise distances provides a particularly good approximation scheme:
\begin{align}\label{eq:delta}
    \delta_n(E) \coloneqq \max_{z_1, \dots, z_n \in E} \Big( \prod_{1 \leq i < j \leq n} |z_i - z_j| \Big)^{\frac{2}{n(n-1)}}.
\end{align}
Any set $\mathcal F_n \subset E$ attaining this maximum is known as a \textit{Fekete set}. It can be shown that the quantities $\delta_n(E)$ form a decreasing sequence and thus converges to what is known as the \textit{transfinite diameter}: $\tau(E) \coloneqq \lim\limits_{n\to\infty} \delta_n(E)$.  

We let $\ell_n(z)$ denote the node polynomial corresponding to a Fekete set and note that 
\begin{align}\label{eq:Fekete-asymptotic}
    |\ell_n(z)| \delta_n(E)^{\frac{n(n-1)}{2}} 
    =  
    \max_{z_0,\dots,z_{n} \in E \colon z_0 = z} 
    \prod_{0\leq i < j \leq n}
    |z_i - z_j|
    \leq \delta_{n+1}(E)^{\frac{n(n+1)}{2}}.
\end{align}
Therefore, rearranging \cref{eq:Fekete-asymptotic}, we obtain $\lim_{n\to\infty} \|\ell_n\|_{L^\infty(E)}^{1/n} \leq \tau(E)$. In fact, this inequality can be replaced with equality, showing that Fekete sets allow us to describe the asymptotic behaviour of the node polynomials on the domain of approximation. 

To extend these results, it is useful to recast the maximisation problem \cref{eq:delta} into the following minimisation problem, describing the minimal logarithmic energy attained by $n$ particles lying in $E$ with the repelling force $1/|z_i - z_j|$ between particles $i$ and $j$ lying at positions $z_i$ and $z_j$, respectively:
\begin{align}\label{eq:En}
    \mathcal E_n(E) \coloneqq 
    \min_{z_1,\dots,z_n \in E}
    \sum_{1 \leq i < j \leq n}
    \log \frac{1}{|z_i - z_j|}
    = \frac{n(n-1)}{2} \log \frac{1}{\delta_n(E)}.
\end{align}
Fekete sets can therefore be seen as minimal energy configurations and described by the normalised counting measure $\nu_n \coloneqq \frac{1}{n} \sum_{j = 1}^n \delta_{z_j}$ where $\mathcal F_n = \{z_j\}_{j = 1}^n$.

The minimisation problem \cref{eq:En} may be extended for general unit Borel measures $\mu$ supported on $E$ by defining the logarithmic potential and corresponding total energy by
\[
    U^\mu(z) \coloneqq \int \log\frac{1}{|z - \xi|} \mathrm{d}\mu(\xi) 
    \qquad \text{and} \qquad
    I(\mu)\coloneqq \iint \log\frac{1}{|z - \xi|}
    \mathrm{d}\mu(\xi)\mathrm{d}\mu(z).
\]
The infimum of the energy over the space of unit Borel measures supported on $E$, known as the \textit{Robin constant} for $E$, will be denoted $-\infty < V_E \leq +\infty$. The \textit{capacity} of $E$ is defined as $\mathrm{cap}(E)\coloneqq e^{-V_E}$ and is equal to the transfinite diameter \cite{Embree1999}. Using a compactness argument, it can be shown that there exists an \textit{equilibrium measure} $\omega_E$ with $I(\omega_E) = V_E$ and, in the case $V_E <\infty$, by the strict convexity of the integral, $\omega_E$ is unique \cite{bk:Saff1997}. Moreover, if $V_E < \infty$ (equivalently, if $\mathrm{cap}(E) > 0$), then $U^{\omega_E}(z) \leq V_E$ for all $z \in \mathbb C$, with equality holding on $E$ except on a set of capacity zero (we say this property holds \textit{quasi-everywhere}).

Moreover, if $\mathrm{cap}\,E > 0$, then it can be shown that the normalised counting measures, $\nu_n$, corresponding to a sequence of Fekete sets weak-$\star$ converges to $\omega_E$. Since $U^{\nu_n}(z) = \frac{1}{n}\log \frac{1}{|\ell_n(z)|}$, the weak-$\star$ convergence allows one to conclude that
\begin{align}
    \lim_{n\to\infty} \|\ell_n\|_{L^\infty(E)}^{1/n} &= \mathrm{cap}(E), 
    \qquad \text{and} \qquad 
    \lim_{n\to\infty} |\ell_n(z)|^{1/n} = e^{-U^{\omega_E}(z)} \eqqcolon \mathrm{cap}(E) e^{g_E(z)}
    \label{eq:limitF}
\end{align}
uniformly on compact subsets of $\mathbb C \setminus E$. Here, we have defined the \textit{Green's function} $g_E(z) \coloneqq V_E - U^{\omega_E}(z)$, which describes the asymptotic behaviour of the node polynomials corresponding to Fekete sets. We therefore wish to understand the Green's function $g_E$. 

\subsubsection{Construction of the Green's Function}
\label{pf:Green}
Now we restrict our attention to the particular case where $E \subset \mathbb R$ is a union of finitely many compact intervals of non-zero length.

\begin{figure}[htb]
    \centering
   
    \includegraphics[width=.9\textwidth]{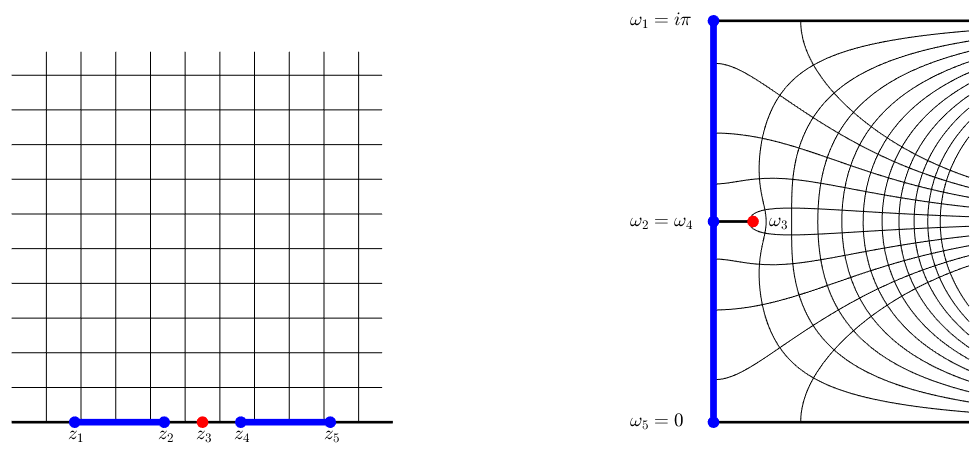}
    
    \caption{The Schwarz–Christoffel mapping $G_E$ with $E = [z_1,z_2] \cup [z_4, z_5]$ which maps the upper half plane (left) onto the infinite slit strip $\{\omega \in \mathbb C \colon \mathrm{Re}\, \omega > 0, \, \mathrm{Im}\, \omega \in (0,\pi)\}$ (right), is continuous on $\{ z \in \mathbb C \colon \mathrm{Re} z \geq 0\}$ and maps the intervals $[z_1,z_2]$, $[z_4,z_5]$ to $[\omega_1,\omega_2], [\omega_4,\omega_5] \subset i [0,\pi]$, respectively. We also plot the image of an $10 \times 10$ equi-spaced grid. A parameter problem is solved in order to obtain $z_3$ and thus $\omega_3$ and $\omega_2 = \omega_4$ whereas the other constants are fixed. Here, we take $z_1 = -1, z_2 = -\ep, z_4 = \ep, z_5 = 1, \omega_1 = i\pi, \omega_5 = 0$ with $\ep = 0.3$.}
    \label{fig:G}
\end{figure}

It can be shown that the Green's function $g_E$ satisfies the following Dirichlet problem on $\mathbb C \setminus E$ \cite{Ransford1995}:
\begin{subequations}\label{eq:Green}
\begin{equation}
    \Delta g_E(z) = 0 \quad \text{ on } \mathbb C \setminus E,
\end{equation}
\begin{equation}\label{eq:log_blowup}
    g_E(z) \sim \log |z| \quad \text{ as } |z| \to \infty, 
\end{equation}
\begin{equation}\label{eq:zeroOnE}
   g_E(z) = 0 \quad \text{ on } E.
\end{equation}
\end{subequations}
In fact, it can be shown that \cref{eq:Green} admits a unique solution \cite{Ransford1995} and thus \cref{eq:Green} is an alternative definition of the Green's function. Using this characterisation, it is possible to explicitly construct the Green's function $g_E$ as follows. In the upper half plane, $g_E(z) = \mathrm{Re}(G_E(z))$ where 
$G_E\colon\{ z \in\mathbb C \colon \mathrm{Im}(z) \geq 0 \} \to \{ z\in\mathbb C \colon \mathrm{Re}(z) \geq 0, \mathrm{Im}(z) \in [0,\pi]\}$
is a conformal mapping on 
$\{z \colon \mathrm{Im}(z) > 0\}$
such that $G_E(E) = i[0,\pi]$, $G_E( \min E ) = i \pi$, and $G_E( \max E ) = 0$. Using the symmetry of $E$ with respect to the real axis, we may extend $\mathrm{Re}(G_E(z))$ to the whole complex plane via the Schwarz reflection principle. Then, one can easily verify that this analytic continuation satisfies \cref{eq:Green}. Since the image of $G_E$ is a (generalised) polygon, $z \mapsto G_E(z)$ is an example of a Schwarz–Christoffel mapping \cite{DriscollSC}. See Figure~\ref{fig:G} for the case $E =  [-1,-\ep]\cup[\ep,1]$.

\begin{figure}[htb]
    \centering
   
     \begin{subfigure}{.5\textwidth}
      \centering
      \includegraphics[width=.9\textwidth]{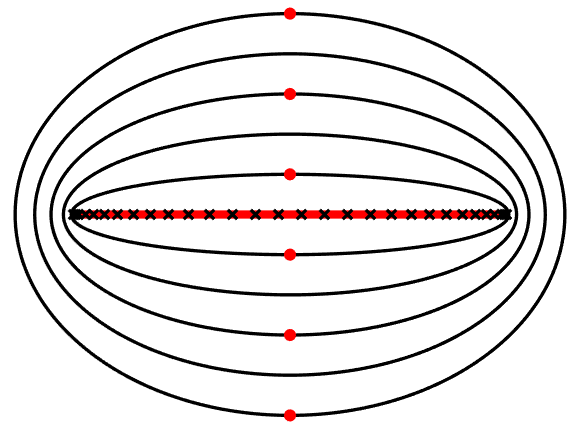}
      \caption{Metal: $E = [-1,1]$.}
      \label{fig:equipot-a}
    \end{subfigure}%
    \begin{subfigure}{.5\textwidth}
      \centering
      \includegraphics[width=.9\textwidth]{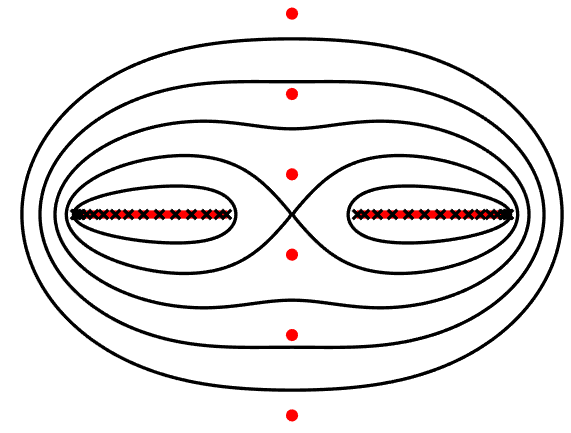}
      \caption{Insulator: $E = [-1,-\ep]\cup[\ep,1]$,  $\ep = 0.30$ ($2$~s.f.).}
     \label{fig:equipot-b}
    \end{subfigure}
    
    \caption{Equi-potential curves $\mathscr C_{r_k} \coloneqq \{z\in \mathbb C \colon e^{g_E(z)} = r_k \}$ for both metals \textsc{(a)} and insulators \textsc{(b)} where $\frac{1}{2}(r_k - r_k^{-1}) = \frac{k \pi}{\beta}$ for $k \in \{ 1,2,3,4,5 \}$ and $\beta = 10$. In the case of metals \textsc{(a)}, the equi-potential curves agree with Bernstein ellipses. We also plot the poles of $F^\beta(\,\cdot\,)$ which determine the maximal admissible integration contours: for \textsc{(a)}, we can take contours $\mathscr C_r$ for all $r < r_1$ and, for \textsc{(b)}, the contour $\mathscr C_{r_2}$ can be used for all positive Fermi-temperatures (we have chosen the gap carefully so that $\mathscr C_{r_2}$ self-intersects at $\mu$). Shown in black crosses are $30$ Fej\'er points in each case. To create these plots we consider an integral formula for the Green's function $z\mapsto g_{E}(z)$ \cite{phd:Etter2019} and use the Schwarz-Christoffel \textsc{matlab} toolbox \cite{Driscoll1996,DriscollSC} to approximate these integrals.}
    \label{fig:equipot}
\end{figure}

We shall briefly discuss the construction of the Schwarz–Christoffel mapping $G_E$ for $E = [-1,\ep_-]\cup[\ep_+,1]$. We define the \textit{pre-vertices} $z_1 = -1, z_2 = \ep_-, z_4 = \ep_+, z_5 = 1$ and wish to construct a conformal map $G_E$ with $G_E(z_k) = \omega_k$ as in Figure~\ref{fig:G}. For simplicity, we also define $z_0 \coloneqq -\infty$ and $z_6 \coloneqq \infty$ and observe that because the image is a polygon, $\mathrm{arg} \, G_E^\prime(z)$ must be constant on each interval $(z_{k-1}, z_k)$ and  
\begin{align}\label{eq:argDiff}
    \mathrm{arg} \, G_E^\prime(z_k^+) - \mathrm{arg} \, G_E^\prime(z_k^-) = (1 - \alpha_k) \pi
\end{align}
where $z_k^- \in (z_{k-1}, z_k)$, $z_k^+ \in  (z_{k}, z_{k+1})$, and $\alpha_k \pi$ is the interior angle of the infinite slit strip at vertex $\omega_k$ (that is, $\alpha_1 = \alpha_2 = \alpha_4 = \alpha_5 = \frac{1}{2}$ and $\alpha_3 = 2$). After defining $z^\alpha \coloneqq |z|^\alpha e^{i \alpha \,\mathrm{arg}\,z}$ where $\mathrm{arg}\,z \in (-\pi,\pi]$, we can see that for $z \in (z_{k-1}, z_k)$, we have $\mathrm{arg} \prod_{j = k}^5 (z - z_j)^{\alpha_j - 1} = \sum_{j = k}^{5} (\alpha_j - 1)\pi$ and so the jump in the argument of $z \mapsto \prod_{j = 1}^5 (z - z_j)^{\alpha_j - 1}$ is $(1 - \alpha_k)\pi$ at $z_k$ as in \cref{eq:argDiff}. Therefore, integrating this expression, we obtain
\begin{align}
\label{eq:SC-mapping}
    G_E(z) = A + B \int^z_1 
    \frac
        {\zeta - z_3}
        {\sqrt{\zeta + 1}\sqrt{\zeta - \ep_-}\sqrt{\zeta - \ep_+}\sqrt{\zeta - 1}}
    \mathrm{d}\zeta.
\end{align}
Since $G_E(1) = A$, we take $A = 0$ (to ensure \cref{eq:zeroOnE} holds). Moreover, since the real part of the integral is $\sim \log|z|$ as $|z| \to \infty$, we apply \cref{eq:log_blowup} to conclude $B = 1$. Finally, we can choose $z_3$ such that $\mathrm{Re}\,G_E(z) = 0$ for all $z \in E$. That is,
\begin{align}
\label{eq:z3}
   z_3 \in (\ep_-, \ep_+) \, \colon \qquad 
    \int_{\ep_-}^{\ep_+} 
    \frac
        {\zeta - z_3}
        {\sqrt{\zeta + 1}\sqrt{\zeta - \ep_-}\sqrt{\ep_+-\zeta}\sqrt{1 - \zeta}}
    \mathrm{d}\zeta = 0.
\end{align}
For more details, see \cite{phd:Etter2019}. We use the Schwarz–Christoffel toolbox \cite{DriscollSC} in \textsc{matlab} to evaluate \cref{eq:SC-mapping} and plot Figure~\ref{fig:equipot}. 

For the simple case $E \coloneqq [-1,1]$, by the same analysis, we can disregard $z_2,z_3,z_4$ and $\omega_2,\omega_3,\omega_4$ and integrate the corresponding expression to obtain the closed form $G_{[-1,1]}(z) = \log ( z + \sqrt{z - 1}\sqrt{z + 1})$. 

A similar analysis allows one to construct conformal maps from the upper half plane to the interior of any polygon. For further details, rigorous proofs and numerical considerations, see \cite{Driscoll2002}.

\subsubsection{Interpolation Nodes}
\label{sec:nodes}
The only difficulty in obtaining \cref{eq:limitF} in practice is the fact that Fekete sets are difficult to compute. An alternative, based on the Schwarz–Christoffel mapping $G_E$, are \textit{Fej\'er points}. For equally spaced points 
$\{\zeta_j\}_{j=1}^{n}$
on the interval $i [0,\pi]$, the $n^\text{th}$ Fej\'er set is defined by 
$\{ G_E^{-1}(\zeta_j)\}_{j=1}^n$. 
Fej\'er sets are also \textit{asymptotically optimal} in the sense that \cref{eq:limitF} is satisfied where $\ell_n$ is now the node polynomial corresponding to $n$-point Fej\'er set. 

Another approach is to use \textit{Leja points} which are generated by the following algorithm: for fixed $z_1,\dots,z_n$, the next interpolation node $z_{n+1}$ is constructed by maximising $\prod_{j = 1}^n |z_j - z|$ over all $z \in E$. Sets of this form are also asymptotically optimal \cite{Taylor2010:Leja} for any choice of $z_1 \in E$. Since we have fixed the previous nodes $z_1,\dots,z_{n}$, the maximisation problem for constructing $z_{n+1}$ is much simpler than that of \cref{eq:delta}.

More generally, if the normalised counting measure corresponding to a sequence of sets $\{z_j\}_{j=1}^n \subset E$ weak-$\star$ converges to the equilibrium measure $\omega_E$, then the corresponding node polynomials satisfy \cref{eq:limitF}.

For the simple case where $E=[-1,1]$, many systems of zeros or maxima of sequences of orthogonal polynomials are asymptotically optimal in the sense of \cref{eq:limitF}. 
In fact, since the equilibrium measure for $[-1,1]$ is the arcsine measure \cite{Saff2010}
\[
    \mathrm{d}\mu_{[-1,1]}(x) = \frac{1}{\pi} \frac{1}{\sqrt{1-x^2}} \mathrm{d}x,
\]
any sequence of sets with this limiting distribution is asymptotically optimal. 
An example of particular interest are the \textit{Chebyshev points} $\{ \cos \frac{j\pi}{n} \}_{0 \leq j \leq n}$ given by the $n+1$ extreme points of the Chebyshev polynomials defined by $T_n( \cos\theta ) = \cos n \theta$. 

\subsubsection{Asymptotically Optimal Polynomial Approximations}
\label{sec:optimal}
Suppose that $E$ is the union of finitely many compact intervals of non-zero length and $O\colon E \to \mathbb C$ extents to an analytic function in an open neighbourhood of $E$. On defining $\mathscr C_\gamma \coloneqq \{ z \in \mathbb C \colon g_E(z) = \gamma \}$, we denote by $\gamma^\star$ the maximal constant for which $O$ is analytic on the interior of $\mathscr C_{\gamma^\star}$. We let $P_N^\star$ be the best $L^\infty(E)$-approximation to $O$ in the space of polynomials of degree at most $N$ and suppose that $I_N$ is a polynomial interpolation operator in $N+1$ points satisfying \cref{eq:Green}. Then, the Green's function $g_E$ determines the asymptotic rate of approximation for not only polynomial interpolation but also for best approximation:
\begin{align}\label{eq:optimal_poly_approx}
    \lim_{N\to\infty} \| O - P_N^\star \|_{L^\infty(E)}^{1/N}
    = \lim_{N\to\infty} \| O - I_NO \|_{L^\infty(E)}^{1/N}
    = e^{-\gamma^\star}.
\end{align}

For a proof that the asymptotic rate of best approximation is given by the Green's function see \cite{Saff2010}. The result for polynomial interpolation uses the Hermite integral formula and \cref{eq:limitF}, see \cref{eq:Hermite} and \cref{eq:lim_ell}, below. 

\subsection{Linear body-order approximation}
\label{sec:body-order}

In this section, we use the classical logarithmic potential theory from \S\ref{pf:pot-theory} to prove the approximation error bounds for interpolation. However, we first show that polynomial approximations lead to body-order approximations:

\begin{proof}[Proof of Proposition~\ref{prop}]
    We first simplify the notation by absorbing the effective potential and two-centre terms into the three-centre summation:
    \begin{align}
        \Ham(\s)_{k_1 k_2} = \sum_m \Ham_{k_1 k_2 m}, 
        \quad \textrm{where} \quad 
        \Ham_{k_1k_2 m} \coloneqq \begin{cases}
            \frac{1}{2} h( \s_{k_1k_2} ) + \delta_{k_1k_2} v_{k_1} \mathrm{Id}_{\ctNumorbitals}, &\text{if } m \in \{k_1,k_2\}, \\ 
            t(\s_{k_1 m}, \s_{k_2 m}), &\text{if } m \not\in \{k_1,k_2\}.
        \end{cases}
    \end{align}

    Now, supposing that $I_XO(z) = \sum_{j = 0}^{|X|-1} c_j z^j$, we obtain
    \begin{align}
        I_XO_\ell(\s) &= 
        \mathrm{tr}
       \sum_{j = 0}^{|X|-1} c_{j} \sum_{\ell_1,\dots,\ell_{j-1}} 
        \Ham_{\ell \ell_1} \Ham_{\ell_1 \ell_2} \dots \Ham_{\ell_{j-1} \ell} \nonumber\\ 
        &=  
        \mathrm{tr}
        \sum_{j = 0}^{|X|-1} c_{j} \sum_{
            \above 
                {\ell_1,\dots,\ell_{j-1}}
                {m_1,\dots,m_{j}}
        }
        \Ham_{\ell \ell_1 m_1} \Ham_{\ell_1 \ell_2 m_2} \dots \Ham_{\ell_{j-1} \ell m_j}, \label{pf:body-order}
    \end{align}
    where the first two terms in the outer summation are $c_0$ and $c_1 \Ham_{\ell\ell}$. 
    Now, for a fixed body-order $(n+1)$, and $k_1 < \dots < k_{n}$ with $k_l \not=\ell$, we construct $V_{nN}(\s_\ell; \s_{\ell k_1}, \dots, \s_{\ell k_n})$ by collecting all terms in \cref{pf:body-order} with $0 \leq j \leq |X|-1$ and $\{ \ell, \ell_1, \dots, \ell_{j-1}, m_1, \dots, m_j\} = \{\ell, k_1, \dots, k_n\}$. In particular, the maximal body-order in this expression is $2(|X|-1)$ for three-centre models and $|X|-1$ in the two-centre case.
    
    More explicitly, using the notation \cref{eq:restriction}, we have
    \begin{align}
        V_{nN}(\s_\ell; \s_{\ell k_1}, \dots, \s_{\ell k_n}) 
        &= \mathrm{tr}
        \sum_{j = 0}^{|X|-1} c_{j} \sum_{
            \above 
                {\ell_1,\dots,\ell_{j-1},m_1,\dots,m_j}
                {\{\ell,\ell_1,\dots,\ell_{j-1},m_1,\dots,m_j\} = \{\ell,k_1,\dots,k_n\}}
        }
        \Ham_{\ell \ell_1 m_1} \Ham_{\ell_1 \ell_2 m_2} \dots \Ham_{\ell_{j-1} \ell m_j} \label{pf:body-order2}\\
        &= \mathrm{tr}
        \sum_{K \subseteq \{k_1,\dots,k_n\}} 
        (-1)^{n-|K|} 
        I_X O\big(\Ham\big|_{\ell;K}\big)_{\ell\ell}.\label{pf:body-order3}
    \end{align}
    Here, we have applied an inclusion-exclusion principle to ensure that we are not only summing over sites in $\{k_1,\dots,k_n\}$ but we select at least one of each site in this set. Indeed, if we choose $\ell_1, \dots, \ell_{j-1}, m_1, \dots, m_j$ such that $\{\ell,\ell_1, \dots, \ell_{j-1}, m_1, \dots, m_j\} = \{\ell\}\cup K_0$, then the expression $\Ham_{\ell \ell_1 m_1} \Ham_{\ell_1 \ell_2 m_2} \cdots \Ham_{\ell_{j-1} \ell m_j}$ appears in each term of \cref{pf:body-order3} with $K \supseteq K_0$ exactly once (with a $\pm$ sign). Therefore, the number of times $\Ham_{\ell \ell_1 m_1} \Ham_{\ell_1 \ell_2 m_2} \dots \Ham_{\ell_{j-1} \ell m_j}$ appears is exactly
    \[
        \sum_{l = 0}^{ n - |K_0| } (-1)^{n-|K_0|-l} 
        \genfrac{(}{)}{0pt}{0}
            { n - |K_0| }
            {l}
        = \begin{cases}
            1 & \text{if } |K_0| = n,\\
            0 &\text{otherwise}.
        \end{cases}
    \]
    That is, \cref{pf:body-order3} only contains the terms in the summation \cref{pf:body-order2}.
\end{proof}

\begin{proof}[Proof of Theorem~\ref{thm:bodyOrder_interpolation}]
    We let  
    $\ell_N(x) \coloneqq \prod_{j} (x - x_j^N)$ 
    be the node polynomial for 
    $X_N \coloneqq \{x_j^N\}_{j=0}^N$.
    %
    Again, we fix the configuration $\s$ and consider $\Ham \coloneqq \Ham(\s)$.
    
    Supposing that $\mathscr C$ is a simple closed positively oriented contour encircling 
    $\sigma(\Ham)$, 
    we apply the Hermite integral formula \cref{eq:Hermite:prelim-2} to obtain:
    \begin{align}
        \big| 
            O^\beta_\ell(\s) - I_{X_N} O^\beta_\ell(\s) 
        \big|
        &\leq \| 
            O^\beta( \Ham ) - I_{X_N} O^\beta( \Ham ) 
        \|_{\ell^2 \to \ell^2} 
        = \sup_{z\in \sigma(\Ham)} 
        \big| 
            O^\beta(z) - I_{X_N} O^\beta( z )  
        \big| \nonumber\\
        &\leq \sup_{z\in \sigma(\Ham)} 
        \left| \frac{1}{2\pi i} \oint_{\mathscr C}
        \frac
            {\ell_N(z)}
            {\ell_N(\xi)} 
        \frac
            {O^\beta(\xi)}
            {\xi - z}
        \mathrm{d}\xi \right| 
        \leq C \sup_{z \in \sigma(\Ham), \,
                \xi \in \mathscr C}
        \left|
            \frac
                {\ell_N(z)}
                {\ell_N(\xi)}
        \right|,
        \label{eq:Hermite}
    \end{align}
    where
    \begin{align}
    \label{eq:constant-prefactor}
        C \coloneqq \frac{\mathrm{len}(\mathscr C)}{2\pi}
        \frac
            {\max_{\xi \in \mathscr C} |O^\beta(\xi)|}
            {
                \mathrm{dist}\big(\mathscr C, \sigma(\Ham)\big)
            }.  
    \end{align}
    
    At this point we apply standard results of classical logarithmic potential theory (see, \S\ref{pf:pot-theory} or \cite{ch:Levin2006}) and conclude by noting that if the interpolation points are asymptotically distributed according to the equilibrium distribution corresponding to 
    $E\coloneqq I_- \cup I_+$, 
    then after applying \cref{eq:limitF}, we have
    \begin{align}\label{eq:lim_ell} 
        \lim_{N\to\infty}\left|
            \frac
                {\ell_N(z)}
                {\ell_N(\xi)}
        \right|^{\frac{1}{N}} 
        = e^{g_E(z) - g_E(\xi)}.
    \end{align}
    Here, the equilibrium distribution and the Green's function $g_E(z)$ are concepts introduced in \S\ref{pf:pot-theory} and \S\ref{pf:Green}. 
    
    Therefore, by choosing the contour
    $\mathscr C \coloneqq \{ \xi \in \mathbb C \colon g_E(\xi) = \gamma \}$ for $0 < \gamma < g_E(\mu + i\pi \beta^{-1})$, 
    the asymptotic exponents in the approximation error is $\gamma$.
    The maximal asymptotic convergence rate is given by $g_E(\mu + i\pi \beta^{-1})$ since $\mathscr C$ must be contained in the region of analyticity of $O^\beta$ and the first singularity of $O^\beta$ is at $\mu + i\pi \beta^{-1}$ (for $O^\beta = F^\beta$ or $G^\beta$).
    Examples of the equi-potential level sets $\mathscr C$ are given in Figure~\ref{fig:equipot}. 
    
    Using the Green's function results of \S\ref{pf:Green}, $g_E( \mu + i \pi \beta^{-1} ) = \mathrm{Re}\, G_E( \mu + i \pi \beta^{-1} )$ where $G_E$ is the integral \cref{eq:SC-mapping}. The asymptotic behaviour of this maximal asymptotic convergence rate for the separate $\beta \to \infty$ and $\mathsf{g}\to0$ limits can be found in \cite{phd:Etter2019,Shen2001}. Here, we consider the $\beta^{-1} + \mathsf{g} \to 0$ limit where the gap remains symmetric about the chemical potential $\mu$. 
    
   To simplify the notation we consider $I_- \cup I_+ = [-1,\ep_-] \cup [\ep_+,1]$ where $\ep_\pm = \mu \pm \frac{1}{2}\mathsf{g}$. By choosing to integrate \cref{eq:SC-mapping} along the contour composed of the intervals $[1,\mu]$ and $[\mu, \mu + i\pi \beta^{-1}]$, we obtain
    \begin{align}\label{eq:GE_expand}
        G_E(\mu + i\pi\beta^{-1}) = G_E(\mu) + \int_\mu^{\mu + i\pi\beta^{-1}} 
                \frac
                {\zeta - z_3}
                {\sqrt{\zeta +1} \sqrt{\zeta - \ep_-} \sqrt{\zeta - \ep_+} \sqrt{\zeta - 1}}
            \mathrm{d}\zeta.
    \end{align}
    Since $g_E(\mu) \sim \mathsf{g}$ as $\mathsf{g} \to 0$ \cite{phd:Etter2019}, we only consider the remaining term in \cref{eq:GE_expand}.

    For $\zeta \in \mu + i [0,\pi \beta^{-1} ]$, we have $c^{-1} \leq |\sqrt{\zeta \pm 1}| \leq c$, and so the integral in \cref{eq:GE_expand} has the same asymptotic behaviour as the following 
    \begin{align}
        \int_{\mu}^{\mu + i \pi \beta^{-1} }  
        \frac
            {\zeta - z_3}
            {\sqrt{\zeta - \ep_-} 
            \sqrt{\zeta - \ep_+}} 
        \mathrm{d}\zeta
        &=
        \mathsf{g} \int_{\frac{1}{2}}^{\frac{1}{2}+\frac{i\pi\beta^{-1}}{\mathsf{g}}}
            \frac{\sqrt{\zeta}}{\sqrt{\zeta - 1}} \mathrm{d}\zeta
        + (\ep_- - z_3) \int_{\frac{1}{2}}^{\frac{1}{2}+\frac{i\pi\beta^{-1}}{\mathsf{g}}}
            \frac{1}{\sqrt{\zeta}\sqrt{\zeta-1}}
            \mathrm{d}\zeta \label{eq:epminus} 
        %
        %
    \end{align}
    where we have used the change of variables $\widetilde{\zeta} = \frac{\zeta - \ep_-}{\ep_+-\ep_-}$. 
    Since the integrands are uniformly bounded along the domain of integration, \cref{eq:epminus} is $\sim \beta^{-1}$ as $\beta \to \infty$.
    
    The constant pre-factor in \cref{eq:constant-prefactor} is inversely proportional to the distance 
    $\mathrm{dist}\big( \mathscr C, \sigma(\Ham) \big)$
    between the contour $\mathscr C = \{ g_E = \gamma \}$ and the spectrum $\sigma(\Ham)$. In particular, since $g_E$ is uniformly Lipschitz with constant $L>0$ on the compact region bounded by $\mathscr C$, we have: there exists $\lambda \in \sigma(\Ham)$ and $\xi \in \mathscr C$ such that
    \[
        \mathrm{dist}\big( 
            \mathscr C, \sigma(\Ham)
        \big)
        = |\xi - \lambda| \geq \frac{1}{L} |g_E(\xi) - g_E(\lambda) | = \frac{1}{L} \gamma. 
    \]
    Therefore, choosing $\gamma$ to be a constant multiple of $g_E(\mu + i\pi\beta^{-1})$, we conclude that the constant pre-factor $C$ satisfies $C \sim (\mathsf{g} + \beta^{-1})^{-1}$ as $\mathsf{g} + \beta^{-1} \to 0$.

To extend the body-order expansion results to derivatives (in particular, to forces), we write the quantities of interest using resolvent calculus, apply Lemma~\ref{lem:der_resolvent} to bound the derivatives of the resolvent, and use the Hermite integral formula \cref{eq:Hermite} to conclude: for $\mathscr C_1$, $\mathscr C_2$ simple closed positively oriented contours encircling the spectrum $\sigma\big( \Ham(\s) \big)$ and $\mathscr C_1$, respectively, we have
\begin{align}
\label{eq:forces-body-order-1}
    \bigg|
        \frac
            { \partial O_\ell(\s) }
            {\partial \s_m}
        - 
        \frac
            { \partial I_{X_N} O_\ell(\s) }
            {\partial \s_m}
    \bigg|
    &= \frac{1}{2\pi} 
    \bigg|
        \oint_{\mathscr{C}_1} 
        \big(
            O(z) - I_{X_N}O(z)
        \big)
        \frac
            {\partial \big( 
                \Ham(\s) - z
            \big)^{-1}_{\ell\ell}}
            {\partial \s_m}
        \mathrm{d}z
    \bigg| \nonumber\\
    &= \frac{1}{4\pi^2} 
    \bigg|
        \oint_{\mathscr{C}_1} 
        \oint_{\mathscr{C}_2} 
        \frac
            {\ell_N(z)}
            {\ell_N(\xi)}
        \frac
            {O(\xi)}
            {\xi - z}
        \frac
            {\partial \big( 
                \Ham(\s) - z
            \big)^{-1}_{\ell\ell}}
            {\partial \s_m}
        \mathrm{d}\xi
        \mathrm{d}z
    \bigg| \nonumber\\
    &\leq C e^{-\eta r_{\ell m}}
    \sup_{z \in \mathscr C_1, \xi \in \mathscr C_2} 
    \bigg|
        \frac
            {\ell_N(z)}
            {\ell_N(\xi)}
    \bigg|.
\end{align}
We conclude by choosing appropriate contours $\mathscr C_l = \{g_E = \gamma_l\}$ for $l=1,2$ and applying \cref{eq:lim_ell}. 
\end{proof}

\subsubsection{The Role of the Point Spectrum}
\label{pf:point-spectrum}
To begin this section, we sketch the proof of Proposition~\ref{prop:pert_spec}:

\begin{proof}
[Proof of Proposition~\ref{prop:pert_spec}]
    \textit{(i) Sup-norm perturbations.} We suppose that $\sup_k \big[ |\bm r_k - \bm r^{\mathrm{ref}}_k| + |v_k - v_k^{\mathrm{ref}}| \big] \leq \delta$ for $\delta>0$ sufficiently small such that
    \begin{align*}
        \big|
            h( \s_{\ell k} ) - h( \s^{\mathrm{ref}}_{\ell k} )
        \big|
        &= \big|
            \nabla h( \xi_{\ell k} ) \cdot ( \bm r_{\ell k} - \bm r_{\ell k}^{\mathrm{ref}} )
        \big|
        \leq C \delta e^{-\frac{1}{2} \gamma_0 r_{\ell k}}, \quad \text{and}
        \\
        \big|
            t( \s_{\ell m}, \s_{km}) 
            - t( \s_{\ell m}^{\mathrm{ref}}, \s_{km}^{\mathrm{ref}})
        \big|
        &= \big|
            \nabla_1 t( \xi^{(1)}_{\ell m}, \zeta^{(1)}_{km} ) 
            \cdot ( \bm r_{\ell m} - \bm r_{\ell m}^{\mathrm{ref}} )
            +
            \nabla_2 t( \xi^{(2)}_{\ell m}, \zeta^{(2)}_{k m} ) 
            \cdot ( \bm r_{k m} - \bm r_{k m}^{\mathrm{ref}} )
        \big| \\
        &\leq C \delta e^{-\frac{1}{2} \gamma_0(r_{\ell m} + r_{k m})},
    \end{align*}
    where, $\xi_{\ell k} \in [\bm r_{\ell k}, \bm r_{\ell k}^{\mathrm{ref}}]$, $\xi^{(l)}_{\ell m} \in [\bm r_{\ell m}, \bm r_{\ell m}^{\mathrm{ref}}]$, and $\zeta^{(l)}_{k m} \in [\bm r_{km}, \bm r_{km}^{\mathrm{ref}}]$. Therefore, if $\psi \in \ell^2$, we have
    \begin{align}
        \left\| 
            \big( 
                \Ham(\s) - \Ham(\s^{\mathrm{ref}}) 
            \big) 
            \psi
        \right\|_{\ell^2}^2
        &\leq C \delta^2 \sum_{\ell k}
        e^{-\frac{1}{2} \gamma_0 r_{\ell k}} |\psi_k|^2 
        \leq C \delta^2 \|\psi\|_{\ell^2}^2.  
    \end{align}
    Therefore, applying standard results from perturbation theory \cite[p.~291]{bk:Kato1995}, we obtain
    \[
        \mathrm{dist}\Big(
            \sigma\big( \Ham(\s) \big), 
            \sigma\big( \Ham(\s^{\mathrm{ref}}) \big)
        \Big)
        \leq \left\|
            \Ham(\s) -  
            \Ham(\s^{\mathrm{ref}})
        \right\|_{\ell^2 \to \ell^2}
        \leq C\delta.
    \]

    \textit{(ii) Finite rank perturbations.} The finite rank perturbation result has been presented in \cite{OrtnerThomas2020:pointdefs} in a slightly different setting. We sketch the main idea here for completeness. 
    
    Since the essential spectrum is stable under compact (in particular, finite rank) perturbations \cite{bk:Kato1995}, the set 
    \[
        \sigma\big( \Ham(\s) \big) 
        \setminus B_\delta\big( \sigma\big( \Ham(\s^{\mathrm{ref}}) \big) \big) 
        \subseteq \sigma_\mathrm{disc}\big( \Ham(\s) \big) 
        \setminus B_\delta\big( \sigma_\mathrm{ess}\big( \Ham(\s^{\mathrm{ref}}) \big) \big) 
    \]
    is both compact and discrete and therefore finite.
\end{proof}

\begin{proof}[Proof of Theorem~\ref{thm:bodyOrder_interpolation_discrete}]
    Suppose that $\mathscr C$ is a simple closed contour encircling the spectrum 
    $\sigma\big(\Ham(\s)\big)$
    and $(\lambda_s, \psi_s)$ are normalised eigenpairs corresponding to the finitely many eigenvalues outside $I_- \cup I_+$. Therefore, we have
       \begin{align}\label{eq:point-defs}
            O^\beta_\ell(\s) - I_{X_N} O^\beta_\ell(\s) &= 
        \frac{1}{2\pi i}
            \oint_{\mathscr C} 
            \big(
                O^\beta(z) - I_{X_N} O^\beta(z)
            \big) 
            \,\mathrm{tr}\big[\big( 
                z - \Ham(\s) 
            \big)^{-1}\big]_{\ell\ell} \mathrm{d}z \nonumber\\ 
        &\qquad\qquad+
        \sum_s \big( 
            O^\beta(\lambda_s) - I_{X_N} O^\beta(\lambda_s) 
        \big) 
        \big|
            [\psi_s]_{\ell}
        \big|^2.
    \end{align}
    The first term of \cref{eq:point-defs} may be treated in the same way as in the proof of Theorem~\ref{thm:bodyOrder_interpolation}. Moreover, derivatives of this term may be treated in the same way as in \cref{eq:forces-body-order-1}. It is therefore sufficient to bound the remaining term and its derivative.
    
    Firstly, we note that the eigenvectors corresponding to isolated eigenvalues in the spectral gap have the following decay \cite{ChenOrtnerThomas2019:locality}: for $\mathscr C^\prime$ a simple closed positively oriented contour (or system of contours) encircling the $\{\lambda_s\}$, we have
    \begin{align}
        \sum_s
        \big|
            [\psi_s]_{\ell}
        \big|^2
        &= \frac{1}{2\pi} \bigg| 
            \oint_{\mathscr{C}^\prime}
                \big[ 
                    \big( 
                        \Ham(\s) - z
                    \big)^{-1} 
                \big]_{\ell\ell}
            \mathrm{d}z
        \bigg| \nonumber\\ 
        &= \frac{1}{2\pi} \bigg| 
            \oint_{\mathscr{C}^\prime}
                \big[ 
                    \big( 
                        \Ham(\s) - z
                    \big)^{-1} 
                    - 
                    \big( 
                        \Ham(\s^\mathrm{ref}) - z
                    \big)^{-1}
                \big]_{\ell\ell}
            \mathrm{d}z
        \bigg| \nonumber\\ 
        &\leq C e^{-\gamma_{\mathrm{CT}} [|\bm{r}_\ell| - R_\mathrm{def}]}, \label{eq:decay-evector}
    \end{align}
    where $\gamma_\mathrm{CT}$ is the Combes-Thomas constant from Lemma~\ref{lem:CT} with $\mathfrak{d} = \mathrm{dist}\big(\mathscr C^\prime, \sigma(\Ham(\s))\big)$. The constant pre-factor in \cref{eq:decay-evector} depends on the distance between the contour and the defect spectrum $\sigma\big( \Ham(\s) \big)$. Similar estimates hold for the derivatives. For full details on the derivation of \cref{eq:decay-evector}, see \cite[(5.18)--(5.21)]{ChenOrtnerThomas2019:locality}.
    
    Therefore, combining \cref{eq:decay-evector} and the Hermite integral formula, we conclude as in the proof of Theorem~\ref{thm:bodyOrder_interpolation}.
    %
    %
\end{proof}

\subsection{Non-linear body-order approximation}\label{pf:BOP}
In this section, we prove Theorem~\ref{thm:BOP} by applying the recursion method to reformulate the problem into a semi-infinite linear chain and replacing the far-field with vacuum.

\subsubsection{Recursion Method}\label{sec:recursion}
In the following we briefly introduce the recursion method \cite{Haydock1972:recursion, Haydock1972:recursionII}, a reformulation of the Lanczos process \cite{Lanczos1950}, which generates a tri-diagonal (Jacobi) operator $T$ \cite{Teschl:jacobioperators} whose spectral measure is $D_\ell$ and the corresponding sequence of orthogonal polynomials \cite{Freud:OP}. This process provides the basis for constructing approximations to the LDOS giving rise to nonlinear approximation schemes satisfying Theorem~\ref{thm:BOP}.

Recall that $D_\ell$ is the LDOS satisfying \cref{eq:LDOS}. We start by defining $p_0 \coloneqq 1$, $a_0 \coloneqq \int x \mathrm{d}D_\ell(x)$ and $b_1 p_1(x) \coloneqq x - a_0$ where $b_1$ is the normalising constant to ensure $\int p_1(x)^2 \mathrm{d}D_\ell(x) = 1$. 
Then, supposing we have defined $a_0, a_1, b_1, \dots, a_n, b_{n}$ and the polynomials $p_0(x), \dots, p_n(x)$, we set 
\begin{gather}\label{eq:recurrence}
    b_{n+1} p_{n+1}(x) \coloneqq (x - a_n) p_n(x) - b_n p_{n-1}(x), \qquad \text{with} \\
    \int p_{n+1}(x)^2 \mathrm{d}D_\ell(x) = 1, 
    \qquad 
    a_{n+1} \coloneqq \int x p_{n+1}(x)^2 \mathrm{d}D_\ell(x).
\end{gather}
Then, $\{p_n\}$ is a sequence of orthogonal polynomials with respect to $D_\ell$ 
(i.e.~$\int p_n p_m \mathrm{d}D_\ell = \delta_{nm}$) 
and we have
\begin{align}\label{eq:tridiag}
    T_N \coloneqq
    \Big(
        \int x p_n p_m \mathrm{d}D_\ell 
    \Big)_{
            0 \leq n,m \leq N
    }
    = 
    \begin{pmatrix}
        a_0 & b_1    &        &      \\
        b_1 & a_1    & \ddots &      \\
            & \ddots & \ddots & b_N  \\
            &        & b_N    & a_N  
    \end{pmatrix}
\end{align}
(see Lemma~\ref{lem:quadrature} for a proof). Moreover, we denote by $T$ the infinite symmetric tridiagonal matrix on $\mathbb N_0$ with diagonal $(a_n)_{n \in \mathbb N_0}$ and off-diagonal $(b_n)_{n\in\mathbb N}$.

\begin{remark}\label{rem:recurrence-modified}
    It will also prove convenient for us to renormalise the orthogonal polynomials by defining $P_n(x) \coloneqq b_n p_n(x)$ and $b_0 \coloneqq 1$. That is, 
    \begin{gather}\label{recurrence-modified}
        P_0(x) = 1, \quad 
        P_1(x) = x - a_0, \quad \text{and} \quad
        P_{n+1}(x) = \frac{x-a_n}{b_n} P_n(x) - \frac{b_n}{b_{n-1}}P_{n-1}(x), \quad \text{for }n\geq 1\\
        \label{recursion_coeff}
       b_{n+1}^2 = \int P_{n+1}(x)^2 \mathrm{d}D_\ell(x),
       \quad \text{and} \quad 
       a_{n+1} = \frac
            {\int x P_{n+1}(x)^2 \mathrm{d}D_\ell(x),}
            {b_{n+1}^2}.
    \end{gather}
    One advantage of this formulation is that it explicitly defines the coefficients $\{b_n\}$. 
\end{remark}


Therefore, if we have the first $2N+1$ moments $\Ham_{\ell\ell}, \dots, (\Ham^{2N+1})_{\ell\ell}$, it is possible to evaluate $Q_{2N+1}(\Ham)_{\ell\ell}$ (that is, $\int Q_{2N+1} \mathrm{d}D_\ell$) for all polynomials $Q_{2N+1}$ of degree at most $2N+1$, and thus compute $T_N$. In particular, for a fixed observable of interest $O$, we may write 
\begin{align}\label{eq:theta}
    \ctNonlin\big( \Ham_{\ell\ell}, \dots, [\Ham^{2N+1}]_{\ell\ell} \big) \coloneqq O(T_N)_{00}.
\end{align}

\begin{remark}
    In Appendix~\ref{app:generalBOP} we introduce more complex bond order potential (BOP) schemes based on the recursion method and show that they also satisfy Theorem~\ref{thm:BOP}.
\end{remark}

\subsubsection{Error Estimates} Equation~\cref{eq:theta} states that the nonlinear approximation scheme given by $\ctNonlin$ simply approximates the LDOS with the spectral measure of $T_N$ corresponding to $\bm e_0 \coloneqq (1,0,\dots,0)^\mathrm{T}$. We now show that $[(T_N)^n]_{00} = [T^n]_{00} = [\Ham^n]_{\ell\ell}$ for all $n \leq 2N+1$ and thus we may apply \cref{eq:general_estimate} to conclude. 

By the orthogonality, we have $[T^0]_{ij} = \int p_i(x) x^0 p_j(x) \mathrm{d}D_\ell(x) = \delta_{ij}$. Therefore, assuming $[T^n]_{ij} = \int p_i(x) x^n p_j(x) \mathrm{d}D_\ell(x)$, we can conclude 
\begin{align}
    [T^{n+1}]_{ij} = \sum_{k} [T^n]_{ik} T_{kj} 
    &= \int p_i(x) x^n \big[ 
        b_j p_{j-1}(x) + a_j p_j(x) + b_{j+1} p_{j+1}(x)
    \big]
    \mathrm{d}D_\ell(x) \\
    &= \int p_{i}(x) x^{n+1} p_{j}(x) \mathrm{d}D_\ell(x).
\end{align}
Here, we have applied \cref{eq:recurrence} directly. In particular, if $i=j=0$, we obtain $[T^n]_{00} = [\Ham^n]_{\ell\ell}$.

\subsubsection{Analyticity} To conclude the proof of Theorem~\ref{thm:BOP}, we show that $\ctNonlin$ as in \cref{eq:theta} extents to an analytic function on some open set $U \subset \mathbb C^{2N+1}$. Throughout this section, we use the rescaled orthogonal polynomials $\{P_n\}$ from Remark~\ref{rem:recurrence-modified}.

For a polynomial $P(x) = \sum_{j=0}^m c_j x^j$, we use the notation $\mathcal{L} P(z_1,\dots,z_m) \coloneqq c_0 + \sum_{j=1}^m c_j z_j$ for the linear function satisfying $P(x) = \mathcal{L}P(x,x^2,\dots,x^m)$. To extend the recurrence coefficients from \cref{recurrence-modified}, we start by defining
\begin{gather}
    b_0 = 1, \quad a_0(z_1) \coloneqq z_1, \quad 
    P_1(x;z_1)\coloneqq x - a_0(z_1) = x - z_1, \nonumber\\
    b_1^2(z_1,z_2) \coloneqq \mathcal{L}\big(x \mapsto P_1(x;z_1)^2\big)(z_1,z_2) = z_2 - z_1^2, \label{eq:rational_01}
    \\ 
    a_1(z_1,z_2,z_3) \coloneqq 
    \frac
        {\mathcal{L}\big(x\mapsto xP_1(x;z_1)^2\big)(z_1,z_2,z_3)}
        {b_1^2(z_1,z_2)} = 
    \frac
        {z_3 - 2 z_1z_2 + z_1^3}
        {z_2 - z_1^2}. \nonumber
\end{gather}

To simplify the notation, we write $\bm z_{1:m}$ for the $m$-tuple $(z_1,\dots,z_m)$. Given $a_0(z_1), \dots, a_n( \bm z_{1:2n+1})$ and $b_1(\bm z_{1:2}), \dots, b_n(\bm z_{1:2n})$, we define $P_{n+1}(x;\bm z_{1:2n+1})$ to be the polynomial in $x$ satisfying the same recursion as \cref{recurrence-modified} but as a function of $\bm z_{1:2n+1}$: 
\begin{align*}
    P_{n+1}(x;\bm z_{1:2n+1}) 
    = \frac
        {x-a_n(\bm z_{1:2n+1})}
        {b_n(\bm z_{1:2n})} 
    P_n(x;\bm z_{1:2n-1}) - 
    \frac
        {b_n(\bm z_{1:2n})}
        {b_{n-1}(\bm z_{1:2n-2})}
    P_{n-1}(x;\bm z_{1:2n-3}).
\end{align*}
With this notation, we define
\begin{align}
    b_{n+1}^2(\bm z_{1:2n+2}) 
    &\coloneqq \mathcal{L}\big(
        x \mapsto P_{n+1}(x;\bm z_{1:2n+1})^2
    \big)(\bm z_{1:2n+2}), 
        \label{eq:b_complex}
    \\ 
    a_{n+1}(\bm z_{1:2n+3}) 
    &\coloneqq 
    \frac
        {\mathcal{L}\big(
        x\mapsto xP_{n+1}(x;\bm z_{1:2n+1})^2
    \big)(\bm z_{1:2n+3})}
        {b_{n+1}^2(\bm z_{1:2n+2}) }
    .
         \label{eq:a_complex}
\end{align}
Since $P_{n+1}(x) = P_{n+1}(x;\Ham_{\ell\ell},\dots,[\Ham^{2n+1}]_{\ell\ell})$, we have extended the definition of the recursion coefficients \cref{recursion_coeff} to functions of multiple complex variables. 

We now show that $a_n(\bm z_{1:2n+1})$ and $b_n^2(\bm z_{1:2n})$ are rational functions. As a preliminary step, we show that both $P_{n+1}^2$ and $\frac{P_{n+1} P_n}{b_n}$ are polynomials in $x$ with coefficients given by rational functions of $a_n,b_n^2$ and all previous recursion coefficients. This statement is clearly true for $n = 0$: $P_1^2 = (x - a_0)^2$ and $\frac{P_1 P_0}{b_0} = x - a_0$. Therefore, by induction and noting that
\begin{align}
    P_{n+1}^2 &= \left( \frac{x-a_n}{b_n} P_n \right)^2 - 2 (x - a_n) \frac{P_nP_{n-1}}{b_{n-1}} + \frac{b_n^2}{b_{n-1}^2} P_{n-1}^2 \\
    \frac{P_{n+1} P_n}{b_n} &= \frac{x - a_n}{b_n^2} P_{n}^2 - \frac{P_nP_{n-1}}{b_{n-1}},
\end{align}
we can conclude. Therefore, by \cref{eq:rational_01} and \cref{eq:b_complex} and \cref{eq:a_complex}, we can apply another induction argument to conclude that $a_{n+1}(\bm z_{1:2n+3})$ and $b_{n+1}^2(\bm z_{1:2(n+1)})$ are rational functions.

We fix $N$ and define the following complex valued tri-diagonal matrix
\begin{align}\label{eq:complexTri} 
T_N(\bm z)\coloneqq 
\begin{pmatrix}
    a_0(z_1) & b_1^2(\bm z_{1:2}) &                  &          &      \\
    1        & a_1(\bm z_{1:3}) & b_2^2(\bm z_{1:4}) &           &      \\
             & 1           & \ddots             & \ddots   &  \\
            &                   &      \ddots       &  \ddots    &   b_N^2(\bm z_{1:2N})  \\
             &                 &                   & 1           & a_N(\bm z_{1:2N+1})  
\end{pmatrix}.
\end{align}
If $z_j = [\Ham^j]_{\ell\ell}$ for each $j = 1,\dots,N$, \cref{eq:complexTri} is similar to $T_N$ from \cref{eq:tridiag}.

Now, on defining $U \coloneqq \{ \bm z \in \mathbb C^{2N+1} \colon b_n^2(\bm z_{1:2n}) \not= 0 \,\, \forall n=1,\dots,N\}$, the mapping $U \to \mathbb C^{(N+1)\times (N+1)}$ given by $\bm z \mapsto T_N(\bm z)$ is analytic. Therefore, for appropriately chosen contours $\mathscr C_{\bm z}$ encircling $\sigma\big(T_N(\bm z) \big)$, we have
\begin{align}\label{pf:g_analytic}
    \ctNonlin(\bm z) \coloneqq O\big(T_N(\bm z)\big)_{00}
    = -\frac{1}{2\pi i} \oint_{\mathscr{C}_{\bm z}} O(\omega) 
    \Big[
        \big( T_N(\bm z) - \omega \big)^{-1}
    \Big]_{00} \mathrm{d}\omega.
\end{align}
In particular, $\ctNonlin$ is an analytic function on $\big\{ \bm z \in U \colon O \text{ analytic in an open neighbourhood of } \sigma\big( T_N(\bm z) \big) \big\}$.

\begin{remark}
    Since $\mathbb C^{2N+1} \setminus U$ is the zero set for some (non-zero) polynomial $P$ in $2N+1$ variables, it has $(2N+1)$-dimensional Lebesgue measure zero \cite{CliffordRossi1965:complexVariables}.
\end{remark}

\begin{remark}
    In Appendix~\ref{app:quadrature} we show that the eigenvalues of $T_N(\bm z)$ are distinct for $\bm z$ in some open neighbourhood, $U_0 \subset U$, of $\mathbb R^{2N+1}$, which leads to the following alternative proof. On $U_0$, the eigenvalues and corresponding left and right eigenvectors can be chosen to be analytic: there exist analytic functions $\ep_j, \psi_{j}, \phi_{j}^\star$ for $j = 0,\dots,N$ such that
    \[
        T_N(\bm z) \psi_j(\bm z) = \ep_j(\bm z) \psi_j(\bm z),
        \quad 
        \phi^\star_j(\bm z) T_N(\bm z)  = \ep_j(\bm z) \phi_j^\star(\bm z),
        \quad \text{and} \quad
        \phi^\star_i(\bm z) \psi_j(\bm z) = \delta_{ij}.
    \]
    (More precisely, we apply \cite[Theorem~2]{Greenbaum2020} to obtain analytic functions $\psi_j,\phi^\star_j$ of each variable $z_0,\dots,z_{2N+1}$ separately and then apply Hartog's theorem \cite{Krantz2001} to conclude that $\psi_j,\phi^\star_j$ are analytic as functions on $U \subset \mathbb C^{2N+1}$).
    Therefore, the nonlinear method discussed in this section can also be written in the following form
    \begin{gather}
        \ctNonlin 
        = \sum_{j=0}^N [\psi_j]_{0} [\phi^\star_j]_0  \cdot \big( O\circ \ep_j \big)
    \end{gather}
    which is an analytic function on $\{ \bm z \in U_0 \colon O \text{ analytic at } \ep_j(\bm z) \text{ for each } j\}$ (as it is a finite combination of analytic functions only involving products, compositions and sums).
\end{remark}

\subsection{Self-consistent tight binding models} \label{eq:sctb}
%

We start with the following preliminary lemma:
\begin{lemma}\label{lem:ell_infty}
    Suppose that $T \colon \ell^2(\Lambda) \to \ell^2(\Lambda)$ is an invertible bounded linear operator with matrix entries $T_{\ell k}$ satisfying
    $
        \big| T_{\ell k} \big| \leq c_T e^{-\gamma_T r_{\ell k}}
    $
    for some $c_T, \gamma_T > 0$.
    Then, there exists an invertible bounded linear operator $\overline{T} \colon \ell^\infty(\Lambda) \to \ell^\infty(\Lambda)$ extending $T\colon \ell^2(\Lambda) \to \ell^2(\Lambda)$ (that is, $\overline{T}\big|_{\ell^2(\Lambda)} = T$). 
\end{lemma}

\begin{proof}
    First, we denote the inverse of $T$ and its matrix entries by $T^{-1}\colon \ell^2(\Lambda) \to \ell^2(\Lambda)$ and $T^{-1}_{\ell k}$, respectively. Then, applying the Combes-Thomas estimate to $T$ yields the off-diagonal decay estimate $|T^{-1}_{\ell k}| \leq C e^{-\gamma_{\mathrm{CT}} r_{\ell k}}$ for some $C, \gamma_\mathrm{CT} > 0$ \cite{Thomas2020:scTB}.
    
    Due to the off-diagonal decay properties of the matrix entries, the operators $\overline{T}, \overline{T}^{-1} \colon \ell^\infty(\Lambda) \to \ell^\infty(\Lambda)$ given by 
    \[
        [\overline{T}\phi]_\ell \coloneqq \sum_{k \in \Lambda} T_{\ell k} \phi_k
        \qquad \text{and} \qquad
        [\overline{T}^{-1}\phi]_\ell \coloneqq \sum_{k \in \Lambda} T^{-1}_{\ell k} \phi_k
    \]
    are well defined bounded linear operators with norms $\sup_\ell \sum_{k \in \Lambda} |T_{\ell k}|$ and $\sup_\ell \sum_{k \in \Lambda} |T^{-1}_{\ell k}|$, respectively. To conclude, we note that 
    \begin{align}
         [\overline{T}\overline{T}^{-1}\phi]_{\ell} &= \sum_{k}\sum_m T_{\ell k} T^{-1}_{km} \phi_m 
         = \sum_m [TT^{-1}]_{\ell m} \phi_m = \phi_\ell
    \end{align}
    and so $\overline{T}^{-1}$ is the inverse of $\overline{T}$. Here, we have exchanged the summations over $k$ and $m$ by applying the dominated convergence theorem: $\big|\sum_k T_{\ell k}T^{-1}_{km} \phi_m\big| \leq C e^{-\frac{1}{2}\min\{\gamma_T, \gamma_\mathrm{CT}\} r_{\ell m}} \|\phi\|_{\ell^\infty}$ is summable over $m \in \Lambda$.
\end{proof}

Throughout the following proofs, we denote by $B_r(\rho)$ the open ball of radius $r$ about $\rho$ with respect to the $\ell^\infty$-norm. Moreover, we briefly note that the stability operator can be written as the product $\mathscr L(\rho) \coloneqq \mathscr F(\rho) \nabla v(\rho)$ where \cite{Thomas2020:scTB}
    \begin{align}\label{eq:stability-op}
        \mathscr F(\rho)_{\ell k} \coloneqq \frac{1}{2\pi i}
        \oint_{\mathscr{C}} F^\beta(z) 
        \Big[ \big( \Ham( \s(\rho) ) - z \big)^{-1}_{\ell k} \Big]^2 \mathrm{d}z
    \end{align}
    where $\mathscr C$ is a simple closed contour encircling the spectrum $\sigma\big( \Ham(\s(\rho)) \big)$. 
    
\begin{proof}[Proof of Theorem~\ref{thm:nonlin}]
Since $\rho \mapsto F^\beta(\s(\rho))$ is $C^2$, and 
$\big(I - \mathscr L(\rho^\star)\big)^{-1}$ 
is a bounded linear operator, we necessarily have that 
$\big(I - \mathscr L(\rho)\big)^{-1}$ 
is a bounded linear operator for all $\rho \in B_r(\rho^\star)$ for some $r > 0$.

By applying Theorem~\ref{thm:bodyOrder_interpolation}, together with the assumption \asEP, we obtain 
\begin{align}\label{clickhere}
    \big|\big[
        \mathscr L(\rho) - {\mathscr L}_N(\rho)
    \big]_{\ell k}\big|
    &\leq \sum_m \bigg|\bigg[ 
        \frac
            {\partial F^\beta_\ell(\s)}
            {\partial v_m}
        - 
        \frac
            {\partial I_N F^\beta_\ell(\s)}
            {\partial v_m}
    \bigg]
        \frac
            {\partial v(\rho)_m}
            {\partial \rho_k}
    \bigg|\\
    &\leq C  \Big[ \sum_m e^{-\eta \, r_{\ell m}} e^{-\gamma_v\, r_{mk}} \Big]   e^{-\frac{1}{2}\gamma_N N} \\
    &\leq C e^{-\frac{1}{2}\min\{ \eta, \gamma_v \} r_{\ell k}} e^{-\frac{1}{2}\gamma_N N}
\end{align}
%
%
%
%
for all $\rho \in B_r(\rho^\star)$. 
As a direct consequence, we have
$\| {\mathscr L}(\rho) - \mathscr L_N(\rho) \|_{\ell^2 \to \ell^2}
\leq C e^{-\frac{1}{2}\gamma_N N}$
and we may choose $N$ sufficiently large such that $\| \mathscr L(\rho) - \mathscr L_N(\rho) \|_{\ell^2 \to \ell^2} < \| (I - \mathscr L(\rho))^{-1} \|^{-1}_{\ell^2 \to \ell^2}$. In particular, for such $N$, the operator $I - {\mathscr L}_N(\rho) \colon \ell^2 \to \ell^2$ is invertible with inverse bounded above in operator norm independently of $N$.
%
%
%

We now show that $I - \mathscr L_N(\rho)$ satisfies the assumptions of Lemma~\ref{lem:ell_infty}. Using \cref{eq:stability-op} and \asEP, together with the Combes-Thomas estimate (Lemma~\ref{lem:CT}), we conclude
\[
    \big| \mathscr L_N(\rho)_{\ell k} \big| 
    \leq C \sup_{z \in \mathscr C} |I_N F^\beta(z)| \sum_{m \in \Lambda} e^{-2\gamma_\mathrm{CT} r_{\ell m}} e^{-\gamma_v r_{mk}}
    \leq C e^{-\frac{1}{2}\min\{ 2\gamma_\mathrm{CT}, \gamma_v \} r_{\ell k} }
\]
for all $\rho \in B_r(\rho^\star)$. 
In particular, $I - \mathscr L_N(\rho)$ extends to a invertible bounded linear operator $\ell^\infty \to \ell^\infty$ and thus its inverse $\big(I - \mathscr L_N(\rho)\big)^{-1} \colon \ell^\infty \to \ell^\infty$ is bounded. 

Now, the mapping $\rho \mapsto \rho - I_NF^\beta\big(\s(\rho)\big)$ between $\ell^\infty \to \ell^\infty$ is continuously differentiable on $B_r(\rho^\star)$ and the derivative at $\rho^\star$ is invertible (i.e.~$\big(I - \mathscr L_N(\rho^\star)\big)^{-1}\colon\ell^\infty \to \ell^\infty$ is a well defined bounded linear operator). Since the map $\rho \mapsto I_N{F}^\beta\big(\s(\rho)\big)$ is $C^2$, its derivative 
${\mathscr L}_N$ 
is locally Lipschitz about $\rho^\star$ and so there exists $L>0$ such that
\begin{align*}
    \big\| 
        \big(
            I - {\mathscr L}_N(\rho^\star)
        \big)^{-1} 
        \big(
            {\mathscr L}_N(\rho_1) - {\mathscr L}_N(\rho_2)
        \big) 
    \big\|_{\ell^\infty \to \ell^\infty}
    \leq L \|\rho_1 - \rho_2 \|_{\ell^\infty}
    \qquad \text{for } \rho_1, \rho_2 \in B_r(\rho^\star).
\end{align*}
Moreover, by Theorem~\ref{thm:bodyOrder_interpolation}, we have
\begin{align*}
    \big\| 
        \big(
            I - {\mathscr L}_N(\rho^\star)
        \big)^{-1} 
        \big(
            \rho^\star - I_N {F}^\beta(\s(\rho^\star))
        \big) 
    \big\|_{\ell^\infty} 
    &\leq C
    \big\|
        F^\beta(\s(\rho^\star)) - I_N {F}^\beta(\s(\rho^\star)) 
    \big\|_{\ell^\infty} 
    \eqqcolon b_{N}
\end{align*}
where $b_{N} \lesssim e^{-\gamma_N N}$. In particular, we may choose $N$ sufficiently large such that $2b_{N} L < 1$ and $t^\star_{N} \coloneqq \frac{1}{L}( 1 - \sqrt{1 - 2 b_{N} L} ) < r$. 
    
Thus, the Newton iteration with initial point $\rho^0 \coloneqq \rho^\star$, defined by
\[
    \rho^{i+1} = \rho^i - 
    \big(
        I - {\mathscr L}_N(\rho^i)
    \big)^{-1} 
    \big(
        \rho^i - I_N{F}^\beta(\s(\rho^i))
    \big),
\]
converges to a unique fixed point 
$\rho_N = I_N{F}^\beta(\s(\rho_N))$ 
in $B_{t^\star_{N}}(\rho^\star)$ 
\cite{Zhengda1993:NewtonIteration, Yamamoto1986}. That is, 
$\| \rho_N - \rho^\star \|_{\ell^\infty} \leq t_N^\star \leq 2 b_{N}$. 
Here, we have used the fact that $1 - \sqrt{1 - x} \leq x$ for all $0 \leq x \leq 1$. 

Since $\rho_N \in B_r(\rho^\star)$, we have $I - \mathscr L_N(\rho_N) \colon \ell^2 \to \ell^2$ is invertible and thus Lemma~\ref{lem:stable} also holds.
\end{proof}    

\begin{proof}[Proof of Proposition~\ref{cor:SCF}]
We proceed in the same way as in the proof of Theorem~\ref{thm:nonlin}. In particular, since $\rho_N$ is stable, if 
$\|\rho^0 - \rho_N\|_{\ell^\infty}$ 
is sufficiently small, 
$(I - \mathscr L_N(\rho^0))^{-1}$ 
is a bounded linear operator on $\ell^2$. Moreover, by the exact same argument as in the proof of Theorem~\ref{thm:nonlin}, $I - \mathscr L_N(\rho^0)\colon\ell^\infty \to \ell^\infty$ defines an invertible bounded linear operator. 
Also, $I - \mathscr L_N(\rho)$ is Lipschitz in a neighbourhood about $\rho^0$ and
\begin{align*}
    \big\| 
        \big(
            I - {\mathscr L}_N(\rho^0)
        \big)^{-1} 
        \big(
            \rho^0 - I_N {F}^\beta(\s(\rho^0))
        \big) 
    \big\|_{\ell^\infty} 
    &\leq C
    \big\|
        \rho^0 - \rho_N - 
        \big( 
            I_N {F}^\beta(\s(\rho^0)) - I_N {F}^\beta(\s(\rho_N))
        \big)
    \big\|_{\ell^\infty}  \\
    &\leq C \| \rho^0 - \rho_N \|_{\ell^\infty}.
\end{align*}
Here, we have used:
\begin{align}
    \big| 
        I_N {F}^\beta_\ell(\s(\rho^0)) - I_N {F}^\beta_\ell(\s(\rho_N)) 
    \big|
    &= \frac{1}{2\pi} \Big|
        \oint_{\mathscr C} I_N F^\beta(z) 
        \big[
            \mathscr R_z(\rho^0) - \mathscr R_z(\rho_N)
        \big]_{\ell\ell}
    \Big| \nonumber\\
    &\leq C \sum_{k \in \Lambda} e^{-2\gamma_\mathrm{CT} r_{\ell k}}
    |v(\rho^0)_k - v(\rho_N)_k| \nonumber\\
    &\leq C \sum_{k \in \Lambda} 
    e^{-2\gamma_\mathrm{CT} r_{\ell k}}
   \Big|
    \int_0^1
        \sum_{m \in \Lambda}
         \frac
            {\partial v(t\rho^0 + (1 - t) \rho_N)_k}
            {\partial \rho_m} 
        \big[
            \rho^0 - \rho_N
        \big]_m
   \mathrm{d}t
   \Big| \nonumber\\
   &\leq C \sum_{m \in\Lambda} e^{-\frac{1}{2}\min\{2\gamma_\mathrm{CT}, \gamma_v\} r_{\ell m} } 
   \big|\big[
        \rho^0 - \rho_N
    \big]_m\big| \nonumber\\
    &\leq C \|\rho^0 - \rho_N\|_{\ell^\infty}. \label{eq:IN-error}
\end{align}

Therefore, as long as $\| \rho^0 - \rho_N\|_{\ell^\infty}$ is sufficiently small, we may apply the Newton iteration starting from $\rho^0$ to conclude.
\end{proof}

\begin{proof}[Proof of Corollary~\ref{cor:sc-observables}]
    As a direct consequence of \cref{eq:IN-error}, we have 
    \begin{align}
        \big|
            O_\ell^\mathrm{sc}(\s) - I_N O_\ell\big( \s(\rho_N) \big)
        \big|
        &\leq \big| 
            O_\ell\big( \s(\rho^\star) \big) -  I_N O_\ell\big( \s(\rho^\star) \big)
        \big| + 
        \big| 
            I_N O_\ell\big( \s(\rho^\star) \big) -  I_N O_\ell\big( \s(\rho_N) \big)
        \big| \\
        &\leq C \big[ e^{-\gamma_N N} + \|\rho_N - \rho^\star\|_{\ell^\infty} \big]\\
        &\leq C e^{-\gamma_N N}.
    \end{align}
    Here, we have applied the standard convergence result (Theorem~\ref{thm:bodyOrder_interpolation}) with fixed effective potential.
\end{proof}

\appendix

\section{Notation} Here we summarise the key notation: 
\begin{itemize}
    \item $\Lambda$ : finite or countable index set,

    \item $\s_\ell = (\bm r_\ell, v_\ell, \z_\ell)$ : state of atom $\ell$ where $\bm r_\ell \in \mathbb R^d$ denotes the atomic position, $v_\ell$ the effective potential, and $\z_\ell$ the atomic species, 
    
    \item $\s = \{\s_\ell\}_{\ell \in \Lambda}$ : configuration,
    
     \item $\bm{r}_{\ell k} \coloneqq \bm{r}_k - \bm{r}_\ell$ and $r_{\ell k} \coloneqq |\bm{r}_{\ell k}|$ : relative atomic positions,
    
    \item $\delta_{ij}$ : Kronecker delta ($\delta_{ij} = 0$ for $i\not=j$ and $\delta_{ii} = 1$), 
    
    \item $\mathrm{Id}_n$ : $n \times n$ identity matrix, 
    
    \item $|\,\cdot\,|$ : absolute value on $\mathbb R^d$ or $\mathbb C$,
    
    \item $|\,\cdot\,|$ : Frobenius matrix norm on $\mathbb R^{n\times n}$,
    
     \item $\nabla h = (\nabla h_{ab})_{1 \leq a,b \leq n}$ : gradient of $h \colon \mathbb R^d \to \mathbb R^{n\times n}$,
    
    \item $M^\mathrm{T}$ : transpose of the matrix $M$,
    
    \item $\mathrm{Tr}$ : trace of an operator,
    
    \item $f \sim g$ as $x \to x_0 \in \mathbb R\cup\{\pm\infty\}$ or $\mathbb C \cup \{\infty\}$ if there exists an open neighbourhood $N$ of $x_0$ and positive constants $c_1,c_2 > 0$ such that $c_1 g(x) \leq f(x) \leq c_2 g(x)$ for all $x \in N$,
    
    \item $C$ : generic positive constant that may change from one line to the next,
    
     \item $f \lesssim g$ : $f \leq C g$ for some generic positive constant,
    
    \item $\mathbb N_0 = \{ 0, 1, 2, \dots \}$ : Natural numbers including zero,
    
     \item $\delta(\,\cdot\,)$ : Dirac delta, distribution satisfying $\int f(x) \mathrm{d}\delta(x) = f(0)$,

    \item $\|f\|_{L^\infty(X)} \coloneqq \sup_{x\in X} |f(x)|$ : sup-norm of $f$ on $X$,
    
    \item $\mathrm{dist}(z, A) \coloneqq \inf_{a\in A} |z - a|$ : distance between $z\in\mathbb C$ and the set $A\subset \mathbb C$,
    
    \item $a + b S \coloneqq \{a + bs \colon s \in S\}$,
    
    \item $[\psi]_\ell$ : the $\ell^\text{th}$ entry of the vector $\psi$,
    
    \item $\|\psi\|_{\ell^2}\coloneqq \left(
        \sum_{k} |[\psi]_k|^2
    \right)^{1/2}$ : $\ell^2$-norm of $\psi$,
    
    \item $\mathrm{tr}\,M \coloneqq \sum_\ell M_{\ell\ell}$ : trace of matrix $M$,
    
    \item $\|M\|_\mathrm{max} \coloneqq \max_{\ell, k} |M_{\ell k}|$ : max-norm of the matrix $M$,
    
    \item $\sigma(T)$ : the spectrum of the operator $T$,
    
    \item $\sigma_\mathrm{disc}(T) \subset \sigma(T)$ : isolated eigenvalues of finite multiplicity,
    
    \item $\sigma_\mathrm{ess}(T) \coloneqq \sigma(T) \setminus \sigma_\mathrm{disc}(T)$ : essential spectrum,
    
    \item $\|T\|_{X \to Y} \coloneqq \sup_{x \in X, \|x\|_X = 1} \|Tx\|_Y$ : operator norm of $T \colon X \to Y$,
    
    \item $\nabla v$ : Jacobian of $v \colon \mathbb R^\Lambda \to \mathbb R^\Lambda$,
    
    \item $[a, b] \coloneqq \{ (1-t) a + t b \colon t \in [0,1] \}$ : closed interval between $a,b \in \mathbb R^d$ or $a,b \in \mathbb C$,
    
    \item $\int_a^b \coloneqq \int_{[a,b]}$ : integral over the interval $[a,b]$ for $a,b \in \mathbb C$,
    
    \item $\mathrm{len}(\mathscr C)$ : length of the simple closed contour $\mathscr C$,

    \item $\mathrm{supp} \, \nu$ : support of the measure $\nu$, set of all $x$ for which every open neighbourhood of $x$ has non-zero measure,
    
    \item $\mathrm{conv}\, A \coloneqq \{ t a + (1-t) b \colon a,b \in A, t \in [0,1] \}$ : convex hull of $A$.
\end{itemize}

\section{Locality: Truncation of the Atomic Environment}
\label{app:locality}

\label{sec:sparse_linear}
%

We have seen that analytic quantities of interest may be approximated by body-order approximations. However, each polynomial depends on the whole atomic configuration $\s$. In this section, we consider the truncation of the approximation schemes to a neighbourhood of the central site $\ell$ and prove the exponential convergence of the corresponding sparse representation.

\subsection{Banded Approximation}

One intuitive approach is to restrict the interaction range globally and consider the following \textit{banded approximation}:
\begin{align}\label{eq:banded}
    \widetilde{\Ham}^{r_\mathrm{c}}(\s)_{km} \coloneqq \begin{cases}
    h(\s_{km}) + \sum\limits_{
        \genfrac{}{}{0pt}{2}
            {m^\prime \not\in \{k,m\}\colon}
            {r_{km^\prime}, r_{mm^\prime} \leq r_\mathrm{c}}
        }
        t(\s_{km^\prime}, \s_{mm^\prime})
        +\delta_{km} v_k \mathrm{Id}_{\ctNumorbitals}
            &\text{if } r_{km} \leq r_\mathrm{c} \\
        0
            &\text{otherwise}.
    \end{cases}
\end{align}
Therefore, approximating $O_\ell(\s)$ with a function depending on the first $N$ moments $[\widetilde{\Ham}^{r_\mathrm{c}}]_{\ell\ell}$ (e.g.~applying Theorem~\ref{thm:bodyOrder_interpolation_discrete} or \ref{thm:BOP} to $\widetilde{\Ham}^{r_\mathrm{c}}$) results in an approximation scheme depending only on finitely many atomic sites in a neighbourhood of $\ell$. This can be seen from the fact that  
\begin{align}\label{eq:hopping}
    [(\widetilde{\Ham}^{r_\mathrm{c}})^n]_{\ell\ell}
    = \sum_{
        \genfrac{}{}{0pt}{1}
            {\ell_1,\dots,\ell_{n-1}}
            {r_{\ell\ell_1}, r_{\ell_1\ell_2}, \dots, r_{\ell_{n-1}\ell} \leq r_c}
    }
    \Ham_{\ell \ell_1} \Ham_{\ell_1\ell_2}\dots \Ham_{\ell_{n-1}\ell}.
\end{align}
Moreover, we obtain appropriate error estimates by combining Theorem~\ref{thm:bodyOrder_interpolation_discrete} or \ref{thm:BOP} with the following estimate:
\begin{prop}\label{prop:sparse_nonlinear}
    Suppose $\s$ satisfies Definition~\ref{def:config}. Fix $0 < \beta \leq \infty$ and suppose that, if $\beta = \infty$, then $\mathsf{g}, \mathsf{g}^\mathrm{def} >0$. Then, we have
    \[
    \Big|
        O_\ell(\s) - \mathrm{tr} \, O\big(\widetilde{\Ham}^{r_\mathrm{c}}(\s)\big)_{\ell\ell}
    \Big|
    \lesssim e^{-\frac{1}{2}\gamma_0 r_\mathrm{c}}.
    \]
     Suppose $\gamma_N(r_\mathrm{c})$ and $\gamma_N^\mathrm{def}(r_\mathrm{c})$ are the rates of approximation from Theorems~\ref{thm:bodyOrder_interpolation_discrete} and \ref{thm:BOP} when applied to $\widetilde{\Ham}^{r_\mathrm{c}}$. Then $\gamma_N(r_\mathrm{c}) \to \gamma_N$ and $\gamma_N^\mathrm{def}(r_\mathrm{c}) \to \gamma_N^\mathrm{def}$ as $r_\mathrm{c} \to \infty$, with an exponential rate.
\end{prop}
\begin{proof}
We first note that
    \begin{align}\label{pf:bandedApprox}
        \left[ 
            \Ham(\s) - \widetilde{\Ham}^{r_\mathrm{c}}(\s)
        \right]_{km}
        &= \begin{cases}
            \Ham(\s)_{km} &\text{if } r_{km} > r_\mathrm{c} \\
            \sum_{
                \above
                    {m^\prime \colon}
                    {r_{km^\prime} > r_\mathrm{c} \text{ or } r_{mm^\prime} > r_\mathrm{c}}
            }
            t(\s_{km^\prime}, \s_{mm^\prime} )
            &\text{if } r_{km} \leq  r_\mathrm{c}.
        \end{cases}
    \end{align}
    Therefore, applying \asTB, we obtain
    \begin{align}\label{eq:error-banded-truncation}
        \big|[ 
            \Ham(\s) - \widetilde{\Ham}^{r_\mathrm{c}}(\s)
        ]_{km}\big|
        \lesssim e^{-\frac{1}{2}\gamma_0 r_\mathrm{c}} \sum_{m^\prime} e^{-\frac{1}{2}\gamma_0 (r_{km^\prime} + r_{mm^\prime})}
        \lesssim e^{-\frac{1}{2}\gamma_0 r_\mathrm{c}} e^{-\frac{1}{4}\gamma_0 \,r_{km}}.
    \end{align}

    To conclude we choose a suitable contour $\mathscr{C}$ and apply the Combes-Thomas estimate (Lemma~\ref{lem:CT}) together with \cref{eq:error-banded-truncation}:  
    \begin{align}
         \Big|
            O_\ell(\s) - \mathrm{tr} \, O\big(\widetilde{\Ham}^{r_\mathrm{c}}(\s)\big)_{\ell\ell}
        \Big|
        &\leq \Bigg|\frac{\mathrm{tr}}{2\pi} 
            \oint_{\mathscr C} O(z)  
            \Big[
                (\Ham(\s) - z)^{-1} 
                \big( \Ham(\s) - \widetilde{\Ham}^{r_\mathrm{c}}(\s) \big)
                (\widetilde{\Ham}^{r_\mathrm{c}}(\s) - z)^{-1}
            \Big]_{\ell\ell} \mathrm{d}z
        \Bigg| \\
        &\lesssim \max_{z \in \mathscr C} |O(z)| 
        e^{-\frac{1}{2}\gamma_0 r_{\mathrm{c}} }
        \sum_{km} e^{-\gamma_\mathrm{CT} (r_{\ell k} + r_{m\ell}) } e^{-\frac{1}{4}\gamma_0 r_{km}} 
        \lesssim e^{-\frac{1}{2}\gamma_0 r_{\mathrm{c}}}
        \label{eq:banded-proof}
    \end{align}
    
    As a direct consequence of \cref{eq:error-banded-truncation}, we have also have $\| \Ham(\s) - \widetilde{\Ham}^{r_\mathrm{c}}(\s) \|_{\ell^2 \to \ell^2} \lesssim e^{-\frac{1}{2}\gamma_0 r_{\mathrm{c}}}$ and so 
    $\mathrm{dist}\big( \sigma(\Ham), \sigma(\widetilde{\Ham}^{r_\mathrm{c}}) \big) \lesssim e^{-\frac{1}{2}\gamma_0 r_{\mathrm{c}}}$ \cite{bk:Kato1995}. This means that for sufficiently large $r_\mathrm{c}$, we obtain the same rates of approximation when applying Theorems~\ref{thm:bodyOrder_interpolation_discrete} and \ref{thm:BOP} to $\widetilde{\Ham}^{r_\mathrm{c}}$. 
\end{proof}

\subsection{Truncation}
One downside of the banded approximation is that the truncation radius depends on the maximal polynomial degree (e.g.~see \cref{eq:hopping}). In this section, we consider truncation schemes that only depend on finitely many atomic sites independent of the polynomial degree:
\begin{align}\label{eq:truncation-op-1}
    \widetilde{\Ham}^{r_\mathrm{c}} \coloneqq \Ham\big|_{\ell; \Lambda \cap B_{r_c}(\ell)}
\end{align}
where the restriction of the Hamiltonian has been introduced in \cref{eq:restriction}.

On defining the quantities
\begin{align}\label{eq:sparse}
    %
    %
    %
    %
    I_N \widetilde{O}_\ell(\s) 
    \coloneqq \mathrm{tr} \big[ 
        I_N O\big(\widetilde{\Ham}^{r_\mathrm{c}} \big)
    \big]_{\ell\ell},
\end{align}
where the operators $I_N$ are given by Theorem~\ref{thm:bodyOrder_interpolation}, we obtain a sparse representation of the $N$-body approximation depending only on finitely many atomic sites, independently of the maximal body-order $N$.

\begin{prop}\label{thm:trun-nbh_int}
    Suppose $\s$ satisfies Definition~\ref{def:config}. Fix $0 < \beta \leq \infty$ and suppose that, if $\beta = \infty$, then $\mathsf{g}, \mathsf{g}^\mathrm{def} >0$. Then,
    \[
        \big|
            I_N O^\beta_\ell(\s) - I_N \widetilde{O}^\beta_\ell(\s)
        \big| 
        \lesssim e^{- \frac{1}{4} \min\{\gamma_\mathrm{CT}, \gamma_0\} r_\mathrm{c} }
    \]
    where $O^\beta = F^\beta$ or $G^\beta$ and $\gamma_\mathrm{CT}$ is the constant from Lemma~\ref{lem:CT} applied to $\Ham(\s)$.
\end{prop}
\begin{proof}
    Applying the Hermite integral formula \cref{eq:Hermite:prelim} directly, we conclude that $I_NO^\beta(z)$ is bounded uniformly in $N$ along a suitably chosen contour $\mathscr C \coloneqq \{ g_E = \gamma \}$ (examples of such contours are given in Figure~\ref{fig:equipot}). 
    It is important to note that the contour $\mathscr C$ must be chosen to encircle both $\sigma(\Ham)$ and $\sigma( \widetilde{\Ham}^{r_\mathrm{c}} )$. 
    In the following, we let $\gamma_\mathrm{CT}$ be the Combes-Thomas exponent from Lemma~\ref{lem:CT} corresponding to $\Ham$.

    Similarly to \cref{eq:banded-proof}, we obtain
    \begin{align}
        \left|
            I_N O^\beta_\ell(\s) - I_N \widetilde{O}^\beta_\ell(\s)      
        \right| 
        &\lesssim \max_{z \in \mathscr C} |I_NO(z)|
        \sum_{km} e^{-\gamma_\mathrm{CT} r_{\ell k}}
        \big|\big[ \Ham(\s) - \widetilde{\Ham}^{r_\mathrm{c}}(\s) \big]_{km} \big|
        \label{eq:T3} \\
        &\lesssim \sum_{
            \above
                {k,m\colon}
                {r_{\ell k} \geq r_\mathrm{c} \textrm{ or } 
                r_{\ell m} \geq r_\mathrm{c}}
        }
        e^{-\gamma_\mathrm{CT} r_{\ell k}}
        e^{-\frac{1}{2}\gamma_0 r_{km}}
        + \sum_{
            \above
                {k,m\colon}
                {r_{\ell k}, r_{\ell m} < r_\mathrm{c}}
        }
        e^{-\gamma_\mathrm{CT} r_{\ell k}} \sum_{
            \above
                { m^\prime \colon }
                {r_{\ell m^\prime} \geq r_\mathrm{c}}
        }
        e^{-\gamma_0(r_{km^\prime} +r_{mm^\prime})} \nonumber\\
        &\lesssim e^{-\frac{1}{2} \min\{\gamma_\mathrm{CT}, \frac{1}{2}\gamma_0\} r_\mathrm{c} }. \nonumber
    \end{align}
    This concludes the proof.
\end{proof}

The fact that the exponents of Proposition~\ref{thm:trun-nbh_int} are independent of the defect states within the band gap is in the same spirit to the improved locality estimates of \cite{ChenOrtnerThomas2019:locality}.

\begin{remark}[Divide-and-conquer Methods]
This truncation scheme is closely related to the divide-and-conquer method for solving the electronic structure problem \cite{Yang1991:divideconquer}. In this context the system is split into many subsystems that are only related through a global choice of Fermi level. In our notation, this method consists of constructing $N_{\mathrm{DAC}}$ smaller Hamiltonians 
$\widetilde{\Ham}^{r_\mathrm{c},\ell_j}$ centred on the atoms $\ell_j$
(for $j = 1,\dots,N_{\mathrm{DAC}}$) and approximating the quantities $O_\ell(\s)$ for $\ell$ in a small neighbourhood of $\ell_j$ by calculating $\mathrm{tr}\, O\big( \widetilde{\Ham}^{r_\mathrm{c},\ell_j} \big)_{\ell\ell}$. That is, the eigenvalue problem for the whole system is approximated by solving $N_\mathrm{DAC}$ smaller eigenvalue problems in parallel. In particular, this method leads to linear scaling algorithms \cite{Goedecker1999}. Theorem~\ref{thm:trun-nbh_int} then ensures that the error in this approximation decays exponentially with the distance between $\ell$ and the exterior of the subsystem centred on $\ell_j$. 

A similar error analysis in the context of divide-and-conquer methods in Kohn-Sham density functional theory can be found in \cite{ChenLu2016:divideconquer}.
\end{remark}

\begin{remark}[General truncation operators]
    \hypertarget{assumptionT}{It should be clear from the proof of Proposition~\ref{thm:trun-nbh_int} that more general truncation operators may be used. Indeed,  Proposition~\ref{thm:trun-nbh_int} is satisfied for all truncation operators $\widetilde{\Ham}^{r_\mathrm{c}} = \widetilde{\Ham}^{r_\mathrm{c}}(\s)$ satisfying the following conditions}:
\newcommand{\T}[1]{\textup{(\hyperlink{assumptionT}{T#1})}}
\begin{itemize}
    \item[\textup{(T1)}] For every polynomial $p$, the quantity $p\big(\widetilde{\Ham}^{r_\mathrm{c}}\big)_{\ell\ell}$ depends on at most finitely many atomic sites depending on $r_\mathrm{c}$ but not $p$,
    
    \item[\textup{(T2)}] For all $k,m\in \Lambda$, we have $[\widetilde{\Ham}^{r_\mathrm{c}}]_{km} \to \Ham_{km}$ as $r_\mathrm{c} \to \infty$,
    
    \item[\textup{(T3)}] There exists $c_0 > 0$ such that for all $\gamma, r_\mathrm{c} > 0$,
    \[
        \sum_{km} e^{-\gamma r_{\ell k}}
        \left|
            \big[
                \Ham - \widetilde{\Ham}^{r_\mathrm{c}}
            \big]_{k m}
        \right| 
        \leq C e^{- c_0 \min\{\gamma_0, \gamma\} \, r_{\mathrm{c}} }
    \]
    for some $C>0$ depending on $\gamma$ but not on $r_\mathrm{c}$.
\end{itemize}
Due to the exponential weighting of the summation, \T{3}~states that 
$\widetilde{\Ham}^{r_\mathrm{c}}$ 
captures the behaviour of the Hamiltonian in a small neighbourhood of the site $\ell$. Moreover, when making the approximation 
$I_NO\big(
    \Ham
\big)_{\ell\ell}\approx I_NO\big(
    \widetilde{\Ham}^{r_\mathrm{c}}
\big)_{\ell\ell}$,
the number of atomic sites involved is finite by \T{1}.
\end{remark}

\begin{remark}[Non-linear schemes]
One may be tempted to approximate the Hamiltonian with the truncation, $\widetilde{\Ham}^{r_\mathrm{c}}$, and then apply the nonlinear scheme of Theorem~\ref{thm:BOP}. In doing so, we obtain the following error estimates:
\begin{align}
    &\Big|
        O_\ell(\s) - \ctNonlin\big(
            [\widetilde{\Ham}^{r_\mathrm{c}}]_{\ell\ell}, \dots, [(\widetilde{\Ham}^{r_\mathrm{c}})^N]_{\ell\ell}
        \big)
    \Big| \\
    &\qquad\leq 
    \Big|
        O(\Ham)_{\ell\ell} - O\big( \widetilde{\Ham}^{r_\mathrm{c}} \big)_{\ell\ell}
    \Big| 
    + \Big|
        O\big( \widetilde{\Ham}^{r_\mathrm{c}} \big)_{\ell\ell}
        -
        \ctNonlin\big(
            [\widetilde{\Ham}^{r_\mathrm{c}}]_{\ell\ell}, \dots, [(\widetilde{\Ham}^{r_\mathrm{c}})^N]_{\ell\ell}
        \big)
    \Big| \label{eq:sparse_nonlinear_2}\\
    &\qquad \lesssim e^{-\frac{1}{4}\min\{\gamma_0,\gamma_\mathrm{CT}\} r_\mathrm{c}} + e^{-\widetilde{\gamma}_N(r_\mathrm{c}) N}. \label{eq:sparse_nonlinear}
\end{align}

A problem with this analysis is that the constant $\widetilde{\gamma}_N(r_\mathrm{c})$ in \cref{eq:sparse_nonlinear} arises by applying Theorem~\ref{thm:BOP} to $\widetilde{\Ham}^{r_\mathrm{c}}$ rather than the original system $\Ham$. In particular, this means that $\widetilde{\gamma}_N(r_\mathrm{c})$ depends on the spectral properties of $\widetilde{\Ham}^{r_\mathrm{c}}$ rather than $\Ham$. Since spectral pollution is known to occur when applying naive truncation schemes \cite{Lewin2009}, the choice of $\widetilde{\Ham}^{r_\mathrm{c}}$ is important for the analysis. In particular, it is not clear that $\widetilde{\gamma}_N(r_\mathrm{c}) \to \gamma_N$ in general. This is in contrast the the result of Proposition~\ref{prop:sparse_nonlinear}.
\end{remark}

\section{Convergence of Derivatives in the Nonlinear Approximation Scheme}
\label{app:derivatives}
As mentioned in Remark~\ref{rem:derivatives}, the results of this section depend on the ``regularity'' properties of $D_\ell$:
\begin{definition}[Regular $n^\text{th}$-root Asymptotic Behaviour]
\label{def:reg}
    For a unit measure $\nu$ with compact support $E \coloneqq \mathrm{supp}\,\nu \subset \mathbb R$, we say $\nu$ is regular and write $\nu \in \mathbf{Reg}$ if the corresponding sequence of orthonormal polynomials $\{p_n(\,\cdot\,;\nu)\}$ satisfy
    \[
        \lim_{n\to\infty} |p_n(z;\nu)|^{\frac{1}{n}} 
        = e^{g_{E}(z)} 
    \]
    locally uniformly on $\mathbb C \setminus \mathrm{conv}(E)$.
\end{definition}

\begin{remark}\label{rem:nrootasymp}
    The regularity condition says that the $n^\textrm{th}$-root asymptotic behaviour of $|p_n(z;\nu)|$ is minimal: in general, we have \cite[Theorem~1.1.4]{stahl_totik_1992}
    \[
        e^{g_E(z)} \leq \liminf_{n\to\infty} |p_n(z;\nu)|^{\frac{1}{n}}
        \leq \limsup_{n\to\infty} |p_n(z;\nu)|^{\frac{1}{n}}
        \leq e^{g_\nu(z)}
    \]
    where $g_\nu \geq g_E$ is the \emph{minimal carrier Green's function of $\nu$} \cite{stahl_totik_1992}.
\end{remark}

Under the regularity condition of Definition~\ref{def:reg}, we obtain results analogous to \cref{eq:lin-derivatives}:
\begin{theorem}\label{thm:BOP-der}
    Suppose that $\bm u$ satisfies Definition~\ref{def:config} and $\ell \in \Lambda$ is such that $D_\ell \in \mathbf{Reg}$. Then, with the notation of Theorem~\ref{thm:BOP}, we in addition have
    \[
        \bigg|
            \frac{\partial}{\partial \s_m}
            \Big(
                O^\beta_\ell(\s) 
                - \ctNonlin\big( \Ham_{\ell\ell}, [\Ham^2]_{\ell\ell}, \dots, [\Ham^{N}]_{\ell\ell} \big)
            \Big)
        \bigg|
        \lesssim e^{-\frac{1}{2}\gamma_N N} e^{-\eta \,r_{\ell m}}.
    \]
\end{theorem}

More generally, if the regularity assumption is not satisfied, it may still be the case that Theorem~\ref{thm:BOP-der} holds but with reduced locality exponent $\eta$. To formulate this result, we require the notion of \textit{minimal carrier capacity}:
\begin{definition}[Minimal carrier capacity]
    For arbitrary Borel sets $C$, the \textit{capacity of $C$} is defined as
    \[
        \mathrm{cap}(C) \coloneqq \sup \{ \mathrm{cap}(K) \colon K \subset C, \mathrm{compact}\},
    \]
    where $\mathrm{cap}(K)$ is defined as in \S\ref{pf:pot-theory}.

    For a unit measure $\nu$ with compact support $E \coloneqq \mathrm{supp}\,\nu \subset \mathbb R$, the set of \emph{carriers of $\nu$} and the \emph{minimal carrier capacity} are defined as
    \begin{gather}
        \Gamma(\nu) \coloneqq \{ C \subset \mathbb C \colon C\text{ Borel and } \nu( \mathbb C \setminus C ) = 0 \}, \quad \text{and} \\
        c_\nu \coloneqq \inf\{ \mathrm{cap}(C) \colon C \in \Gamma(\nu), C \text{ bounded} \} \leq \mathrm{cap}(E),
    \end{gather}
    respectively.
\end{definition}

Under these definitions, we have the following \cite[p.~8-10]{stahl_totik_1992}:
\begin{remark} For a unit measure $\nu$ with compact support $E \coloneqq \mathrm{supp}\,\nu \subset \mathbb R$, we have 

    \textit{(i)} The set of \emph{minimal carriers} $\Gamma_0(\nu) \coloneqq \{ C \in \Gamma(\nu) \colon \mathrm{cap}(C) = c_\nu, C \subset E\}$ is nonempty,
    
    \textit{(ii)} If $c_\nu >0$, then there exists a \emph{minimial carrier equilibrium distribution $\omega_\nu$}, a (uniquely defined) unit measure with $\mathrm{supp} \,\omega_\nu \subset E$ satisfying
    \[
        g_\nu(z) = - \int \log \frac{1}{|z - t|} \mathrm{d}\omega_\nu(t) - \log c_\nu,
    \]
    
    \textit{(iii)} $g_\nu \equiv g_E$ if and only if $c_\nu = \mathrm{cap}(E)$,
    
    \textit{(iv)} In particular, if $c_\nu = \mathrm{cap}(E)$, then $\nu \in \mathbf{Reg}$ (although the converse is false \cite[Example~1.5.4]{stahl_totik_1992}),
    
    \textit{(v)} Suppose $c_\nu > 0$. Then, on defining $\nu_n$ to be the discrete unit measure giving equal weight to each of the zeros of $p_n(\,\cdot\,;\nu)$, the condition that  
    \[
        \nu_n \rightharpoonup^\star \omega_E,
    \]    
    where $\omega_E$ is the equilibrium distribution for $E$, is equivalent to $\nu\in\mathbf{Reg}$ \cite[Thm.~3.1.4]{stahl_totik_1992}. In particular, this justifies \cref{eq:limitF}.
\end{remark}

We therefore arrive at the corresponding result for $\ell \in \Lambda$ for which the corresponding LDOS has positive minimal carrier capacity:
\begin{prop}\label{prop:BOP}
    Suppose that $\bm u$ satisfies Definition~\ref{def:config} and $\ell \in\Lambda$ such that $c_{D_\ell} > 0$. Then, with the notation of Theorem~\ref{thm:BOP}, we in addition have
    \[
        \bigg|
            \frac{\partial}{\partial \s_m}
            \Big(
                O^\beta_\ell(\s) 
                - \ctNonlin\big( \Ham_{\ell\ell}, [\Ham^2]_{\ell\ell}, \dots, [\Ham^{N}]_{\ell\ell} \big)
            \Big)
        \bigg|
        \lesssim e^{- \frac{1}{2}\gamma_{N} N} e^{-\eta_\ell \,r_{\ell m}}
    \]
    where $\eta_\ell > 0$, 
    \[
        %
        \eta_{\ell} \to \eta \quad
        \mathrm{as} \text{ } c_{D_\ell} \to \mathrm{cap}(\mathrm{supp}\,D_\ell),
    \]
    and $\eta > 0$ is the constant from Theorem~\ref{thm:BOP-der}.
\end{prop}

The proofs of Theorem~\ref{thm:BOP-der} and Proposition~\ref{prop:BOP} follow from the following estimates on the derivatives of the recursion coefficients $\{a_n, b_n\}$, and the locality of the tridiagonal operators $T_N$, together with the asymptotic upper bounds (i.e.~Definition~\ref{def:reg} or Remark~\ref{rem:nrootasymp}).
\begin{lemma}\label{eq:bound_der_coeff}
    Suppose $\bm u$ satisfies Definition~\ref{def:config}. Then, for a simple closed positively oriented contour $\mathscr C^\prime$ encircling the spectrum $\sigma\big( \Ham(\s) \big)$, there exists $\eta = \eta(\mathscr C^\prime) > 0$ such that 
    \begin{align}
        \bigg|
            \frac
                {\partial b_n}
                {\partial \s_m}
        \bigg|
        &\leq C \|p_n\|^2_{L^\infty(\mathscr C^\prime)} 
        e^{-\eta \, r_{\ell m} } 
        \qquad \text{and} \qquad 
        \bigg|
            \frac
                {\partial a_n}
                {\partial \s_m}
        \bigg|
        \leq C \sum_{l=0}^n \|p_l\|^2_{L^\infty(\mathscr C^\prime)} e^{-\eta \, r_{\ell m} } 
    \end{align}
    where $\eta \sim \mathfrak{d}$ as $\mathfrak{d} \to 0$ where $\mathfrak{d} \coloneqq \mathrm{dist}\big( \mathscr C^\prime, \sigma\big( \Ham(\s^\mathrm{ref}) \big) \big)$.
\end{lemma}

In the following, we denote by $T_\infty$ the infinite symmetric matrix on $\mathbb N_0$ with diagonal $(a_n)_{n \in \mathbb N_0}$ and off-diagonal $(b_n)_{n\in\mathbb N}$.

\begin{lemma}\label{lem:decay:T}
Fix $N \in \mathbb N \cup \{\infty\}$. Suppose that $z \in \mathbb C$ with $\mathfrak{d}_N \coloneqq  \mathrm{dist}\big( z, \sigma(T_N) \big) > 0$. Then, for each $i, j \in \mathbb N_0$, we have
\[
     \left| 
        (T_N - z)^{-1}_{ij} 
    \right| 
    \leq C e^{-\gamma_{|i-j|,N} |i - j|}.
\]

\textit{(i)} For each $r \in \mathbb N$, we have $\gamma_{r,N} \sim \mathfrak{d}_N$ as $\mathfrak{d}_N \to 0$.

\textit{(ii)} We have $\lim_{r\to\infty} \gamma_{r,\infty} = \lim_{N\to\infty} \gamma_{N,N} = g_{\sigma(T_\infty)}(z)$ where $g_{\sigma(T_\infty)}$ is the Green's function for the set $\sigma(T_\infty)$ as defined in \cref{eq:limitF}. 
\end{lemma}

\begin{remark}
    The fact that $g_{\sigma(T_\infty)}$ does not depend on the discrete eigenvalues of $T_\infty$ means that asymptotically the locality estimates do not depend on defect states in the band gap arising due to perturbations satisfying Proposition~\ref{prop:pert_spec}, for example. Indeed, this has been shown more generally for operators with off-diagonal decay \cite{ChenOrtnerThomas2019:locality}. We show an alternative proof using logarithmic potential theory. 
\end{remark}

We will assume Lemmas~\ref{lem:decay:T} and \ref{eq:bound_der_coeff} for now and return to their proofs below. 

We first add on a constant multiple of the identity, $c I$, to the operators $\{T_N\}$ so that the spectra are contained in an interval bounded away from $\{0\}$. Moreover, we translate the integrand by the same constant: $\widetilde{O}(z) \coloneqq O(z - c)$. Then, we extend $T_N$ to an operator on $\ell^2(\mathbb{N}_0)$ by defining $[T_N \psi]_i = \sum_{j=0}^N [T_{N}]_{ij} \psi_j$ for $0\leq i \leq N$ and $[T_N \psi]_i =0$ otherwise. We therefore choose a simple closed contour (or system of contours) $\mathscr C$ encircling $\bigcup_{N} \sigma(T_N)$ so that 
\begin{align}\label{eq:error_derivatives}
    \frac
        {\partial \big[ O_\ell(\s) - O(T_N)_{00} \big]}
        {\partial \s_m} &= 
    \frac{1}{2\pi i}
    \oint_\mathscr{C} \widetilde{O}(z) 
    \frac
        {\partial}{\partial \s_m} 
    \Big[ 
        (T_\infty - z)^{-1}_{0,N+1} b_{N+1} (T_{N} - z)^{-1}_{N0}
    \Big]
    \mathrm{d}z \\
    &= \frac{1}{2\pi i} \oint_\mathscr{C} \widetilde{O}(z) \Big[ 
        \big[ (T_\infty - z)^{-1} 
            \frac
                {\partial T_\infty}
                {\partial \s_m}  
            (T_\infty - z)^{-1}  \big]_{0,N+1} b_{N+1} (T_{N} - z)^{-1}_{N0} \\
         &\qquad\qquad + (T_\infty - z)^{-1}_{0,N+1} 
         \frac
            {\partial b_{N+1}}
            {\partial \s_m}  
        (T_{N} - z)^{-1}_{N0} \\
         &\qquad\qquad + (T_\infty - z)^{-1}_{0,N+1} 
         b_{N+1} 
         \big[ 
            (T_N - z)^{-1} 
            \frac
                {\partial T_N}
                {\partial \s_m}  
            (T_N - z)^{-1}  
        \big]_{N0}
    \Big]
    \mathrm{d}z. 
\end{align}

Therefore, applying Lemma~\ref{eq:bound_der_coeff}, a simple calculation reveals
\begin{align}
    \bigg|
        \frac
            {\partial \big[ O_\ell(\s) - O(T_N)_{00} \big]}
            {\partial \s_m}
    \bigg| &\leq C
    \sum_{n=0}^\infty 
    \Big[ \Big|
        \frac
            {\partial a_n}
            {\partial \s_m} 
    \Big| + 
    \Big|
        \frac
            {\partial b_n}
            {\partial \s_m} 
    \Big|
    \Big]
    e^{-\min\{\gamma_{n,N}, \gamma_{n,\infty}\} n}  e^{-\min\{\gamma_{N,N}, \gamma_{N+1,\infty}\} N}
    \label{eq:der_error2} \\
    &\leq C 
    \bigg[ 
        \sum_{n=0}^\infty 
        \sum_{l=0}^n \|p_l\|^2_{L^\infty(\mathscr C^\prime)} 
        e^{-\min\{\gamma_{n,N}, \gamma_{n,\infty}\} n} 
    \bigg]
    e^{-\min\{\gamma_{N,N}, \gamma_{N+1,\infty}\} N}
    e^{-\eta \, r_{\ell m} } 
\end{align}
where $\gamma_{r,N} = \gamma_{r,N}(\mathscr C)$ is the constant from Lemma~\ref{lem:decay:T}. We therefore may conclude by choosing $\mathscr C^\prime \coloneqq \{ g_E = \gamma \}$ if $D_\ell \in \mathbf{Reg}$ and $\mathscr C^\prime \coloneqq \{ g_{D_\ell} = \gamma \}$ otherwise for some constant $\gamma > 0$ sufficiently small such that the summation in the square brackets converges. 

\begin{proof}[Proof of Lemma~\ref{eq:bound_der_coeff}]
The proof follows from the following identities:
\begin{align}\label{eq:der_b}
        \frac
            {\partial ( b_{n}^2 ) }
            {\partial \s_m}
        &= 
        \oint_{\mathscr C} b_n^2 p_{n}(z)^2 
        \Big[
            (\Ham - z)^{-1} \frac
                    {\partial \Ham(\s)}
                    {\partial \s_m}  (\Ham - z)^{-1}
        \Big]_{\ell\ell} \frac{\mathrm{d}z}{2\pi i} \quad \text{and} \\
        \label{eq:der_a}
        \frac
            {\partial a_{n} }
            {\partial \s_m} &= \oint_{\mathscr C}
        \Big( 
        (z - a_n) p_n(z)^2
        +\sum_{k=0}^{n-1} (-1)^{n-k}
        (2z - 3a_k) p_k(z)^2
        \Big)
        \Big[
            (\Ham - z)^{-1} \frac
                    {\partial \Ham(\s)}
                    {\partial \s_m}  (\Ham - z)^{-1}
        \Big]_{\ell\ell} \frac{\mathrm{d}z}{2\pi i}.
\end{align}

To do this, it will be convenient to renormalise the orthogonal polynomials as in Remark~\ref{rem:recurrence-modified} (that is, we consider $P_n(x) \coloneqq b_n p_n(x)$). Moreover, we define $b_{-1} \coloneqq 1$. 
    %
    %
    %
   %
   %
%
Using the shorthand $\partial \coloneqq \frac{\partial}{\partial \s_m}$, we therefore obtain: $\partial b_{-1} = \partial b_0 = 0$,  $\partial P_{-1}(x) = \partial P_0(x) = 0$, and 
\begin{gather}\label{recurrence-modified_der}
    \partial P_{n+1}(x) 
    = \frac{x-a_n}{b_n} \partial P_n(x) - 
    \frac{b_n}{b_{n-1}} \partial P_{n-1}(x)
    - \partial \Big( \frac{a_n}{b_n} \Big) P_n(x)
    - \partial \Big( \frac{b_n}{b_{n-1}} \Big) P_{n-1}(x),
\end{gather}
for all $n \geq 0$. 

By noting $\partial P_1(x) = - \partial a_0$ and applying \cref{recurrence-modified_der}, we can see that $\partial P_{n}$ is a polynomial of degree $n-1$ for all $n \geq 0$. Therefore, since $P_{n}$ is orthogonal to all polynomials of degree $n-1$, we have
\begin{align}
    \partial( b_{n}^2 ) &= 
    2 \int P_{n}(x) \partial P_{n}(x) \mathrm{d}D_\ell + 
    \oint_{\mathscr C} P_{n}(z)^2 
    \Big[
        (\Ham - z)^{-1} \frac
                {\partial \Ham}
                {\partial \s_m}  (\Ham - z)^{-1}
    \Big]_{\ell\ell} \frac{\mathrm{d}z}{2\pi i} \\
    &= 
    \oint_{\mathscr C} P_{n}(z)^2 
    \Big[
        (\Ham - z)^{-1} \frac
                {\partial \Ham}
                {\partial \s_m}  (\Ham - z)^{-1}
    \Big]_{\ell\ell} \frac{\mathrm{d}z}{2\pi i}
\end{align}
which concludes the proof of \cref{eq:der_a}.

    %
    %

To prove a similar formula for the derivatives of $a_{n}$, we first state a useful identity which will be proved after the conclusion of the proof of \cref{eq:der_a}:
\begin{align}\label{eq:x_partial_p}
    x\partial P_n(x) = \sum_{k=0}^n c_{nk} P_k(x), 
    \quad \text{where} \quad 
    c_{nn} = \sum_{k=0}^{n-1} \Big( a_k \frac{\partial b_k}{b_k} - \partial a_k \Big). 
\end{align}
Therefore, we have
\begin{align}
    \partial a_{n} &= \frac{1}{b_{n}^2} \Big( \oint_{\mathscr C} z P_n(z)^2 
    \Big[
        (\Ham - z)^{-1} \frac
                {\partial \Ham}
                {\partial \s_m}  (\Ham - z)^{-1}
    \Big]_{\ell\ell} \frac{\mathrm{d}z}{2\pi i}
    + 2\int x P_{n}(x) \partial P_n(x) \mathrm{d}D_\ell(x) 
    \Big)
    - a_n \frac{\partial( b_n^2 )}{b_n^2} \nonumber\\
    &= \frac{1}{b_{n}^2} \oint_{\mathscr C} (z - a_n) P_n(z)^2 
    \Big[
        (\Ham - z)^{-1} \frac
                {\partial \Ham}
                {\partial \s_m}  (\Ham - z)^{-1}
    \Big]_{\ell\ell} \frac{\mathrm{d}z}{2\pi i}
   + \sum_{k=0}^{n-1}  \Big( a_k \frac{\partial (b_k^2)}{b_k^2} - 2\partial a_k \Big). \label{eq:der_a2}
\end{align}
Applying \cref{eq:x_partial_p} for $k \leq n-1$, we can see that $\partial a_n$ can be written as 
\[
    \partial a_n = 
    \oint_{\mathscr C} 
    \Big(
        (z - a_n) p_n(z)^2 + \sum_{k=0}^{n-1}( d_{1,k} z + d_{0,k} ) p_k(z)^2 
    \Big)
    \Big[
        (\Ham - z)^{-1} \frac
                {\partial \Ham}
                {\partial \s_m}  (\Ham - z)^{-1}
    \Big]_{\ell\ell} \frac{\mathrm{d}z}{2\pi i}.
\]
for some coefficients $d_{1,k}, d_{0,k}$. Using \cref{eq:x_partial_p} and assuming the result for $k \leq n-1$, we have
\begin{align}
    d_{1,k} z + d_{0,k} &= a_k - 2(z - a_k) - 2 \sum_{l = k+1}^{n-1} (-1)^{l-k}(2z - 3a_k) \\
    &= -2 z + 3a_k - (-1)^{k}\big( (-1)^{k+1} + (-1)^{n-1} \big) (2z - 3a_k) \\
    &= (-1)^{n-k}(2z - 3a_k).
\end{align}
for all $k \leq n-1$.
\end{proof}

\begin{proof}[Proof of \cref{eq:x_partial_p}]
We have
\begin{align}
    x \partial P_{n}(x) &= 
     \frac{x}{b_{n-1}} x\partial P_{n-1}(x) 
     - \partial \Big( \frac{a_{n-1}}{b_{n-1}} \Big) x P_{n-1}(x)
     +\mathrm{l.o.t.}
     %
    %
    %
    \\
    &= \frac{1}{b_{n-1}} c_{n-1,n-1} x P_{n-1}(x)  
    - \partial \Big( \frac{a_{n-1}}{b_{n-1}} \Big) b_{n-1} P_n(x)
     +\mathrm{l.o.t.} \\
    &=  c_{n-1,n-1} P_{n}(x)  
    - \partial \Big( \frac{a_{n-1}}{b_{n-1}} \Big) b_{n-1} P_n(x)
     +\mathrm{l.o.t.}
\end{align}
where $\mathrm{l.o.t.}$ (``lower order term'') denotes a polynomial of degree strictly less than $n$ that changes from one line to the next. That is, since $c_{11} = -\partial a_0 = \partial\big( \frac{a_0}{b_0}\big) b_0$, we apply an inductive argument to conclude 
\[
    c_{nn} = c_{n-1,n-1} - \partial \Big( \frac{a_{n-1}}{b_{n-1}} \Big) b_{n-1}
    = -\sum_{k=0}^{n-1} \partial \Big( \frac{a_k}{b_k} \Big) b_k
    = \sum_{k=0}^{n-1} \Big( a_k \frac{\partial b_k}{b_k} - \partial a_k \Big)= \sum_{k=0}^{n-1} \Big( \frac{a_k}{2}  \frac{\partial (b_k^2)}{b_k^2} - \partial a_k \Big).
\]
\end{proof}

\begin{proof}[Proof of Lemma~\ref{lem:decay:T}]
    The first statement is the Combes-Thomas resolvent estimate (Lemma~\ref{lem:CT}) for tridiagonal operators (which, in particular, satisfy the off-diagonal decay assumptions of Lemma~\ref{lem:CT}). 
    
    To obtain the asymptotic estimates of \textit{(ii)}, we apply a different approach based on the banded structure of the operators. Since $T_N$ is tri-diagonal, $[(T_N)^n]_{ij} = 0$ if $|i - j| > n$. Therefore, for any polynomial $P$ of degree at most $|i-j|-1$, we have \cite{Benzi2013}
    \begin{align}\label{pf:tridiag_locality}
        | (T_N - z)^{-1}_{ij} | 
        &=  \left| \big[ (T_N - z)^{-1} - P(T_N) \big]_{ij} \right| 
        \leq 
        \Big\| \frac{1}{\,\cdot - z} - P\Big\|_{L^\infty( \sigma(T_N) )}.
    \end{align}
    We may apply the results of logarithmic potential theory (see \cref{eq:optimal_poly_approx}), to conclude. Here, it is important that $\left|\sigma(T_\infty) \setminus \sigma(T_N)\right|$ remains bounded independently of $N$ so that, asymptotically, \cref{pf:tridiag_locality} has exponential decay with exponent $g_{\sigma(T_\infty)}$.
    
    The proof that $\left|\sigma(T_\infty) \setminus \sigma(T_N)\right|$ is uniformly bounded can easily be shown when considering the sequence of orthogonal polynomials generated by $T_\infty$. A full proof is given in parts~\textit{(ii)} and~\textit{(iv)} of Lemma~\ref{lem:quadrature}. 
\end{proof}

\section{Quadrature Method}
\label{app:quadrature}
The quadrature method as outlined in this section was introduced in \cite{Nex1978} to approximate the LDOS. For a comparison of various nonlinear approximation schemes, see \cite{Haydock1984:comparison} and \cite{Glanville1988}. The former is a practical comparison of quadrature and BOP methods, while the later also discusses the maximum entropy method \cite{Mead1984}.

We now give an alternative proof of Theorem~\ref{thm:BOP} by introducing the quadrature method \cite{Nex1978}. 

Recall that $D_\ell$ is the {local density of states} (LDOS) satisfying \cref{eq:LDOS} and $\{p_n\}$ is the corresponding sequence of orthogonal polynomials generated via the recursion method:
\[
    [p_n(\Ham)p_m(\Ham)]_{\ell\ell} 
    = \int p_n(x) p_m(x) \mathrm{d}D_\ell(x) = \delta_{nm}
\]  
(see the proof of Lemma~\ref{lem:quadrature}, below).

We use the set of zeros of $p_{N+1}$, denoted by $X_N = \{ \ep_0,\dots,\ep_N \}$, as the basis for the following quadrature rule:
\begin{gather}\label{eq:quadrature}
    O(\Ham)_{\ell\ell} = \int O(x)  \mathrm{d}D_\ell(x)
    \approx \int I_{X_N} O(x)  \mathrm{d}D_\ell(x)
    = \sum_{j = 0}^N w_j O(\ep_j), \qquad \text{where}\\
    w_j = \int\ell_j(x)  \mathrm{d}D_\ell(x) = \ell_j(\Ham)_{\ell\ell}, \quad \text{and} \quad 
    \ell_j(x) = \prod_{i\not=j} \frac{x - \ep_i}{\ep_j - \ep_i}.\nonumber
\end{gather}
Here, $\ell_j$ is the polynomial of degree $N$ with $\ell_j(\ep_i) = \delta_{ij}$. 

The following lemma highlights the fundamental properties of Gauss quadrature and allows us to show that the approximation scheme given by
\begin{align}\label{eq:non-lin:quad}
    \ctNonlin^\mathrm{q}\Big( \Ham_{\ell\ell}, [\Ham^2]_{\ell\ell}, \dots, [\Ham^{2N+1}]_{\ell\ell} \Big) \coloneqq
    \sum_{j=0}^N w_j O(\ep_j). 
\end{align}
satisfies Theorem~\ref{thm:BOP}:
\begin{lemma}\label{lem:quadrature}
Suppose that $\{p_n\}$ is the sequence of polynomials generated by the recursion method \cref{eq:recurrence}, $X_N$ is the set of zeros of $p_{N+1}$, and $\{w_j\}$ are the weights satisfying 
$\int I_{X_N} O(x) \mathrm{d}D_\ell(x) = \sum_{j=0}^N w_j O(\ep_j)$.
Then,
\renewcommand{\theenumi}{\roman{enumi}}%
\begin{enumerate}
    \item $\{p_n\}$ is orthonormal with respect to $D_\ell$: $\int p_n(x) p_m(x) \mathrm{d}D_\ell(x) = [p_n(\Ham)p_m(\Ham)]_{\ell\ell} = \delta_{nm}$, 

    \item $X_N = \sigma(T_N)$ where $T_N$ is given by \cref{eq:tridiag},

    \item $X_N\subset \mathbb R$ is a set of $N+1$ distinct points,
    
    \item If $[a,b] \cap \mathrm{supp}\, D_\ell = \emptyset$, then the number of points in $X_N \cap [a,b]$ is at most one,
    
    \item If $P_{2N+1}$ is a polynomial of degree at most $2N+1$, then
    $
        P_{2N+1}(\Ham)_{\ell\ell} = \sum_{j=0}^N w_j P_{2N+1}(\ep_j),
    $
    
    \item The weights $\{w_j\}$ are positive and sum to one.
\end{enumerate}
\end{lemma}
\begin{proof}
    The idea behind the proofs are standard in the theory of Gauss quadrature (e.g.~see \cite{Freud:OP}) but, for the convenience of the reader, they are collected together in \ref{sec:quad}.
\end{proof}

\begin{remark}\label{rem:quadrature}
    The quadrature rule discussed in this section can be seen as the exact integral with respect to the following approximate LDOS
    \[
        D_\ell^{N,\mathrm{q}} \coloneqq \sum_{j = 0}^N w_j \delta(\,\cdot - \ep_j).
    \]
    This measure has unit mass by Lemma~\ref{lem:quadrature}~\textit{(vi)}, and, by Lemma~\ref{lem:quadrature}~\textit{(v)}, the first $2N+1$ moments of $D_\ell^{N,\mathrm{q}}$ are given by $[\Ham^n]_{\ell\ell}$ for $n = 1,\dots, 2N+1$.
\end{remark}

In the following two sections we prove error estimates and show that the functional form is analytic on an open set containing $\big(\Ham_{\ell\ell}, \dots, [\Ham^{2N+1}]_{\ell\ell}\big)$.

\subsection{Error Estimates.} 
Applying Remark~\ref{rem:quadrature}, together with \cref{eq:general_estimate}, we have: for every polynomial $P_{2N+1}$ of degree at most $2N+1$, 
\begin{align}\label{eq:Gauss-error}
   \Big| O_\ell(\bm r) - \sum_{j=0}^N w_j O(\ep_j) \Big|
   %
   %
    %
    &\leq 2 \| O - P_{2N+1} \|_{L^\infty( \sigma(\Ham) \cup X_N)}.
\end{align}
Now, since $\sigma\big(\Ham\big) \subset I_- \cup \{\lambda_j\} \cup I_+$ where $\{\lambda_j\}$ is a finite set, we may apply part \textit{(iv)} of Lemma~\ref{lem:quadrature} to conclude that the number of points in $X_N \setminus \big( I_- \cup I_+ \big)$ is bounded independently of $N$. Accordingly, we may apply \cref{eq:optimal_poly_approx} with $E = I_- \cup I_+$, to obtain the following asymptotic bound
\[
    \lim_{N\to\infty} 
    \Big| O_\ell(\bm r) - \sum_{j=0}^N w_j O(\ep_j) \Big|^{1/N}
    \leq e^{- \gamma^\star} 
    \quad \text{where} \quad 
    O \text{ is analytic on } \{ z \colon g_{E}(z) < \gamma^\star\}.
\]
In particular, we obtain the stated asymptotic behaviour.

\begin{remark}[Spectral pollution]\label{rem:spectral_pollution}
    While $\sigma(\Ham) \subset \liminf_{N\to\infty} \sigma(T_N)$, we do not claim that the sequence $\sigma(T_N)$ is free from spurious eigenvalues. That is, there may exist sequences $\lambda_N \in \sigma(T_N)$ such that $\lambda_N \to \lambda$ along a subsequence and $\lambda \not\in \sigma(\Ham)$. Indeed, there exist measures supported on a union of disjoint intervals $[a,b]\cup[c,d]$ for which the corresponding sequences of orthogonal polynomials suffer from spurious eigenvalues at every point of the gap $(b,c)$ \cite{Denisov2003, stahl_totik_1992}. In this paper, we only require the much milder property that the number of eigenvalues in the gap remains uniformly bounded in the limit $N\to\infty$. 
    
    For a more general discussion of spectral pollution, see \cite{Lewin2009,CancesEhrlacherMaday2012}.
\end{remark}

\subsection{Analyticity.} To conclude the proof of Theorem~\ref{thm:BOP}, we show that $\ctNonlin^\mathrm{q}$ as defined in \cref{eq:non-lin:quad} is analytic in a neighbourhood of $(\Ham_{\ell\ell}, [\Ham^2]_{\ell\ell}, \dots, [\Ham^{2N+1}]_{\ell\ell})$. Recall that in \cref{eq:complexTri} we have extended the definition of $T_N$ to an analytic function on $U \coloneqq \{ \bm z \in \mathbb C^{2N+1} \colon b_{n}^2(\bm z_{1:2n}) \not= 0 \,\, \forall n = 1,\dots,N \}$.

We define $X_N(\bm z)$ to be the set of eigenvalues of $T_N(\bm z)$. Since $X_N = X_N\big(\Ham_{\ell\ell}, \dots, [\Ham^{2N+1}]_{\ell\ell}\big)$ is a set of $N+1$ distinct points (Lemma~\ref{lem:quadrature}~\textit{(iii)}), there exists a continuous choice of eigenvalues $X_N(\bm z) = \{ \ep_0(\bm z), \dots, \ep_N(\bm z)\}$ such that $X_N(\bm z)$ is a set of $N+1$ distinct points in a neighbourhood, $U_0$, of $(\Ham_{\ell\ell},\dots,[\Ham^{2N+1}]_{\ell\ell})\in U$ and each $\ep_j$ is analytic on $U_0$ \cite{bk:Kato1995,Tsing1994}. With this in hand, we define $\ctNonlin^\mathrm{q} \colon U_0 \to \mathbb C$ by 
\begin{align}
    \label{eq:g_quad}
    \ctNonlin^\mathrm{q}(\bm z) &\coloneqq 
    \mathcal{L} \big( x \mapsto I_{X(\bm z)}O(x) \big)(\bm z_{1:N})
    =\sum_{j=0}^N 
    \mathcal{L}\Big( x \mapsto
        \prod_{i\not=j} \frac{x-\ep_i}{\ep_j - \ep_i}
    \Big) \cdot 
    O \circ \ep_j,
\end{align}
which is analytic on $\{ \bm z \in U_0 \colon O \text{ analytic at } \ep_j(\bm z) \,\,\forall j=0,\dots,N\}$.

\subsection{Proof of Lemma~\ref{lem:quadrature}}\label{sec:quad}
\textit{Proof of (i).} First note that $\int p_0 p_1 \mathrm{d}D_\ell = 0$. We assume that $p_0, \dots, p_n$ are mutually orthogonal with respect to $D_\ell$, and note that,
\begin{align}
    b_1 = b_1 \int p_1^2 \mathrm{d}D_\ell 
    &= \int (x - a_0) p_1(x) \mathrm{d}D_\ell(x)
    = \int x p_0(x) p_1(x) \mathrm{d}D_\ell(x), \qquad\text{and} \nonumber
    \\
    b_{n} = b_{n} \int p_{n}^2 \mathrm{d}D_\ell
    &= \int  \big( (x-a_{n-1})p_{n-1}(x)p_{n}(x) - b_{n-1} p_{n-2}(x)p_{n}(x) \big) \mathrm{d}D_\ell(x) \nonumber\\
    &= \int  x p_{n-1}(x)p_{n}(x) \mathrm{d}D_\ell(x)
    \qquad \text{for } n \geq 2. \label{eq:bn}
\end{align}
Therefore, we conclude by noting
\begin{align}
    b_{n+1}\int p_{n+1} p_j \mathrm{d}D_\ell &=
    \int \big( (x-a_n)p_n(x)p_j(x) - b_n p_{n-1}(x)p_j(x) \big) \mathrm{d}D_\ell(x) \\
    &= \begin{cases}
        \int x p_n(x)^2 \mathrm{d}D_\ell - a_n
            &\text{if } j = n, \\
         \int x p_n(x)p_{n-1}(x) \mathrm{d}D_\ell - b_n
            &\text{if } j = n-1 \\
        0   &\text{if } j \leq n-2,\\
    \end{cases}
\end{align}
and applying \cref{eq:bn}. Equation~\cref{eq:bn} also justifies the tri-diagonal structure \cref{eq:tridiag}.

    \textit{Proof of (ii).} We may rewrite the recurrence relation \cref{eq:recurrence} as 
    $x \bm p(x) = T_N \bm p(x) + b_{N+1}p_{N+1}(x) \bm{e}_N$ 
    where 
    $\bm p(x) \coloneqq \big(1, p_1(x), \dots, p_{N}(x) \big)^T$, 
    $[\bm e_N]_j = \delta_{jN}$, and $T_N$ is the tri-diagonal matrix \cref{eq:tridiag}. In particular, each $\ep_j \in X_N$ is an eigenvalue of $T_N$ (with eigenvector $\bm p(\ep_j)$). 
    
    \textit{Proof of (iii).} Since $T_N$ is symmetric, the spectrum is real. Now, for each $\ep_j \in X_N = \sigma(T_N)$, the matrix 
    $(T_N - \ep_j)_{\neg N\neg 0}$
    formed by removing the $N^\textrm{th}$ row and $0^\textrm{th}$ column is lower-triangular with diagonal $(b_1,\dots,b_N)$. Since each $b_i > 0$, $(T_N - \ep_j)_{\neg N\neg 0}$ has full rank and thus $\ep_j$ is a simple eigenvalue of $T_N$.
    
    \textit{Proof of (iv).} Suppose that (after possibly relabelling) $\ep_0, \ep_1 \in X_N \cap [a,b]$. After defining $R(x) \coloneqq \prod_{j = 2}^N (x - \ep_j)$, a polynomial of degree $N-1$, and noting $(x - \ep_0) (x - \ep_1) > 0$ on $\mathrm{supp}\,D_\ell$, we obtain
    \[
        \int p_{N+1}(x) R(x) \mathrm{d}D_\ell(x) = \int R(x)^2 (x - \ep_0) (x - \ep_1) \mathrm{d}D_\ell(x) > 0,
    \]
    contradicting part \textit{(i)}.

    \textit{Proof of (v).} We may write $P_{2N+1} = p_{N+1} q_N + r_N$ where $q_N, r_N$ are polynomials of degree at most $N$ and note that $[p_{N+1}(\Ham)q_N(\Ham)]_{\ell\ell} = 0$ by \textit{(i)} and $P_{2N+1}(\ep_j) = r_N(\ep_j)$ since $X$ is the set of zeros of $p_{N+1}$. Therefore, 
    \begin{align}
        \int P_{2N+1}(x) \mathrm{d}D_\ell(x)
        &= \int \Big[ p_{N+1}(x) q_N(x) + r_N(x) \Big] \mathrm{d}D_\ell(x) \\
        &= \int r_N(x) \mathrm{d}D_\ell(x) 
        = \int I_X r_N(x) \mathrm{d}D_\ell(x) \label{eq:poly_int_poly}\\
        &= \sum_{j=0}^N w_j r_N(\ep_j)
        = \sum_{j=0}^N w_j P_{2N+1}(\ep_j).
    \end{align}
    In \cref{eq:poly_int_poly} we have used the fact that polynomial interpolation in $N+1$ distinct points is exact for polynomials of degree at most $N$. 

   \textit{Proof of (vi).} $\ell_j(x)^2$ is a polynomial of degree $2N$ and so, by \textit{(v)}, we have
    \[
        0 \leq \int \ell_j(x)^2 D_\ell(x) \mathrm{d}x 
        = \sum_{i=0}^N w_i \ell_j(\ep_i)^2
        = w_j.
    \]
    Moreover, $\sum_{j=0}^N \ell_j(x)$ is a polynomial of degree $N$ equal to one on $X_N$ (a set of $N+1$ distinct points) and so $\sum_{j=0}^N \ell_j(x) \equiv 1$. Finally, $\sum_{j=0}^N w_j = \int \big( \sum_{j=0}^N \ell_j(x) \big)  D_\ell(x) \mathrm{d}x = 1$.

\section{Numerical Bond-Order Potentials (BOP)}
\label{app:generalBOP}
In mathematical terms, the idea behind BOP methods is to replace the local density of states (LDOS) with an approximation using only the information from the truncated tri-diagonal matrix $T_N$ (and possibly additional hyper-parameters). Since the first $N$ coefficients contain the same information as the first $2N+1$ moments $\Ham_{\ell\ell}, \dots, [\Ham^{2N+1}]_{\ell\ell}$, this approach is closely related to the method of moments \cite{CyrotLackmann1967moments}. 


Equivalently, the resolvent $[(z - \Ham)^{-1}]_{\ell\ell}$, which can be written conveniently as the continued fraction expansion
\begin{align}\label{eq:CF}
    [(z - \Ham)^{-1}]_{\ell\ell} = 
    \cfrac
        {1}
        {z - a_0 - 
    \cfrac
        {b_1^2}
        {z - a_1 - 
    \cfrac
        {b_2^2}
        {z-a_2-\ddots 
    }}},
\end{align}
is replaced with an approximation $G_\ell^N$ only involving the coefficients from $T_N$. For example, for fixed \textit{terminator} $t_\infty$, we may define
\begin{align}\label{eq:CF_approx}
    G_\ell^N(z) \coloneqq
    \cfrac
        {1}
        {z - a_0 - 
    \cfrac
        {b_1^2}
        {z-a_1 - 
    \dfrac
        {b_2^2}
        {\hspace{.25cm}\ddots\hspace{.25cm} - 
    \cfrac
        {b_N^2}
        {z-a_N - t_{\infty}(z)}
    }}}.
\end{align}

Truncating \cref{eq:CF} to level $N$, which is equivalent to replacing the far-field of the linear chain with vacuum and choosing $t_\infty = 0$, results in a rational approximation to the resolvent and thus a discrete approximation to the LDOS. We have seen that truncation of the continued fraction in this way leads to an approximation scheme satisfying Theorem~\ref{thm:BOP}.

Alternatively, the far-field may be replaced with a constant linear chain with $a_{N+j} = a_\infty$ and $b_{N+j} = b_{\infty}$ for all $j \geq 1$ leading to the square root terminator 
$t_\infty(z) = \frac{b_\infty^2}{z-a_\infty - t_\infty(z)}$
\cite{bk:finnis,Turchi1982:asympototiccoeffs,Haydock1972:recursion}.

More generally, one may choose any ``approximate'' local density of states $\widetilde{D}_\ell$ and construct a corresponding terminator that encodes the information from $\widetilde{D}_\ell$ \cite{Haydock1985:generalterminator, Luchini1987}. 
For example, $\widetilde{D}_\ell(x) \coloneqq \frac{1}{b_\infty \pi} \sqrt{1 - \big(\frac{x - a_\infty}{2b_\infty}\big)^2}$ results in the square root terminator.
While we are unaware of any rigorous results, there is numerical evidence \cite{Haydock1985:generalterminator} to suggest that the error in the approximation scheme is related to the smoothness of the difference $D_\ell - \widetilde{D}_\ell$.

Equivalently, we may choose any bounded symmetric tri-diagonal (Jacobi) operator $\widetilde{T}_N$ with diagonal $a_0, a_1, \dots, a_N, \widetilde{a}_{N+1}, \dots$ and off-diagonal $b_1,\dots,b_N,\widetilde{b}_{N+1},\dots$. That is, we may evaluate the recursion method exactly to level $N$ and append the far-field boundary condition $\{\widetilde{a}_n, \widetilde{b}_n\}_{n \geq N+1}$ to the semi-infinite linear chain. 
This approach also includes the case $t_\infty = 0$ as in \S\ref{pf:BOP} by choosing $\widetilde{a}_n = \widetilde{b}_n = 0$ for all $n$.

With this in hand, we define 
\begin{align}\label{eq:general:BOP}
    {O}_\ell^{N,\mathrm{BOP}}(\s) \coloneqq O(\widetilde{T}_N)_{00} = \int O\, \mathrm{d}\widetilde{D}_\ell^{N,\mathrm{BOP}}
\end{align}
where $\widetilde{D}_\ell^{N,\mathrm{BOP}}$ is the appropriate spectral measure corresponding to $\widetilde{T}_N$.


\subsection{Error Estimates} Since $[(\widetilde{T}_N)^n]_{00} = [(T_N)^n]_{00} = [(T_\infty)^n]_{00}$ is independent of the far-field coefficients $\{\widetilde{a}_j, \widetilde{b}_j\}$ for all $n \leq 2N+1$, we can immediately see that the first $2N+1$ moments of $\widetilde{D}_\ell^{N,\mathrm{BOP}}$ agree with those of $D_\ell$. In particular, we may immediately apply \cref{eq:general_estimate} to obtain error estimates that depend on $\mathrm{supp}\big( D_\ell - \widetilde{D}_\ell^{N,\mathrm{BOP}}\big)$. 

Therefore, as long as the far-field boundary condition is chosen so that there are only finitely many discrete eigenvalues in the band gap independent of $N$, the more complicated BOP schemes converge at least as quickly as the $t_\infty = 0$ case. Intuitively, if the far-field boundary condition is chosen to capture the behaviour of the LDOS (e.g.~the type and location of band-edge singularities), then the integration against the signed measure $D_\ell - \widetilde{D}_\ell^{N,\mathrm{BOP}}$ as in \cref{eq:general_estimate} may lead to improved error estimates. A rigorous error analysis to this affect is left for future work. 

\subsection{Analyticity} Since $\widetilde{T}_N$ is bounded and symmetric, the spectrum $\sigma(\widetilde{T}_N)$ is contained in a bounded interval of the real line. In particular, we can apply the same arguments as in \cref{pf:g_analytic} to conclude that \cref{eq:general:BOP} defines a nonlinear approximation scheme given by an analytic function on an open subset of $\mathbb C^{2N+1}$.

\section{Kernel Polynomial Method \& Analytic Bond Order Potentials}
\label{app:KPM}

We first introduce the Kernel Polynomial Method (KPM) for approximating the LDOS \cite{SilverRoder1994,VoterKressSilver1996,Silver1996}. In this section, we scale the spectrum and assume that $\sigma(\Ham) \subset [-1,1]$.

For a sequence of \textit{kernels} $K_N(x,y)$, we define the approximate quantities of interest 
\begin{align}
    O_\ell^N \coloneqq \int K_N \star O \, \mathrm{d}D_\ell 
    \eqqcolon \iint K_N(x,y) O(y) \,\mathrm{d}y \, \mathrm{d}D_\ell(x).
\end{align}
Under the choice 
$K_N(x,y) \coloneqq \frac{2}{\pi} \sqrt{1-y^2} \sum_{n=0}^N  U_n(x) U_n(y)$
(where $U_n$ denotes the $n^\text{th}$ Chebyshev polynomial of the second kind), we arrive at a projection method similar to that discussed in \S\ref{sec:projection}: if $O(x) = \sum_{m=0}^\infty c_m U_m(x)$, then
\begin{align}
    K_N \star O(x) &= \sum_{m,n} c_m U_n(x) \frac{2}{\pi} \int_{-1}^1 
        U_n(y) U_m(y)
        {\sqrt{1-y^2}}  
    \,\mathrm{d}y 
    = \sum_{n=0}^N c_n U_n(x).
\end{align}
Equivalently, we may consider the corresponding approximate LDOS
\[
    O_\ell^N = \int O(x) D_\ell^N(x) \mathrm{d}x 
    \quad \text{where} \quad 
    D_\ell^N(x) = \frac{2}{\pi} \sqrt{1-x^2} \sum_{n=0}^N U_n(\Ham)_{\ell\ell} U_n(x).
\]

However, truncation of the Chebyshev series in this way leads to artificial oscillations in the approximate LDOS known as Gibbs oscillations \cite{Grafakos2016}. Moreover, without \textit{damping} these oscillations, the approximate LDOS need not be positive. However, on defining
\begin{align}
    K^\mathrm{Fejer}_N(x,y) \coloneqq \frac{1}{N} \sum_{n=1}^N K_n(x,y)
    = \frac{2}{\pi} \sqrt{1-y^2} \sum_{n=0}^N \big(1 - \tfrac{n}{N}\big) U_n(x) U_n(y),
\end{align}
we obtain a positive approximate LDOS \cite{VoterKressSilver1996} where the \textit{damping coefficients} $d_n \coloneqq (1 - \tfrac{n}{N})$ reduce the effect of Gibbs ringing. In practice, one may instead choose the \textit{Jackson kernel} \cite{HammerschmidtEtAl2019}. 

The problem with the above analysis in practice is that the damping factors that we have introduced mean that more moments $[\Ham^n]_{\ell\ell}$ are required in order to obtain good approximations to the LDOS. Instead, analytic BOP methods \cite{Pettifor1989,PettiforDrautz:analyticBOP} compute the first $N$ rows of the tridiagonal operator $T_\infty$, thus obtaining the first $2N+1$ moments exactly. Then, a far-field boundary condition (such as a constant infinite linear chain) is appended to form a corresponding Jacobi operator $\widetilde{T}_N$ as in Appendix~\ref{app:generalBOP}. Now, since higher order moments of $\widetilde{T}_N$ can be efficiently computed, we may evaluate the following approximate LDOS
\begin{align}
    D_\ell^{N,M}(x) &\coloneqq \frac{2}{\pi} \sqrt{1-x^2} \sum_{n=0}^{M} d_n U_n(\widetilde{T}_N)_{00} U_n(x)  
\end{align}
where $d_n$ are damping coefficients and $M > 2N+1$. The damping is chosen so that the lower order moments which are computed exactly and are more important for the reconstruction of the LDOS are only slightly damped. With this choice of kernel, the approximate quantities of interest take the form
\[
    O_\ell^{N,M} = \sum_{n=0}^{2N+1} d_n c_n U_n(\Ham)_{\ell\ell} + \sum_{n = 2N+2}^M d_n c_n U_n(\widetilde{T}_N)_{00}
\]

Efficient implementation of analytic BOP methods can be carried out using the BOPfox program \cite{HammerschmidtEtAl2019}.

        %
%
    %
        %

\bibliography{external/ref}
\bibliographystyle{siam}

\end{document}

%% file: external/preamble.tex
\usepackage[utf8]{inputenc}
\usepackage{ifthen,mathrsfs, amsmath, amssymb, amsthm, braket, bm, graphicx, mathtools,enumitem,color,marginnote}
\usepackage[compress]{cite}
\usepackage{microtype}
\usepackage[english]{babel}
\usepackage[nodayofweek,long]{datetime}
\usepackage[colorlinks=true, allcolors=blue]{hyperref}
\hypersetup{final}
\usepackage[hang,flushmargin]{footmisc}
\usepackage{tikz-cd,upgreek}

\newcommand{\ctNumorbitals}{N_\mathrm{b}}

\newcommand{\ctNonlin}{\Theta}
\newcommand{\Ham}{\mathcal{H}}

\renewcommand{\leq}{\leqslant}\renewcommand{\geq}{\geqslant}
\renewcommand{\above}[2]{\genfrac{}{}{0pt}{}{#1}{#2}}

\newcommand{\ep}{\varepsilon}

\newtheorem*{fact*}{Fact}
\newtheorem{theorem}{Theorem}
\newtheorem*{theorem*}{Theorem}

\newtheorem{definition}{Definition}
\newtheorem{lemma}[theorem]{Lemma}
\newtheorem{prop}[theorem]{Proposition}
\newtheorem{corollary}[theorem]{Corollary}

\theoremstyle{remark}
\newtheorem{remark}{Remark}

\usepackage{cleveref}
\crefformat{equation}{\textup{(#2#1#3)}}
\crefformat{section}{\S#2#1#3} 
\crefformat{subsection}{\S#2#1#3}
\crefformat{subsubsection}{\S#2#1#3}
\crefformat{thm}{Theorem~#2#1#3}

\newenvironment{assumption}[1]{%
    \medskip\par\noindent
    \textbf{#1}\hspace{.1cm}\itshape}
    {\medskip\par}
    
    \usepackage{caption}
    \usepackage{subcaption}
    
    \usepackage{ulem} 

%% file: diagrams/metal_insulator.tex
\tikzset{every picture/.style={line width=0.75pt}} 

\begin{tikzpicture}[x=0.75pt,y=0.75pt,yscale=-1,xscale=1]

\draw  [fill={rgb, 255:red, 0; green, 0; blue, 0 }  ,fill opacity=1 ][line width=2.25]  (413.95,45.1) -- (413.95,55.71) -- (189.28,55.71) -- (189.28,45.1) -- cycle ;
\draw  [fill={rgb, 255:red, 0; green, 0; blue, 0 }  ,fill opacity=1 ][line width=2.25]  (108.61,45.1) -- (108.61,55.71) -- (58.67,55.71) -- (58.67,45.1) -- cycle ;
\draw  [fill={rgb, 255:red, 0; green, 0; blue, 0 }  ,fill opacity=1 ][line width=2.25]  (599.61,45.13) -- (599.61,55.74) -- (471,55.74) -- (471,45.13) -- cycle ;
\draw  [fill={rgb, 255:red, 0; green, 0; blue, 0 }  ,fill opacity=1 ][line width=2.25]  (238.28,112.13) -- (238.28,122.74) -- (59,122.74) -- (59,112.13) -- cycle ;
\draw  [fill={rgb, 255:red, 0; green, 0; blue, 0 }  ,fill opacity=1 ][line width=2.25]  (599.28,112.16) -- (599.28,122.77) -- (385.61,122.77) -- (385.61,112.16) -- cycle ;
\draw  [dash pattern={on 4.5pt off 4.5pt}]  (300.67,15) -- (301.28,150.77) ;
\draw  [draw opacity=0] (651.67,127.45) -- (652,127.45) -- (652,139.78) -- (651.67,139.78) -- cycle ;
\draw  [draw opacity=0] (41.95,127.36) -- (42.29,127.36) -- (42.29,139.7) -- (41.95,139.7) -- cycle ;

\draw (296.25,152.82) node [anchor=north west][inner sep=0.75pt]    {$\mu $};

\end{tikzpicture}

%% file: diagrams/def.tex
\tikzset{every picture/.style={line width=0.75pt}} 

\begin{tikzpicture}[x=0.75pt,y=0.75pt,yscale=-1,xscale=1]

\draw  [fill={rgb, 255:red, 0; green, 0; blue, 0 }  ,fill opacity=1 ][line width=2.25]  (237.39,43.37) -- (237.39,53.98) -- (58.11,53.98) -- (58.11,43.37) -- cycle ;
\draw  [fill={rgb, 255:red, 0; green, 0; blue, 0 }  ,fill opacity=1 ][line width=2.25]  (599.67,43.36) -- (599.67,53.97) -- (386,53.97) -- (386,43.36) -- cycle ;
\draw   (44.13,39.6) -- (251.38,39.6) -- (251.38,57.75) -- (44.13,57.75) -- cycle ;
\draw   (371.13,39.6) -- (609.5,39.6) -- (609.5,57.75) -- (371.13,57.75) -- cycle ;
\draw  [fill={rgb, 255:red, 0; green, 0; blue, 0 }  ,fill opacity=1 ][line width=2.25]  (237.39,108.37) -- (237.39,118.98) -- (58.11,118.98) -- (58.11,108.37) -- cycle ;
\draw  [fill={rgb, 255:red, 0; green, 0; blue, 0 }  ,fill opacity=1 ][line width=2.25]  (599.67,108.36) -- (599.67,118.97) -- (386,118.97) -- (386,108.36) -- cycle ;
\draw  [fill={rgb, 255:red, 0; green, 0; blue, 0 }  ,fill opacity=1 ] (239.75,107.36) -- (240.08,107.36) -- (240.08,119.7) -- (239.75,119.7) -- cycle ;
\draw  [fill={rgb, 255:red, 0; green, 0; blue, 0 }  ,fill opacity=1 ] (257.75,107.36) -- (258.08,107.36) -- (258.08,119.7) -- (257.75,119.7) -- cycle ;
\draw  [fill={rgb, 255:red, 0; green, 0; blue, 0 }  ,fill opacity=1 ] (247.75,107.36) -- (248.08,107.36) -- (248.08,119.7) -- (247.75,119.7) -- cycle ;
\draw  [fill={rgb, 255:red, 0; green, 0; blue, 0 }  ,fill opacity=1 ] (244.75,107.36) -- (245.08,107.36) -- (245.08,119.7) -- (244.75,119.7) -- cycle ;
\draw  [fill={rgb, 255:red, 0; green, 0; blue, 0 }  ,fill opacity=1 ] (241.75,107.36) -- (242.08,107.36) -- (242.08,119.7) -- (241.75,119.7) -- cycle ;
\draw  [fill={rgb, 255:red, 0; green, 0; blue, 0 }  ,fill opacity=1 ] (275.75,107.36) -- (276.08,107.36) -- (276.08,119.7) -- (275.75,119.7) -- cycle ;
\draw  [fill={rgb, 255:red, 0; green, 0; blue, 0 }  ,fill opacity=1 ] (267.75,107.36) -- (268.08,107.36) -- (268.08,119.7) -- (267.75,119.7) -- cycle ;
\draw  [fill={rgb, 255:red, 0; green, 0; blue, 0 }  ,fill opacity=1 ] (293.75,107.36) -- (294.08,107.36) -- (294.08,119.7) -- (293.75,119.7) -- cycle ;
\draw  [fill={rgb, 255:red, 0; green, 0; blue, 0 }  ,fill opacity=1 ] (356,107.36) -- (356.33,107.36) -- (356.33,119.7) -- (356,119.7) -- cycle ;
\draw  [fill={rgb, 255:red, 0; green, 0; blue, 0 }  ,fill opacity=1 ] (349,107.36) -- (349.33,107.36) -- (349.33,119.7) -- (349,119.7) -- cycle ;
\draw  [fill={rgb, 255:red, 0; green, 0; blue, 0 }  ,fill opacity=1 ] (378,107.36) -- (378.33,107.36) -- (378.33,119.7) -- (378,119.7) -- cycle ;
\draw  [fill={rgb, 255:red, 0; green, 0; blue, 0 }  ,fill opacity=1 ] (374,107.36) -- (374.33,107.36) -- (374.33,119.7) -- (374,119.7) -- cycle ;
\draw  [fill={rgb, 255:red, 0; green, 0; blue, 0 }  ,fill opacity=1 ] (380,107.36) -- (380.33,107.36) -- (380.33,119.7) -- (380,119.7) -- cycle ;
\draw  [fill={rgb, 255:red, 0; green, 0; blue, 0 }  ,fill opacity=1 ] (384,107.36) -- (384.33,107.36) -- (384.33,119.7) -- (384,119.7) -- cycle ;
\draw  [fill={rgb, 255:red, 0; green, 0; blue, 0 }  ,fill opacity=1 ] (382,107.36) -- (382.33,107.36) -- (382.33,119.7) -- (382,119.7) -- cycle ;
\draw  [fill={rgb, 255:red, 0; green, 0; blue, 0 }  ,fill opacity=1 ] (606.92,107.45) -- (607.25,107.45) -- (607.25,119.78) -- (606.92,119.78) -- cycle ;
\draw  [fill={rgb, 255:red, 0; green, 0; blue, 0 }  ,fill opacity=1 ] (604.92,107.45) -- (605.25,107.45) -- (605.25,119.78) -- (604.92,119.78) -- cycle ;
\draw  [fill={rgb, 255:red, 0; green, 0; blue, 0 }  ,fill opacity=1 ] (602.92,107.45) -- (603.25,107.45) -- (603.25,119.78) -- (602.92,119.78) -- cycle ;
\draw  [fill={rgb, 255:red, 0; green, 0; blue, 0 }  ,fill opacity=1 ] (600.92,107.45) -- (601.25,107.45) -- (601.25,119.78) -- (600.92,119.78) -- cycle ;
\draw  [fill={rgb, 255:red, 0; green, 0; blue, 0 }  ,fill opacity=1 ] (617.17,107.45) -- (617.5,107.45) -- (617.5,119.78) -- (617.17,119.78) -- cycle ;
\draw  [fill={rgb, 255:red, 0; green, 0; blue, 0 }  ,fill opacity=1 ] (639.17,107.45) -- (639.5,107.45) -- (639.5,119.78) -- (639.17,119.78) -- cycle ;
\draw  [fill={rgb, 255:red, 0; green, 0; blue, 0 }  ,fill opacity=1 ] (29.45,107.36) -- (29.79,107.36) -- (29.79,119.7) -- (29.45,119.7) -- cycle ;
\draw  [fill={rgb, 255:red, 0; green, 0; blue, 0 }  ,fill opacity=1 ] (55.45,107.36) -- (55.79,107.36) -- (55.79,119.7) -- (55.45,119.7) -- cycle ;
\draw  [fill={rgb, 255:red, 0; green, 0; blue, 0 }  ,fill opacity=1 ] (53.45,107.36) -- (53.79,107.36) -- (53.79,119.7) -- (53.45,119.7) -- cycle ;
\draw  [fill={rgb, 255:red, 0; green, 0; blue, 0 }  ,fill opacity=1 ] (49.45,107.36) -- (49.79,107.36) -- (49.79,119.7) -- (49.45,119.7) -- cycle ;
\draw  [fill={rgb, 255:red, 0; green, 0; blue, 0 }  ,fill opacity=1 ] (47.45,107.36) -- (47.79,107.36) -- (47.79,119.7) -- (47.45,119.7) -- cycle ;
\draw   (44.13,104.6) -- (251.38,104.6) -- (251.38,122.75) -- (44.13,122.75) -- cycle ;
\draw   (371.13,104.6) -- (609.5,104.6) -- (609.5,122.75) -- (371.13,122.75) -- cycle ;
\draw  [dash pattern={on 4.5pt off 4.5pt}]  (300.17,10) -- (300.78,145.77) ;
\draw    (254.38,48.6) -- (368.13,48.6) ;
\draw [shift={(371.13,48.6)}, rotate = 180] [fill={rgb, 255:red, 0; green, 0; blue, 0 }  ][line width=0.08]  [draw opacity=0] (7.14,-3.43) -- (0,0) -- (7.14,3.43) -- cycle    ;
\draw [shift={(251.38,48.6)}, rotate = 0] [fill={rgb, 255:red, 0; green, 0; blue, 0 }  ][line width=0.08]  [draw opacity=0] (7.14,-3.43) -- (0,0) -- (7.14,3.43) -- cycle    ;
\draw    (297.88,114.6) -- (345.25,114.6) ;
\draw [shift={(348.25,114.6)}, rotate = 180] [fill={rgb, 255:red, 0; green, 0; blue, 0 }  ][line width=0.08]  [draw opacity=0] (7.14,-3.43) -- (0,0) -- (7.14,3.43) -- cycle    ;
\draw [shift={(294.88,114.6)}, rotate = 0] [fill={rgb, 255:red, 0; green, 0; blue, 0 }  ][line width=0.08]  [draw opacity=0] (7.14,-3.43) -- (0,0) -- (7.14,3.43) -- cycle    ;

\draw (139.3,20.4) node [anchor=north west][inner sep=0.75pt]    {$I_{-}$};
\draw (493.25,19.5) node [anchor=north west][inner sep=0.75pt]    {$I_{+}$};
\draw (295.75,153.82) node [anchor=north west][inner sep=0.75pt]    {$\mu $};
\draw (310.5,52.4) node [anchor=north west][inner sep=0.75pt]    {$\mathsf{g}$};
\draw (305.9,116.4) node [anchor=north west][inner sep=0.75pt]    {$\mathsf{g}^{\mathrm{def}}$};

\end{tikzpicture}